\documentstyle[twoside,epsf,psfig]{article}

\catcode`\@=11
\long\def\@makefntext#1{
\protect\noindent \hbox to 3.2pt {\hskip-.9pt  
$^{{\eightrm\@thefnmark}}$\hfil}#1\hfill}		

\def\@makefnmark{\hbox to 0pt{$^{\@thefnmark}$\hss}}	
	
\def\ps@myheadings{\let\@mkboth\@gobbletwo
\def\@oddhead{\hbox{}
\rightmark\hfil\eightrm\thepage}   
\def\@oddfoot{}\def\@evenhead{\eightrm\thepage\hfil
\leftmark\hbox{}}\def\@evenfoot{}
\def\sectionmark##1{}\def\subsectionmark##1{}}

\oddsidemargin=\evensidemargin
\newcounter{sectionc}\newcounter{subsectionc}\newcounter{subsubsectionc}
\renewcommand{\section}[1] {\vspace{12pt}\addtocounter{sectionc}{1} 
\setcounter{subsectionc}{0}\setcounter{subsubsectionc}{0}\noindent 
	{\tenbf\thesectionc. #1}\par\vspace{5pt}}
\renewcommand{\subsection}[1] {\vspace{12pt}\addtocounter{subsectionc}{1} 
\setcounter{subsubsectionc}{0}\noindent 
{\bf\thesectionc.\thesubsectionc. {\kern1pt \bfit #1}}\par\vspace{5pt}}
\renewcommand{\subsubsection}[1] {\vspace{12pt}\addtocounter{subsubsectionc}{1}
	\noindent{\tenrm\thesectionc.\thesubsectionc.\thesubsubsectionc.
	{\kern1pt \tenit #1}}\par\vspace{5pt}}
\newcommand{\nonumsection}[1] {\vspace{12pt}\noindent{\tenbf #1}
	\par\vspace{5pt}}

\newcounter{appendixc}
\newcounter{subappendixc}[appendixc]
\newcounter{subsubappendixc}[subappendixc]
\renewcommand{\thesubappendixc}{\Alph{appendixc}.\arabic{subappendixc}}
\renewcommand{\thesubsubappendixc}
	{\Alph{appendixc}.\arabic{subappendixc}.\arabic{subsubappendixc}}

\renewcommand{\appendix}[1] {\vspace{12pt}
        \refstepcounter{appendixc}
        \setcounter{figure}{0}
        \setcounter{table}{0}
        \setcounter{lemma}{0}
        \setcounter{theorem}{0}
        \setcounter{corollary}{0}
        \setcounter{definition}{0}
        \setcounter{equation}{0}
        \renewcommand{\thefigure}{\Alph{appendixc}.\arabic{figure}}
        \renewcommand{\thetable}{\Alph{appendixc}.\arabic{table}}
        \renewcommand{\theappendixc}{\Alph{appendixc}}
        \renewcommand{\thelemma}{\Alph{appendixc}.\arabic{lemma}}
        \renewcommand{\thetheorem}{\Alph{appendixc}.\arabic{theorem}}
        \renewcommand{\thedefinition}{\Alph{appendixc}.\arabic{definition}}
        \renewcommand{\thecorollary}{\Alph{appendixc}.\arabic{corollary}}
        \renewcommand{\theequation}{\Alph{appendixc}.\arabic{equation}}
        \noindent{\tenbf Appendix \theappendixc #1}\par\vspace{5pt}}
\newcommand{\subappendix}[1] {\vspace{12pt}
        \refstepcounter{subappendixc}
        \noindent{\bf Appendix \thesubappendixc. {\kern1pt \bfit #1}}
	\par\vspace{5pt}}
\newcommand{\subsubappendix}[1] {\vspace{12pt}
        \refstepcounter{subsubappendixc}
        \noindent{\rm Appendix \thesubsubappendixc. {\kern1pt \tenit #1}}
	\par\vspace{5pt}}

\newcommand{\textlineskip}{\baselineskip=13pt}
\newcommand{\smalllineskip}{\baselineskip=10pt}

\newcommand{\copyrightheading}[1]
	{\vspace*{-2.5cm}\smalllineskip{\flushleft
	{\footnotesize International Journal of Modern Physics D, #1}\\
	{\footnotesize \copyright\kern2pt World Scientific Publishing
	 Company}\\
	 }}

\newcommand{\publisher}[2]{{\begin{center}\footnotesize\smalllineskip 
	Received #1\\
	Revised #2
	\end{center}
	}}

\def\abstracts#1#2#3{{
	\centering{\begin{minipage}{4.5in}\footnotesize\baselineskip=10pt
	\parindent=0pt #1\par 
	\parindent=15pt #2\par
	\parindent=15pt #3
	\end{minipage}}\par}} 

\renewenvironment{thebibliography}[1]
        {\frenchspacing
	 \ninerm\baselineskip=11pt
         \begin{list}{\arabic{enumi}.}
        {\usecounter{enumi}\setlength{\parsep}{0pt}     
	 \setlength{\leftmargin 12.7pt}{\rightmargin 0pt}
         \setlength{\itemsep}{0pt} \settowidth
	{\labelwidth}{#1.}\sloppy}}{\end{list}}

\newcounter{itemlistc}
\newcounter{romanlistc}
\newcounter{alphlistc}
\newcounter{arabiclistc}

\newcommand{\fcaption}[1]{
        \refstepcounter{figure}
        \setbox\@tempboxa = \hbox{\footnotesize Fig.~\thefigure. #1}
        \ifdim \wd\@tempboxa > 5in
           {\begin{center}
        \parbox{5in}{\footnotesize\smalllineskip Fig.~\thefigure. #1}
            \end{center}}
        \else
             {\begin{center}
             {\footnotesize Fig.~\thefigure. #1}
              \end{center}}
        \fi}

\newcommand{\tcaption}[1]{
        \refstepcounter{table}
        \setbox\@tempboxa = \hbox{\footnotesize Table~\thetable. #1}
        \ifdim \wd\@tempboxa > 5in
           {\begin{center}
        \parbox{5in}{\footnotesize\smalllineskip Table~\thetable. #1}
            \end{center}}
        \else
             {\begin{center}
             {\footnotesize Table~\thetable. #1}
              \end{center}}
        \fi}

\def\@citex[#1]#2{\if@filesw\immediate\write\@auxout
	{\string\citation{#2}}\fi
\def\@citea{}\@cite{\@for\@citeb:=#2\do
	{\@citea\def\@citea{,}\@ifundefined
	{b@\@citeb}{{\bf ?}\@warning
	{Citation `\@citeb' on page \thepage \space undefined}}
	{\csname b@\@citeb\endcsname}}}{#1}}

\newif\if@cghi
\def\cite{\@cghitrue\@ifnextchar [{\@tempswatrue
	\@citex}{\@tempswafalse\@citex[]}}
\def\citelow{\@cghifalse\@ifnextchar [{\@tempswatrue
	\@citex}{\@tempswafalse\@citex[]}}
\def\@cite#1#2{{$\null^{#1}$\if@tempswa\typeout
	{IJCGA warning: optional citation argument 
	ignored: `#2'} \fi}}

\def\pmb#1{\setbox0=\hbox{#1}
	\kern-.025em\copy0\kern-\wd0
	\kern.05em\copy0\kern-\wd0
	
\kern-.025em\raise.0433em\box0}

\def\fnt#1#2{\footnotetext{\kern-.3em
	{$^{\mbox{\scriptsize #1}}$}{#2}}}

\def\fpage#1{\begingroup
\voffset=.3in
\thispagestyle{empty}\begin{table}[b]\centerline{\footnotesize #1}
	\end{table}\endgroup}

\def\runninghead#1#2{\pagestyle{myheadings}
\markboth{{\protect\footnotesize\it{\quad #1}}\hfill}
{\hfill{\protect\footnotesize\it{#2\quad}}}}
\font\tenrm=cmr10
\font\tenit=cmti10 
\font\tenbf=cmbx10
\font\bfit=cmbxti10 at 10pt
\font\ninerm=cmr9

\font\eightrm=cmr8

\newcommand{\be}{\begin{equation}} 
\newcommand{\ee}{\end{equation}} 

\begin{document}
\setlength{\textheight}{7.7truein}    

\runninghead{The $r$-mode instability in rotating neutron stars}
{Andersson \& Kokkotas}

\normalsize\textlineskip
\thispagestyle{empty}
\setcounter{page}{1}

\copyrightheading{}		

\vspace*{0.88truein}

\fpage{1}
\centerline{\bf THE $R$-MODE INSTABILITY IN ROTATING NEUTRON STARS}
\vspace*{0.37truein}  

\centerline{\footnotesize NILS ANDERSSON}
\vspace*{0.015truein}
\centerline{ \footnotesize \it
Department of Mathematics, University of Southampton,  
Southampton SO17 1BJ, United Kingdom}
\vspace*{10pt}
\centerline{\footnotesize KOSTAS D. KOKKOTAS}
\vspace*{0.015truein}
\centerline{ \footnotesize \it 
Department of Physics, Aristotle University of Thessaloniki, 
Thessaloniki 54006, Greece}
\vspace*{0.225truein}
\publisher{(received date)}{(revised date)}

\vspace*{0.21truein}
\abstracts{In this review we summarize the current understanding of
the gravitational-wave driven  
instability associated with the so-called $r$-modes in 
rotating neutron stars. We discuss the nature of the $r$-modes, the
detailed mechanics of the instability and its potential
astrophysical significance. In particular we discuss 
results regarding
the spin-evolution 
of nascent neutron stars, the detectability of $r$-mode
gravitational waves and mechanisms limiting the 
spin-rate of accreting neutron stars in binary systems.}{}{}

\vspace*{1pt}\textlineskip	
\section{Introduction}
\vspace*{-0.5pt}

\noindent
The first decade of this new millenium holds great promise for 
gravitational physics. The hope is that forty years of development
will come to fruition
as the new generation of gravitational-wave interferometers (LIGO, VIRGO,
GEO600 and TAMA300) 
come online and reach their projected sensitivities\cite{gwreview}.
These detectors should (finally!) open a new window to the universe and
make the long-heralded field of ``gravitational-wave astronomy''
a reality. For gravitational-wave theorists this means that 
decades of  modelling will finally be tested by observations.
The construction of the new detectors is accompanied by 
attempts to model all likely gravitational-wave sources in 
appropriate detail. This is highly relevant since 
reliable theoretical templates, against which one can match the noisy
data-stream of the detectors, are needed if we want to extract 
reliable astrophysical information from future gravitational-wave data.

In this article we survey the 
territory where stellar pulsation theory meets
gravitational-wave astrophysics.  There has been a resurgence of interest
in this research area (that dates back to
seminal work by Thorne and his colleagues in the late 
1960s\cite{tc67,kst69,ct70}) 
and  
our aim  here is to provide an 
overview of the recent developments. We will focus our 
discussion on the gravitational-wave driven
instability of the so-called $r$-modes. 
Since its serendipitous discovery a couple of years ago\cite{a97,fm97}, the 
$r$-mode
instability has attracted considerable attention.
In this review article we aim to decribe most of 
the recent ideas 
and suggestions, be they speculative or not, and put them in the 
appropriate context. We will try to introduce the relevant concepts
and results from basic principles, in such a way that the article 
should be accessible to readers with little previous knowledge of this field. 
We feel that this is an important task since it may
 help us identify the many
outstanding issues that must be addressed by future work. 

The theory of stellar pulsation 
is richly endowed with interesting phenomena, and ever 
improving observations suggest that most stars exhibit complicated
modes of oscillation. Thus it is natural to try to match 
theoretical models to observed data in order to extract information  about
the dynamics of distant stars. This interplay 
between observations and stellar pulsation theory is known as 
asteroseismology\cite{unno}. It is interesting to speculate that
the advent of gravitational-wave astronomy will permit similar studies
for neutron stars. One would expect most neutron stars to 
oscillate during various phases of their life, 
and actual observations would provide invaluable information 
about the stars internal structure and the supranuclear equation of state\cite{astero1,astero2}. 
Given their compactness, with a mass above $1.3M_\odot$ 
concentrated inside a radius of a mere $10$~km, 
oscillating neutron stars could be interesting 
astrophysical sources of gravitational
waves. But it is not clear that 
astrophysical mechanisms can excite the various oscillation
modes
to a detectable level. It seems likely that only the most violent 
processes, such as
the actual formation of a neutron star following a supernova or
a dramatic starquake following, for example, an internal phase-transition, 
will be of relevance. The strengthening evidence for 
magnetars\cite{duncan,magnetar}, in which a
starquake could release large amounts of energy, is 
also very interesting in this respect. One can estimate that
these events must take place in our immediate neighbourhood 
(the Milky Way or the Local Group) in order 
to be observable. 
 
This does not, however, rule out the possibility that 
neutron star pulsation may be detected by the new generation of 
gravitational-wave detectors. First of all, one should not 
discard the possibility
that unique events in the life of a neutron star may excite the various
modes to a relevant level. More likely to be observationally important
 is the long recognized possibility that 
various pulsation modes
of rapidly rotating neutron stars may be unstable due to the emission of 
gravitational radiation\cite{cfsreview,jlfrev,nslr}. 
Should such an instability operate in a
young neutron star
it may lead to the emission of copious amounts
of gravitational waves\cite{o98}. Of particular recent interest
is the instability of the so-called $r$-modes, 
for which the gravitational waves have been estimated 
to be detectable
for sources in the Virgo cluster (at 15-20~Mpc). 
If we suppose that most
newly born neutron stars pass through a phase where this 
kind of instability is active, several such events  should be observed per year
once the advanced interferometers come into operation. This is a very 
exciting prospect, indeed. 

\section{Nonradial oscillations of rotating stars}

A neutron star has a large number of families of pulsation modes 
with more or less distinct character. 
For the simplest stellar models, the relevant modes are
high frequency pressure $p$-modes and the low frequency 
gravity $g$-modes\cite{unno}.  For a typical nonrotating neutron star
model the fundamental $p$-mode, whose eigenfunction has no nodes
in the star, is usually referred to as the $f$-mode. The $f$-mode
has frequency in the range 2-4~kHz, while the
first overtone lies above $4$~kHz. The $g$-modes depend sensitively on
the internal composition and temperature distribution, but they typically
have frequencies of a few hundred Hz.
The standard mode-classification 
dates back to the seminal work of Cowling\cite{c41}, and is based 
on identifying the main restoring 
force that influences the fluid motion.
As the stellar model is made more detailed and
further restoring forces are included
new families of modes come into play. For example, a neutron star model
with a sizeable solid crust separating a thin ocean from a central
fluid region will have $g$-modes associated with both the
core and the ocean as well as modes associated with
shearing motion in the crust\cite{mcdermott,strohmayer}. 
Of particular interest to relativists is the existence of a class of
modes uniquely associated with the spacetime itself\cite{ks86,ks92,akk96}; 
the so-called
$w$-modes (for gravitational \underline{w}ave). These modes 
 essentially arise because the curvature
of spacetime that is generated by the background density distribution
can temporarily trap impinging gravitational waves. The $w$-modes typically
have high frequencies (above 7~kHz) and damp out in a fraction of a 
millisecond. It is not yet clear whether one should expect these modes
to be excited to an appreciable level during (say) a gravitational
collapse following a supernova. One might argue that they 
provide a natural channel for the release of any initial 
deformation of the spacetime, but there are as yet no solid evidence
indicating a significant level of $w$-mode 
excitation in a realistic scenario\cite{allen1,allen2,kslrev}.

\subsection{Linearised equations of motion}

In this article we will only touch briefly on ``realistic''
neutron star models. Most of our discussion will concern 
(adiabatic) linear 
perturbations of relatively simple
perfect fluid models. We will typically assume that the 
unperturbed pressure $p$ and density $\rho$ are related by a one-parameter
equation of state, $p=p(\rho)$. A useful way of writing this 
is to introduce an effective polytropic index by
\be
\Gamma = {d \log p \over d\log \rho}
\ee
In the special case of a polytrope $\Gamma$ is, of course, 
constant and often expressed as $\Gamma = 1+ 1/n$.

We want to
consider a   star rotating uniformly with frequency $\Omega$.
In the frame of
reference that co-rotates with the star
the equations that govern the fluid motion 
(the Euler equations) can be written
\be 
\partial_t \vec{u} + \vec{u}\cdot \nabla \vec{u} 
+ 2\vec{\Omega} \times \vec{u} + \vec{\Omega} \times(\vec{\Omega} 
\times \vec{r})
 = - {1 \over \rho} \nabla p
-  \nabla \Phi
\label{eulertime}\ee 
where $\vec{u}$ is the fluid 
velocity and $\Phi$ represents the gravitational 
potential. Here it is worth noticing that the 
centrifugal term 
can be rewritten as
\be
\vec{\Omega} \times(\vec{\Omega} \times \vec{r}) = -
{1 \over 2} \nabla (\vec{\Omega} \times \vec{r})^2
\ee
In other words,  hydrostatic equilibrium corresponds to
\be
 {1 \over \rho} \nabla p = -  \nabla \left[  \Phi -  {1 \over 2}
 (\vec{\Omega} \times \vec{r})^2 \right] = -  \nabla \Psi
\label{equil}\ee
From this we can deduce that the level surfaces 
\be
\Psi= \Phi - {1 \over 2} \Omega^2 r^2 \sin^2 |\theta| = \mbox{constant}
\ee
coincide with the isobars (and isopycnic surfaces). Consequently, once 
we have found a convenient way to describe the level surfaces of 
the ``effective gravitational potential''
$\Psi$ we have a complete description of our rotating fluid. 
In the particular case of slow rotation,  
the level surfaces can be described by introducing a new variable
$a$ corresponding to the mean distance of a given constant
$\Psi$ surface to the centre of mass. This new variable is
related to the standard spherical coordinates ($r$ and $\theta$)
through
\be
r = a [ 1 + \epsilon(\theta) ] \ .
\ee
where $a$ ranges from $0$ to $R$ (the  radius of the corresponding 
nonrotating star).
The advantage of this representation is that 
the pressure and the density can now be thought of as functions 
of $a$ only. For a uniform density star
(which incidentally is not a bad approximation for a neutron star)
we have
\be
\epsilon = - { 5\Omega^2 \over 8\pi \rho} P_2 (\cos \theta )
\ee
where $P_2(\cos \theta)$
is the standard Legendre polynomial, 
and for polytropes one can calculate
the required function $\epsilon(\theta)$ 
from the results of 
Chandrasekhar and Lebovitz\cite{ch62}.

Having prescribed a rotating stellar model, we want to study 
small nonradial perturbations away from  equilibrium.
We will consequently assume that the perturbation is characterized
by a displacement vector $\vec{\xi}$, which is small in a suitable sense,
and yields the fluid velocity as
\be 
\delta \vec{u} = \partial_t \vec{\xi} 
\ee
Given that the background model is stationary and symmetric 
with respect to the rotation axis, 
we can always decompose a perturbation into 
modes that depend on the azimuthal angle as $\exp(im\varphi)$ with 
$m$ an integer. 
If we also assume that these modes have a harmonic time-dependence
we can write
\be
\vec{\xi} \to \vec{\xi}e^{i( m\varphi + \omega_r t)}
\label{modedec}\ee
where $\omega_r$ is
the oscillation frequency measured in the rotating frame. 
By recalling that the inertial frame is related to the rotating frame by
\be
d/dt = \partial /\partial t  + \vec{u}\cdot \nabla = 
\partial_t + \Omega  \partial_\varphi
\ee
we see that the inertial frame frequency is given by
\be
\omega_i = \omega_r - m\Omega \ .
\ee

With the Ansatz (\ref{modedec}) 
the linearised Euler equations are
\be 
\partial_t \delta \vec{u}  + 2 \Omega \times  \delta \vec{u} 
 = -\omega_r^2 \vec{\xi} + 2i \omega_r \Omega\times \vec{\xi} = 
{\delta \rho \over \rho^2} \nabla p
- {1 \over \rho} \nabla \delta p - \nabla \delta \Phi
\label{euler}
\ee 
where $\delta \rho$ and $\delta p$ represent the 
Eulerian perturbations
in the density and pressure, respectively, and the 
variation in the gravitational potential is $\delta \Phi$. 
Note that the centrifugal force only enters this equation
through its effect on the background configuration. That this should be the 
case is easily understood from (\ref{equil}): Since 
we are working in an Eulerian framework
the position vector $\vec{r}$ is treated as a constant. 

To fully describe the perturbations we also need  
the equation describing the conservation of mass
\be
\partial_t \delta \rho + \nabla \cdot (\rho \delta \vec{u}) = 0
\label{masscon}\ee
or in integrated form;
\be
\delta \rho + \nabla \cdot (\rho \vec{\xi}) = 0
\ee
as well as 
the perturbed Poisson equation for the gravitational
field
\be 
\nabla^2 \delta \Phi = 4\pi G\delta \rho \ .
\ee
We must also prescribe  
an ``equation of state'' for the perturbations, i.e. a  relation
between the (Lagrangian) perturbations of pressure and density.
Assuming adiabatic perturbations this relation is usually written
\be
{\Delta p \over p} = \Gamma_1 {\Delta \rho \over \rho} 
\ee
where $\Gamma_1$ is the adiabatic index (which need not be
related to the various background quantities). 
In terms of the Eulerian 
perturbations, this corresponds to
\be
{\delta p \over p} = \Gamma_1 {\delta \rho \over \rho} + \Gamma_1
\vec{\xi} \cdot \left[ \nabla \log \rho -
{1 \over \Gamma_1 }\nabla  \log p  \right]
\label{eospert}\ee
At this point it is customary to introduce the so-called 
Schwarzschild discriminant, ${\cal A}_s$. It corresponds to the magnitude of
the vector in the square bracket above, and since $p$ and $\rho$ 
depend only on $a$ in our rotating model we have
\be
{\cal A}_s = {1 \over \rho} {d\rho \over da} -{1 \over \Gamma_1 p} 
{dp \over da}
\ee
As we will see later, the Schwarschild discriminant plays a crucial 
role in stellar pulsation theory. Physically, radial
variation in ${\cal A}_s$ 
corresponds to internal composition or temperature gradients 
in the star. In particular, the special case 
${\cal A}_s =0$, which  
leads to the perturbations obeying the same equation of state as the 
background fluid ($\Gamma_1=\Gamma$), is often used for 
neutron stars.
In the following,
 we will refer to such perturbations as being ``isentropic''. 
 The motivation
for this  assumption is that the temperature of most neutron 
stars is far below
the relevant Fermi temperature, 
and  they can consequently be considered
as essentially having zero temperature\footnote{However, it is worth 
pointing out that 
realistic neutron stars may still have ${\cal A}_s \neq 0$ because of 
radial variation in the chemical composition (say, a varying proton fraction
in the core)\cite{reis}. In view of this, whenever we 
refer to the ``isentropic'' case in this article, we intend this to 
mean stars with {\em no internal entropy or composition gradients}.}.

\subsection{Rotational effects on pulsation modes}

We are now well equipped to discuss 
the effect that rotation has on the various pulsation
modes of a star.
To understand this issue it is helpful
to recall how a mode calculation proceeds in the 
non-rotating case.  For spherically symmetric stars
one can separate the
pulsations into two general classes. We will refer to these as
spheroidal and toroidal perturbations. In relativistic studies
these  two classes are often called polar and axial perturbations
(or even and odd parity perturbations), respectively. 

Spheroidal modes
have displacement vectors of form
\be
{\vec{\xi} \over r}=  \sum_{lm} \left(S_{lm} , H_{lm} \partial_\theta , 
{H_{lm} \over \sin \theta} \partial_\varphi \right) Y_{lm}
\ee
where $Y_{lm}(\theta,\varphi)$ are the standard spherical harmonics.
Spheroidal perturbations are accompanied by variations
in the pressure and the density which can, since they
are scalar quantities, always be expanded in terms of 
the spherical harmonics. Hence, we have
\be
\delta p = \sum_{lm} \delta p_{lm} Y_{lm}  
\ee
and similar for $\delta \rho$.
By analyzing the equations for a spheroidal perturbation
one can deduce that a simple stellar model allows two distinct
classes of pulsation modes (cf. the monograph by Unno et al.\cite{unno}
for details). These have frequencies that are governed by 
\be
\omega^2 \approx {l(l+1) c_s^2 \over r^2}
\ee
where $c_s^2 = {\Delta p / \Delta \rho}$ is the sound speed, and
\be
\omega^2 \approx  -g \xi_r {\cal A}_s
\label{gmode}\ee
where $g=|dp/dr|/\rho$ is the local gravitational acceleration. 
These classes of modes are known as the $p$- and the $g$-modes, 
respectively. 

Toroidal modes have eigenvectors that can be written
\be
{\vec{\xi} \over r} = \sum_{lm}
\left(0 , {T_{lm} \over \sin \theta} \partial_\varphi , 
-T_{lm} \partial_\theta \right) Y_{lm} = \sum_{lm} {1\over \sqrt{l(l+1)}}
T_{lm} \vec{Y}_{lm}^B
\label{toroidal}\ee
where $\vec{Y}_{lm}^B$ are the magnetic multipoles introduced by 
Thorne\cite{th80}.
A non-rotating 
perfect fluid model has toroidal modes with non-zero frequency, 
but in a model with a solid crust there are distinct
toroidal shear modes\cite{mcdermott,strohmayer}.
It should also be mentioned that the relativistic theory 
yields both toroidal and spheroidal $w$-modes\cite{akk96}.

Because of the symmetry of the non-rotating problem, modes 
corresponding to different
$l$ and $m$ decouple. In fact, it is sufficient
to consider the $m=0$ case since the non-zero $m$ solutions then follow
after a trivial rotation. Hence, there is no need to sum over the various 
$l$ and $m$ in the case of perturbed spherical stars. 
The case of rotating stars is 
much more complicated. First of all, the 
symmetry is  broken in such a way that the various $-l\le m\le l$ 
modes become distinct. As a first approximation one finds that\cite{unno}
\be
\omega_i(\Omega) =  \omega_r(\Omega=0) - m\Omega(1+ C_{lm}) + O(\Omega^2)
\label{rotfreq}\ee
according to an inertial observer. Here, $C_{lm}$ is a 
function that depends on the mode-eigenfunction in a non-rotating star.
Secondly, rotation has a significant effect on the 
eigenfunctions by coupling the various $l$-multipoles.
As the rotation rate
 is increased an increasing number of  $Y_{lm}$'s
are needed to  describe a mode. One must also account for coupling
between the spheroidal and toroidal vectors. 
To illustrate this, let us consider
a toroidal  mode that corresponds to a multipole 
$[l,m]$ in the $\Omega=0$ limit. After including
the first rotational correction the eigenfunction of this
mode becomes
\be
{\vec{\xi} \over a} = 
\left(0 , {T_{lm} \over \sin \theta} \partial_\varphi , 
-T_{lm} \partial_\theta \right) Y_{lm} \nonumber  +
\sum_{\nu=l\pm 1} \left(S_{\nu m} , H_{\nu m} \partial_\theta,
{H_{\nu m} \over \sin \theta} \partial_\varphi \right) Y_{\nu m} 
\label{rvec}
\ee
In other words, the $[l,m]$ toroidal mode couples to the $[l\pm 1,m]$
spheroidal ones. The situation is analogous for rotationally
modified spheroidal modes. 


\subsection{Low-frequency modes}

A rotating star has two sets of low-frequency modes, the
spheroidal $g$-modes and the toroidal $r$-modes\cite{pp78,pbr81,sa82,scm81,sm83}.
To understand the origin and nature of these modes
it is helpful to digress on the perturbations of
non-rotating (spherical) stars.
A spherical star has a 
degenerate spectrum of zero-frequency
modes. These modes 
correspond to neutral convective currents, and
the nature of the corresponding fluid perturbations can be understood
in the following way: As follows from
(\ref{euler}) and (\ref{masscon}) stationary,
isentropic, perturbations
of a spherical star are solutions to
\be
\nabla \cdot(\rho \delta \vec{u}) = 0
\label{zerodiv}\ee
and
\be
{\delta \rho \over \rho^2} \nabla p - {1\over \rho } 
\nabla \delta p - \nabla \delta \Phi = 0
\ee
Since these two equations decouple we can consider 
any solution as a superposition of two distinct classes of 
solutions\cite{lf99}:
\begin{eqnarray}
\delta \vec{u} &\ne& 0 \quad , \quad \delta p =\delta \rho = \delta \Phi =0 
\nonumber \\
\delta \vec{u} &=& 0 \quad , \quad \delta p,\delta \rho , \delta \Phi \mbox{ nonzero}
\nonumber
\end{eqnarray}
Since any static self-gravitating perfect fluid must be spherical
the second type of solution simply identifies a 
neighbouring equilibrium model. Hence, all stationary
nonradial perturbations of a spherical star must be of the 
first kind. 

What can we deduce about the fluid velocity for these stationary modes?
First of all, one can readily verify that all toroidal 
displacements satisfy (\ref{zerodiv}). In other words, 
the function $T_{lm}$ is unconstrained. In the case of 
spheroidal perturbations the situation is slightly 
different, and one finds that the following relation between
$S_{lm}$ and $H_{lm}$ must be satisfied;
\be
{d \over dr} (\rho r^3 S_{lm}) - l(l+1)\rho r^2 H_{lm} = 0
\ee
Still, one of the two functions is
left unspecified also in this case. 

Before moving on to  rotating stars, we need
to consider also the case of non-isentropic perturbations.
Then (\ref{eospert})
provides a relation between the radial component
of $\vec{\xi}$ and $\delta p$ and $\delta \rho$.
This will obviously not affect the above conclusion
for toroidal modes: There is still a space of zero-frequency
toroidal modes in a non-isentropic star. But the 
stratification associated with a non-zero ${\cal A}_s$
breaks the degeneracy in the spheroidal case, and 
gives rise to distinct $g$-modes, cf. (\ref{gmode}).
 
Having understood the nature of the zero-frequency
modes in the $\Omega=0$ limit, 
it is easy to see what happens
in the rotating case. For spinning stars
the degeneracy of the zero-frequency modes is broken.  
The Coriolis force provides a weak restoring force 
that gives the toroidal modes genuine dynamics\cite{unno,sa82}. 
This leads to the so-called
$r$-modes. 
The main properties of the $r$-modes can
be understood by carrying out a 
simple exercise. Assume that the mode is
purely toroidal to leading order also in the
rotating case. Then the motion is 
essentially horizontal  and 
$\xi_r$, $\delta\rho$ and $\delta p$ are of higher
order in $\Omega$. Now consider the radial component of the 
equation we get by taking the curl of (\ref{euler});
\be
\partial_t  (\nabla\times \delta \vec{u})_r + 2 \nabla\times
(\vec{\Omega} \times \delta \vec{u}) =
\partial_t  (\nabla\times \delta \vec{u})_r  
+ \delta {u}_h\cdot\nabla_h(2\Omega_r)
= 0
\label{vort}\ee
In this  equation subscripts $r$ and $h$ represent radial and horizontal
components, respectively, and we have neglected terms of order $\Omega^2$.
We now translate this result into the inertial frame and get
\be
{d \over dt} \left[(\nabla\times \delta \vec{u})  + 2 \vec{\Omega} \right]_r =0
\ee
Here we can identify the quantity in the square bracket as the 
radial component of the total vorticity.
This result thus shows that the radial component of the 
vorticity is conserved by the $r$-mode motion\cite{sa82}. Furthermore, by 
inserting the toroidal eigenvector in (\ref{vort}) we readily
find a first approximation of the $r$-mode eigenfrequency
\be
\omega_r \approx  {2m\Omega \over l(l+1)} \ .
\ee

At this level of approximation different radial shells in the star are
effectively uncoupled. Hence, the radial dependency
of the $r$-mode eigenfunctions remain undetermined. To determine the
radial behaviour of the modes we need to extend the 
analysis to order $\Omega^2$. The reason for this can be
understood as follows: From the linearised equations 
we can see that our leading order toroidal perturbation will not
couple to the density and pressure perturbations until at 
order $\Omega^2$. Indeed, if we write down the equations
that describe terms of order $\Omega$ we find that  
they decouple completely from the leading order 
toroidal perturbation. To linear order in $\Omega$ the 
$r$-mode assumption is therefore consistent with $\delta p = \delta \rho = 
\delta \Phi = \xi_r=0$.
As far as the higher order corrections
to the $r$-mode frequency are concerned, one can argue
that the next non-trivial contribution must be of order $\Omega^3$. 
Essentially, the frequency correction should have the 
same effect relative to the leading order frequency irrespective
of the sense of rotation  of the star. It cannot be that the 
$r$-mode frequency is different in magnitude for positive
and negative $\Omega$. If that were the case the frequency
would depend on the orientation of the observer, and change if he/she (say)
 decided to stand on his/her head, which clearly cannot make sense. 
Thus we anticipate that
the  $r$-mode frequency can be written in terms of a
dimensionless eigenvalue $\omega_2$ as
\be
 \omega_r = {2m\Omega \over l(l+1)}\left[ 1 -   
 \omega_2 { R^3\Omega^2\over M} \right] \ .
\label{highfreq}\ee
We also expect that the $r$-mode eigenfunctions will take the form
(\ref{rvec}) with $T_{lm}= O(1)$ and $S_{l\pm 1 m}\sim H_{l\pm 1m} 
\sim O(\Omega^2)$, and that the pressure and density variations
will both be of order $\Omega^2$. 

Once we have cranked through the algebra, the following picture emerges:
For adiabatic non-isentropic perturbations, we have a
Sturm-Liouville problem\cite{pbr81} for the function $T_{lm}$. This 
means that there will be an infinite set of $r$-modes for each 
combination of $l$ and $m$. These modes are suitably
labeled by the number of radial nodes in the respective eigenfunctions. 
A word of caution is in order, however. It is clear that we can 
reliably 
determine only a limited number of $r$-modes within the slow-rotation 
scheme: At some point the frequency correction will become 
large, and the assumption that it is a small 
perturbation will be violated.
From (\ref{highfreq}) we can see that the scheme breaks down 
unless
\be
|\omega_2| << {M\over R^3 \Omega^2}
\label{om2}\ee

\begin{table}
\tcaption{Calculated $r$-mode frequency corrections $\omega_2$ for  constant
density stars ($n=0$) and $n=1$ polytropes. Results are shown both for the 
full calculation (nC) and the Cowling approximation (C). 
The results are for isentropic
perturbations and the fundamental $l=m$ modes.}
\centerline{\footnotesize\smalllineskip
\begin{tabular}{ccccc}
$l$ & \multicolumn{2}{c}{$\omega_2(n=0)$} 
& \multicolumn{2}{c}{$\omega_2(n=1)$}  \\
 & nC & C & nC & C \\
\hline
2 & 0.765  & 0.913  & 0.398 & 0.453 \\
3 & 0.797  & 0.844 & 0.427 & 0.443\\
4 & 0.730  & 0.749 & 0.399 & 0.405\\
\hline\\
\end{tabular}} \label{rfrequ}
\end{table}

Interestingly, the $r$-mode results for the 
isentropic case are rather different.
When ${\cal A}_s=0$ one can determine the eigenfunctions already at
lowest order in the calculation\cite{pbr81}.  
Formally the calculation is taken to order 
$\Omega^2$, but when ${\cal A}_s=0$ one can 
combine the $r$ and $\varphi$ components of (\ref{euler})
to provide an equation relating the leading order 
contributions to $\xi_\theta$ and $\xi_\varphi$.
This yields a rather simple equation for  $T_{lm}$:
\be
\left[ a T_{lm}  + {1\over l(l+1)} {d\over da} (a^2 T_{lm}) \right]
\sin \theta {dP_{lm}\over d \theta} - {d\over da} (a^2 T_{lm})  
\cos \theta P_{lm} = 0
\ee
Using standard relations
for the Legendre functions one can show that
this implies that we must have $T_{lm}=0$ unless $l=m$.
In other words, in the isentropic case $r$-modes can
only exist for $l=m$, and in that case the 
above equation readily yields solutions of form
\be
T_{ll}\sim a^{l-1} \ .
\label{leadK}\ee
If we proceed to a full order $\Omega^2$ mode calculation\cite{pbr81,sa82,aks99,ks99,lmo00,aunp}, we 
find the frequency corrections listed in Table~\ref{rfrequ}.

How can we understand the radical difference between
the isentropic and the non-isentropic cases?
A more detailed study of the $r$-mode eigenfunctions
in the non-isentropic case shows that the fundamental
mode (which has no nodes in the radial eigenfunctions) is well described by 
(\ref{leadK}) also when ${\cal A}_s\ne 0$. Furthermore,
the eigenfrequency of this mode remains virtually unchanged\cite{aks99} 
as ${\cal A}_s \to 0$.
Meanwhile, the frequency corrections for all the other
$r$-modes change dramatically.
This suggests the following 
explanation: A non-isentropic star
has an infinite number of $r$-modes for each $l$ and $m\neq 0$.
These modes are distinct because of the stratification 
associated with the radial variation of
${\cal A}_s$. In this sense these modes are close
relatives to the spheroidal $g$-modes. 
The single $r$-mode that remains for $l=m$
in an isentropic star is likely associated with
the fact that the star is a sphere and not a cylinder\cite{aks99}
(which is relevant since the Coriolis operator has 
cylindrical symmetry).
This ``stratification'' leads to the existence 
of a unique mode also in the isentropic
case and explains why this mode is weakly dependent
on variations in ${\cal A}_s$.
Only one mystery remains: What happens to all the
other $r$-modes as the star becomes isentropic?
We will return to this question in section~7.1.

Detailed calculations show
that the $r$-modes are such that $\Delta p << \delta p$
also to order $\Omega^2$. This means that the modes are well confined
to the potential surfaces of the rotating configuration. 
In principle, this means that one would expect these modes to be 
associated with small variations in the gravitational 
potential. Given this, and the fact that the modes are 
more or less localized in the low density surface region of the star, 
 the Cowling approximation (wherein
$\delta \Phi$ is neglected) should be accurate for the 
$r$-modes. Calculations have verified that
the modes are obtained to within a few percent in 
the Cowling approximation\cite{pbr81,sa82}.
This is illustrated in Table~\ref{rfrequ}.

What do the $r$-mode results imply for rapidly spinning neutron 
stars? A stable star can never spin faster than
the rate at which matter is ejected from the equator.
This corresponds to the Kepler limit, which
is well approximated by 
\be
\Omega_K \approx {2\over3} \sqrt{\pi G \bar{\rho}}
\ee
where $\bar{\rho}$ represents the average density of the
corresponding 
nonrotating 
star. We can compare this estimate to an empirical formula, based on 
fully relativistic calculations for stellar models
using realistic equations of state, which suggests that\cite{fip89,hz,las}
\be
\Omega_K \approx 0.78 \sqrt{\pi G \bar{\rho}}
\ee
Typically, the Kepler limit corresponds to a 
rotation period  in the range\cite{nslr} $0.5-2$~ms.
It has been shown that $O(\Omega^2)$ slow-rotation 
stellar models are accurate to within a few percent
compared to full numerical solutions. Consequently, one would
expect a slow-rotation mode-calculation to be reasonably
accurate even for the fastest spinning neutron stars.
From (\ref{om2}) we immediately see that an $r$-mode
solution ought to be reliable as long as
\be
| \omega_2| << 5 M_{1.4} R_{10}^3 P_{-3}^2 \ .
\ee
We have parameterized this formula in terms of canonical neutron star 
values: 
\begin{eqnarray}
M_{1.4} &=& M/1.4M_\odot \\
R_{10} &=& R/10 \mbox{ km} \\
P_{-3} &=& P/1\mbox{ ms}
\end{eqnarray}
Given this and the data in Table~\ref{rfrequ} we see that the fundamental
$r$-mode as calculated in the slow-rotation approximation 
may be reliable also for the fastest spinning pulsars. 

\subsection{Digression: The fluid motion}

Before we turn our attention to the interplay between $r$-modes,
gravitational radiation and viscosity, 
it is relevant to discuss various 
features of the modes in some more detail. 
  
An important concept in the study of pulsation modes of rotating stars
is the ``pattern speed'' of the mode. Given that every mode is proportional
to $e^{i(m\varphi+\omega t)}$ we can see that, holding $a$ and $\theta$
fixed, surfaces of constant phase are described by
\be 
m\varphi+\omega t = \mbox{ constant}
\ee
After differentiation this leads to
\be
{d\varphi \over dt}= -{\omega \over m} = \sigma
\label{pattern}\ee
which defines the pattern speed $\sigma$ of the mode.
 
Having defined this concept, we make two observations.
First of all we see that the pattern speed for a typical $r$-mode is
\begin{equation}
\sigma_r = - {2 \Omega \over l(l+1) } 
\label{rotpat}\end{equation}
according to an observer rotating with the star.  
On the other hand, an inertial 
observer would find 
\begin{equation}
\sigma_i = \Omega { (l-1)(l+2) \over l(l+1) }  \ .
\label{inpat}\end{equation}
That is, although the modes appear retrograde in the rotating system
an inertial observer would view them as prograde. 

Another useful result concerns the $p$-modes. 
Calculations for nonrotating stars show that
the frequency of these
modes increases (for a given overtone) 
with $m$ slower than linearly. 
This means
that as we increase $m$, the pattern speed of the  $p$-modes will 
decrease, cf. (\ref{pattern}). As a consequence, 
even though the high order $p$-modes
have arbitrarily large frequencies, one can always find a mode with a
very small pattern speed. The implications of this result will become
clear in the following section.
 
\begin{figure}[h]
\hbox to \hsize{\hfill \epsfysize=6cm
\epsffile{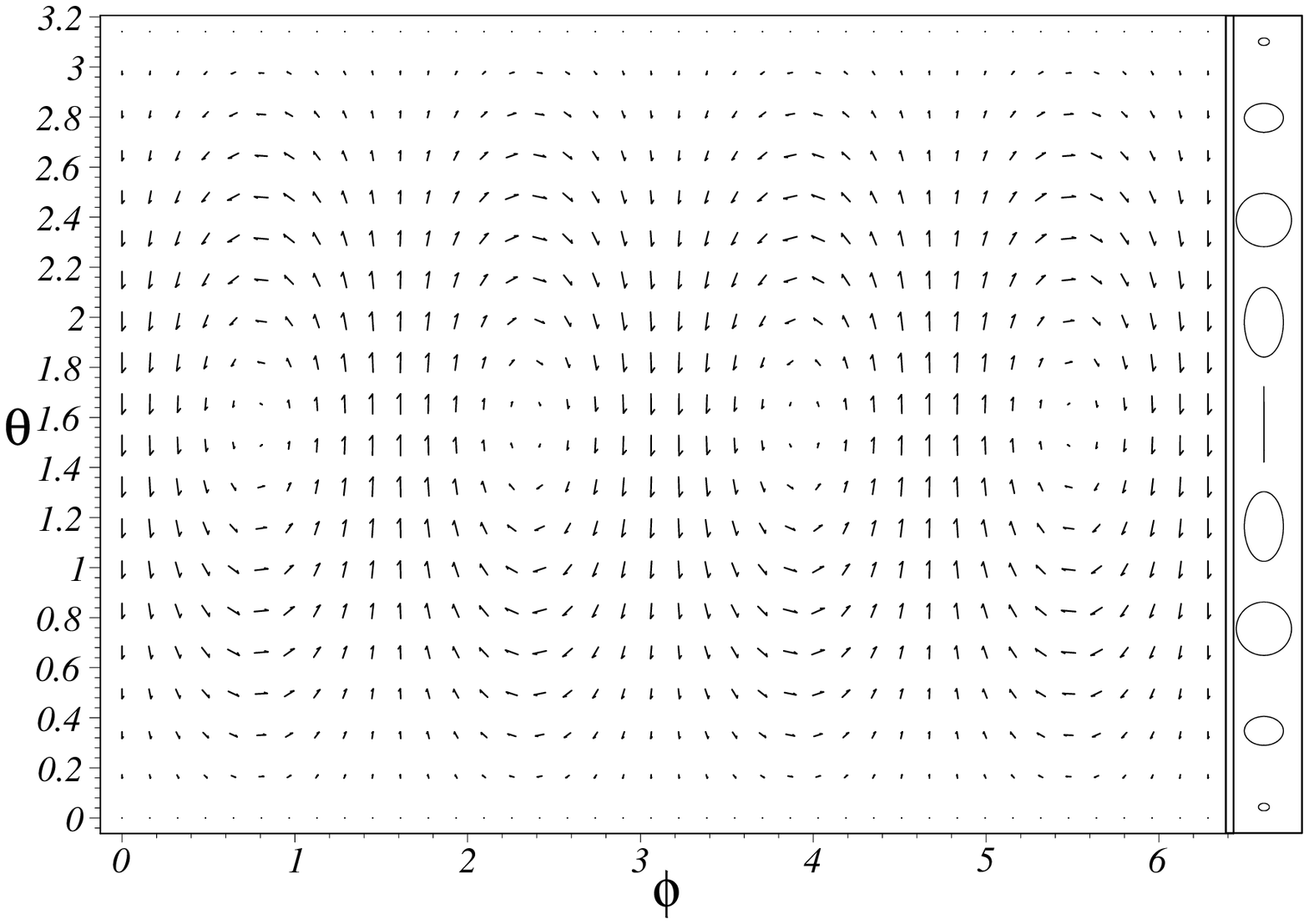} \hfill}
\fcaption{The large left frame shows the velocity field of the
$l=m=2$ $r$-mode at an instant in time. The narrow frame on the right is a 
qualitative picture of the actual motion of individual 
fluid elements (to leading order in the slow-rotation
expansion).}\label{flow}
\end{figure}

Let us now return to the $r$-modes and focus on the detailed fluid motion. 
We will concentrate on the 
$l=m$ $r$-modes of isentropic stars. Introducing a 
suitable dimensionless amplitude $\alpha$, 
we can write the $r$-mode velocity
field (measured by an observer that is co-rotating with the star)
as\cite{o98}
\be
\vec{u} \approx \alpha \Omega R \left( { a \over R} \right)^l \vec{Y}_{ll}^B 
e^{i\omega_r t}
\label{normal}\ee
We can readily illustrate this mode by drawing this
vector field on the two-sphere,  cf. Figure~\ref{flow} and Figure~1
of Saio\cite{sa82}. This gives us an idea of the overall mode pattern. Every
half period of oscillation the motion changes direction, and as we have 
already deduced, the pattern drifts backwards with respect to the 
sense of rotation (according to an observer on the star).

The velocity field in Figure~\ref{flow} does not, however, give us a good 
picture of the motion of the individual fluid elements. 
To understand the nature of the 
$r$-modes better we consider fluid elements on the 
surface of the star. We can 
deduce from  (\ref{normal}) that each fluid element  
moves according to
\begin{eqnarray}
\delta r &\approx & 0 \ , \\
\delta \theta &=& {\xi_\theta \over R} \propto \alpha \sin \theta \cos[m\varphi + \omega_r t] \ , \label{thetaeq} \\
\delta \varphi &=& {\xi_\varphi \over R} \propto \alpha \sin 
\theta \cos \theta 
\sin[m\varphi + \omega_r t] \ .
\label{phieq}\end{eqnarray}
From these equations we see that 
\be
{\delta \theta^2 \over \sin^2 \theta} + {\delta \varphi^2 \over
\sin^2\theta \cos^2\theta} \propto \alpha^2
\ee
In other words, to leading order each fluid element moves on an ellipse
the size of which is determined by the mode-amplitude 
$\alpha$. As $\theta \to 0$ or $\pi$ the ellipses approach 
circles and the radius shrinks to zero. 
We can also infer that a fluid element on the equator 
only moves up and 
down in the $\theta$ direction. These features are illustrated
in Figure~\ref{flow}. This is obviously only a first 
approximation, and one would expect higher order terms
(both in $\Omega$ and $\alpha$) to be important for rapidly
spinning stars. 
We will return to briefly discuss the fluid motion and 
possible higher order secular effects
in Sections~7.3-7.4.

It is also worth trying to gain some understanding of the 
 mode-amplitude $\alpha$. Since we are working 
within a perturbative approach we naturally require that $\alpha$
be small, but what exactly does this mean? A useful measure is provided by
the radial displacement. We know that the radial displacement is 
of order $\Omega^2$, and we can approximate it using (\ref{normal})
together with equations (12)--(14) 
from Kokkotas and Stergioulas\cite{ks99}. 
In particular, if we consider a fluid element 
located on the surface of the star, we find that 
\be
{\delta r \over R} \approx 0.07 \alpha \left( {\Omega \over \Omega_K}\right)^2
\label{height}\ee  
This estimates the ``height'' of the surface waves induced by the 
$r$-modes. 
Later on, when we discuss the possible effect that the unstable $r$-modes
may have on the spin of a newly born neutron star, we will 
consider values of $\alpha$ of the order unity. Given that a neutron star
has a radius of  roughly 10~km, such values
of $\alpha$ are obviously by no means ``small'', as the correspond
to waves of amplitude $700$~m on the surface of a star
spinning at the Kepler limit. 
This is a very important insight, 
since it means that our discussion must be considered as (at best) 
an extrapolation into a regime where nonlinear effects 
should be relevant.

Finally, it is worth emphasising that even though the $r$-modes
are relatively exotic to astrophysicists they are very 
familiar to oceanographers and meteorologists. 
The $r$-modes are analogous to the so-called Rossby waves
 in the Earth's atmosphere and oceans.
These waves were first discussed by the
Swedish meteorologist Carl-Gustaf Rossby, and it is the 
association with Rossby waves
that gives the $r$-modes their name.
There are many interesting discussions of terrestrial Rossby 
waves in the literature.  We should try to learn as
much as possible from our colleagues
in, for example, geophysics. Good starting 
points are the monographs
by Greenspan\cite{g64} and Pedlosky\cite{ped}.

\section{Gravitational-wave instabilities in rotating stars}

\subsection{The CFS mechanism}

Many different instabilities can operate in a neutron star.
The most familiar instability is due to the existence of
a maximum mass (beyond which the star must collapse to
form a black hole). Furthermore,
above $\Omega_K$ there are no equilibrium states (unless
the star rotates differentially). Thus the star
must shed some of its angular momentum before it can settle
down into a stationary state. 
There have been many studies
of this process in the literature\cite{barmode0,barmode1,barmode2}.
The so-called bar-mode instability grows essentially on the 
dynamical timescale of the star, and leads to gravitational
waves that may well 
prove to be observable. 

It has long been recognized that the addition of
various dissipation mechanisms may lead to secular
instabilities, proceeding on a relatively long timescale
depending on the magnitude of the added effect.
The archetypal secular instability is associated with 
viscosity. This instability was first studied
by Roberts and Stewartson\cite{roberts} (for more recent results, see
Lindblom\cite{visco}). Essentially,  viscosity 
admits a transition to a lower energy state by violating
the conservation of circulation (which holds in a 
perfect fluid star).
In a similar way, 
rotating neutron stars are generically unstable
due to the emission of gravitational waves (which leads to
angular momentum of the star not being conserved). 
That such an instability would operate was first established by Chandrasekhar\cite{ch69}
for the Maclaurin spheroids. 
This interesting result was subsequently put on a rigorous
footing by Friedman and Schutz\cite{fs78a,fs78b,f78}, who also proved
that the instability is generic: All rotating perfect fluid
neutron stars are unstable!

The mechanism for gravitational-wave instability can be understood
in the following way\cite{fs78a,fs78b}. Consider first a non-rotating star.
Then the mode-problem  leads to eigenvalues for
$\omega^2$, cf. Unno et al\cite{unno}, which in 
turn gives equal values $\pm |\omega|$
for the forwards and backwards propagating modes (corresponding to
$m=\pm|m|$).
These two branches of modes are affected by rotation in different
ways, cf. (\ref{rotfreq}). A backwards moving mode will be dragged forwards by
the stellar rotation, and if the star spins sufficiently
fast the mode will move forwards with respect to the
inertial frame. Meanwhile, the mode is still moving 
backwards in the rotating frame. 
The gravitational waves from such a 
mode carry positive angular momentum 
away from the star, but since the perturbed fluid actually rotates slower than it would in  
absence of the perturbation the angular momentum of the 
retrograde mode is negative. The emission of 
gravitational waves consequently
makes the angular momentum of the mode 
increasingly negative and leads to an instability. 
This class of frame-dragging instabilities is 
usually refered to as Chandrasekhar-Friedman-Schutz
(CFS) instabilities.
It is easy to see that the CFS mechanism is not unique to
gravitational radiation. Any radiative mechanism will do, 
and even though this possibility has not yet attracted much attention, 
one would expect a similar instability to exist
also for electromagnetic waves (see \cite{hl99} 
and comments in section~7.3). 

\begin{figure}[h]
\hbox to \hsize{\hfill \epsfysize=4cm
\epsffile{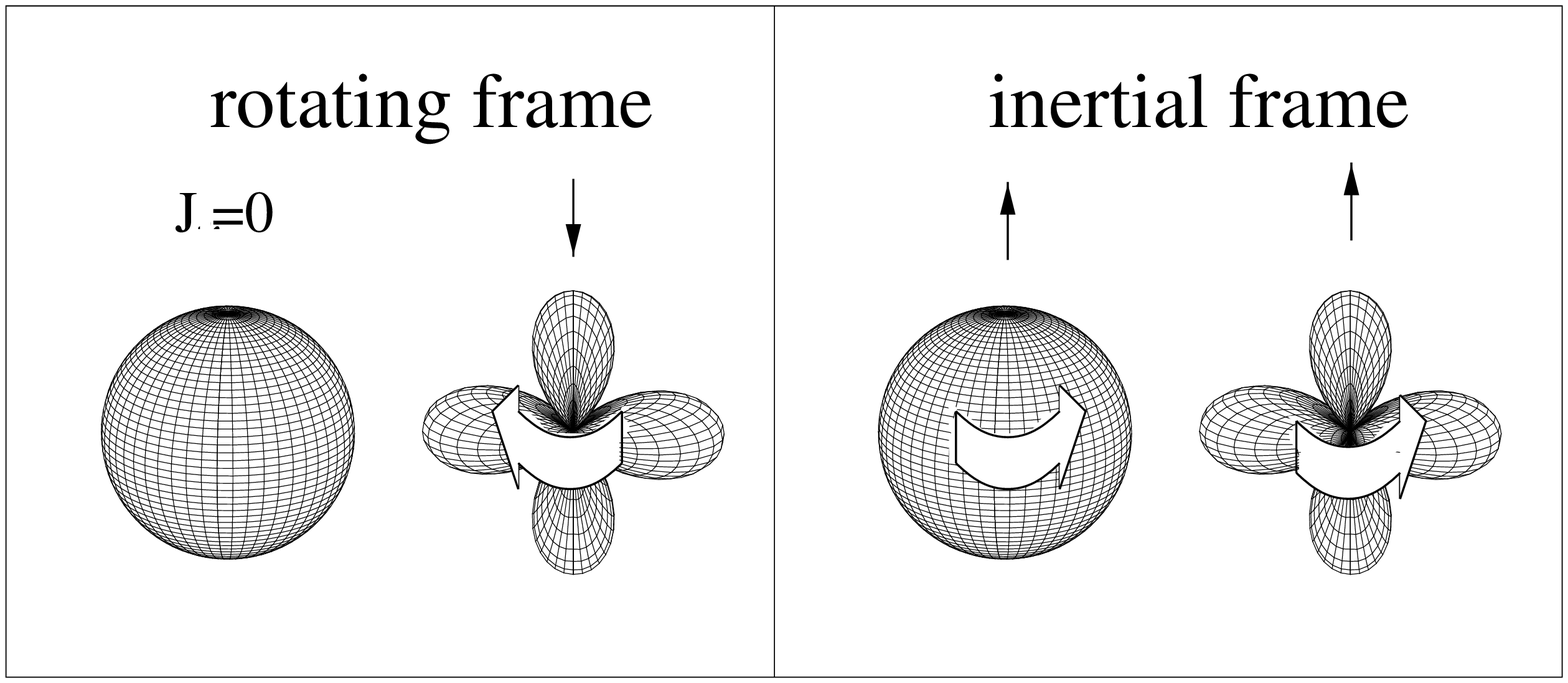} \hfill}
\fcaption{A schematic illustration of the conditions
under which the CFS instability is operating. A perturbed
star can be viewed as a superposition of a uniformly
rotating background and a nonaxisymmetric perturbation.
A mode is unstable if it is retrograde
according to an observer in the fluid (left), but  appears 
prograde in the inertial frame (right).}
\end{figure}

The fact  that the emission of gravitational 
radiation causes a growth in the mode energy in the 
rotating frame, despite
the decrease in the inertial frame energy, may
at first seem a bit strange. However,  
it can be understood from the relation between 
the two energies:
\begin{equation} 
E_r = E_i - \Omega J \ .
\end{equation}
From this we see that $E_r$ may increase if 
both $E_i$  and $J$ decrease\cite{fs78a,fs78b}.
In other words, when the mode radiates away 
angular momentum the star can find a rotational state of lower 
angular momentum {\em and} lower energy.  Under these
conditions the mode 
amplitude may grow.  

Let us take a closer look at some of the relevant ideas
for this mechanism.
The Friedman-Schutz criterion for instability
relies on the so-called canonical energy $E_c$ being negative.
The canonical energy is defined as
\begin{eqnarray}
E_c &=& {1\over 2} \int \left[ \rho |\partial_t \vec{\xi}|^2
-\rho|\vec{u}\cdot\nabla\vec{\xi}|^2 +\Gamma_1 p 
|\nabla \cdot \vec{\xi} |^2 + \vec{\xi}^*\cdot\nabla p
\nabla\cdot\vec{\xi} +  \right. \nonumber \\
&+& \left. \vec{\xi}\cdot\nabla p\nabla\cdot
\vec{\xi}^* + \xi^{i*} \xi^{j*}(\nabla_i\nabla_j p +
\rho\nabla_i\nabla_j \Phi) - { 1 \over 4 \pi G} | \nabla \delta \Phi |^2 
\right]  dV
\end{eqnarray}
which is a conserved quantity. 

Similarly, we can define another conserved quantity
\be
J_c = -\mbox{Re } \int \rho \partial_\varphi \xi^{i*}(
\partial_t {\xi}_i+\vec{u}\cdot\nabla\xi_i) dV
\ee
i.e. a canonical angular momentum.
If $E_c$ is negative at the outset and the system (the star)
is coupled
to another system  (the radiation) in such a way $E_c$
must decrease with time, then the absolute 
value of $E_c$ will increase and the associated mode
is unstable. Generally, an instability can be distinguished 
in a mode-independent way by constructing
(canonical) initial data $[\vec{\xi},\partial_t \vec{\xi}]$ such 
that $E_c$ is negative. 
However, to do this is not at all trivial.
From the computational point of view the most straightforward task 
is to find the so-called neutral modes (which have zero frequency
in the inertial frame), that signal the onset of instability. 
Furthermore, 
if we manage to proceed further and calculate a general 
mode of a rotating star, we
can readily evaluate $E_c$ to assess its stability. 
As was shown by Friedman and Schutz\cite{fs78a,fs78b}, all mode
solutions can be viewed as canonical data, and therefore
it is sufficient to show that the displacement
vector associated with the mode leads to $E_c<0$ to 
demonstrate the presence of an instability.
One can show that the condition $E_c<0$ is equivalent to
the simple notion of a retrograde mode being dragged
forwards, i.e. a change of sign in the pattern speed
as viewed in the inertial frame.
This important conclusion follows immediately
from the relation  
\be 
E_c = -{\omega_i \over m} J_c = \sigma_i J_c
\label{ec}\ee
which is a general property of linear waves. Clearly, $E_c$ 
changes sign when the pattern speed $\sigma_i$ passes through zero.

It is interesting at this point to contrast the gravitational-wave 
driven instability to that due to viscosity. 
For uniformly rotating stars one can show that 
the combination
\be
\delta E - \Omega \delta J = E_c - \Omega J_c = -
{\omega_r \over m} J_c =\sigma_r J_c = E_{c,R}
\label{ecrot}\ee
relating the first order changes in the kinetic
energy and angular momentum to a mode-solution,
is gauge-invariant. 
 $E_{c,R}$ can be viewed as the canonical energy in 
the rotating reference frame. 
Viscosity leads to $E_{c,R}$ being a decreasing
function of time. Comparing (\ref{ec}) to (\ref{ecrot})
we see that the onset of the viscosity driven instability is 
signalled by the vanishing of the pattern speed in the
rotating frame ($\sigma_r=0$).

\subsection{Estimates for the $f$-mode instability}

Many families of neutron star oscillation modes contain members 
that, for a sufficiently high  rotation rate, satisfy the 
CFS instability criterion. We have already pointed out that 
the large $m$ $p$-modes may have arbitrarily
small pattern speeds in a spherical star. 
Combining this result with the qualitative
formula (\ref{rotfreq}) we can deduce that these
$p$-modes will become unstable
already at relatively slow rotation rates. Because it has no 
nodes in its radial eigenfunctions the fundamental $p$-mode
(the $f$-mode) radiates gravitational waves more efficiently
than the various overtones, and consequently  leads 
to the strongest instability for any given $m$.

So why is it that, if all rotating stars are generically 
unstable, we observe millisecond pulsars? This is a
crucial  question, the answer to which follows from 
several important results. First of all, we have seen that 
 the large $m$ modes  become unstable
at the lowest rates of rotation. But as 
Comins\cite{comins1,comins2} have shown 
(for the Maclaurin spheroids) the growth rate of these
modes decreases exponentially with $m$. In order to 
play an astrophysical role a mode must grow fast 
compared to the evolutionary timescale of the star.
In practice, this means that 
one would not expect modes with $m$ significantly larger
than (say) 10 to be relevant. 
Secondly, it turns out that gravitational radiation 
and viscosity compete. Viscosity tends to 
suppress the CFS instability, and since it operates locally
on short length scales it is more effective for high
$m$ modes. Detailed calculations\cite{il91,ll95}
have shown that viscosity suppresses the CFS instability
in the $p$-modes for $m>5$. 

To establish a reliable consensus regarding the stability
properties of rotating neutron stars is  a
high priority issue. However, despite a considerable
amount of work being devoted to the task
this subject may still hide many of its secrets. 
Given that a general criterion for stability
is outstanding, we must  attempt to calculate all the 
various modes of a rotating star to see if any of them 
are unstable. This is a formidable task, but considerable
progress has been made. Currently, 
most of our understanding of unstable modes comes 
from Newtonian calculations, in particular for the 
Maclaurin spheroids. 

Until recently, the conventional wisdow was that the
$f/p$-modes would lead
to the strongest gravitational-wave instability.
This notion dates
back to the very first investigations of the CFS mechanism.
 Friedman and Schutz\cite{fs78b} suggested that
``we can probably safely conjecture that the $p$-modes
of nonisentropic stars are the most important ones for 
stability, since they typically involve larger density
changes than the $g$-modes do, and so will radiate
gravitational waves more effectively''. This would seem
to be a reasonable assumption, and it explains why 
studies of the CFS instability in neutron stars 
remained focussed on the $f/p$-modes for almost
twenty years. 

Before proceeding to discuss the new  ideas 
regarding the CFS-instability that have emerged in the last couple of years, 
we will 
summarise the status of the field as of 
a few years ago. The current understanding of 
the instability in the $f$-modes of a rotating neutron star
is essentially based on  two different bodies of work. 

 Newtonian results for
the $f$-modes of rapidly rotating stars can
be used to estimate the competing influences of 
gravitational radiation reaction and various viscosities.
This  leads to a prediction that the  $m=4$ mode provides the 
most stringent limit on rotation, and that  that the 
CFS instability would limit the rotation of a normal 
fluid star to\cite{il91} 
$\Omega>0.95 \Omega_K$. Similar estimates for superfluid stars, 
suggest that dissipation due to
the so-called mutual friction (see section~4.5) 
will completely suppress the instability\cite{lm95}.

A fully relativistic calculation of pulsation modes of rotating stars
is still outstanding. The best results so far concern neutral
(zero frequency in the inertial frame) modes that would signal the 
onset of instability\cite{sf98}, 
and modes calculated within the
Cowling approximation\cite{ye97,ye99}. 
As could perhaps have been anticipated, these
results indicate that relativistic effects strengthen 
the instability. For example, while Newtonian calculations
predicted that the $m=2$ $f$-mode does not become unstable for 
$\Omega<\Omega_K$ the relativistic results show that this mode
can in fact become unstable and that it may 
provide the strongest constraint
on the rotation rate of a relativistic star.

\section{The $r$-mode instability}

The discovery that the  $r$-modes of a rotating neutron star
are generically unstable due to the emission of gravitational waves\cite{a97,fm97} 
 may have
come as a slight surprise, but
in retrospect the result is rather obvious.
The real surprise is that
this was not realized earlier. After all,
the $r$-modes have been discussed in 
Newtonian stellar pulsation theory for the last 
twenty years or so (following a pioneering 1978
paper by Papaloizou and Pringle\cite{pp78}), 
and their properties are described 
in monographs on the subject\cite{unno}. 
Since the modes are prograde in the
 inertial frame and retrograde in the co-rotating frame,
cf. (\ref{rotpat})-(\ref{inpat}),
they satisfy the Newtonian criterion for the 
CFS-instability. Hence, one can 
easily deduce that the $r$-modes ought to be 
unstable. That this instability is
 generic also in the relativistic case 
was first shown by 
Friedman and Morsink\cite{fm97}, who proved that there 
exist toroidal initial data for which the
canonical energy $E_c$ is negative. 

In perfect fluid stars
the $r$-modes are unstable at all rates of rotation\cite{a97}. This is 
an important observation, because it
indicates that the instability may well
be astrophysically relevant
even though the $r$-modes do not lead to large variations in 
the density and therefore would not intuitively 
be associated with  strong gravitational waves. 
The reason one might expect the $r$-mode instability 
to be competitive with the more familiar instability in the 
$f$-modes is that an $f$-mode has nonzero
frequency in a spherical star. Thus, the star must
be spun up to a critical rotation rate ($\Omega_c$)
in order for the $f$-mode to go unstable, cf. (\ref{rotfreq}).
Because the $r$-modes are unstable at all rates of rotation
 
their growth times may become rather 
short already at spin rates below $\Omega_c$, which
could then lead to the
$r$-mode instability being stronger than the 
$f$-mode one. The first attempt to calculate the $r$-mode growth 
times\cite{a97}  suggested that this would be the
case. The growth times for the $r$-modes were estimated to be very
short indeed. It was suggested that the amplitude of an unstable $r$-mode 
would grow 
by a factor of two in a second for a typical neutron star spinning with a 
period of 1.6~ms. However, these estimates
came with a serious disclaimer: The method used to
determine the modes, and infer the growth times, was not designed
to study low-frequency modes. 

Nevertheless, these first estimates provided a strong
motivation for more detailed work. As a first
step towards understanding this new instability better
it seemed sensible to obtain estimates that allowed
an immediate comparison with the established $f$-mode results
(recall that these estimates
suggest that the $f$-mode instability becomes active
at rotation rates above $0.95 \Omega_K$). 
Such a comparison would establish whether the $r$-mode 
instability is of relevance astrophysically or if 
it is just 
a peculiarity associated with general relativity. 
To make the desired comparison one must
calculate the characteristic growth time of the modes, as well as
the damping times due to various dissipation mechanisms  
that may be relevant. As in the case of the 
$f$-mode, the stability
of the $r$-modes depends on the competing influences
of gravitational radiation reaction (that drives the mode)
and viscous damping. In order to establish the 
astrophysical relevance of the instability we must 
confirm two things: First, the unstable
mode must grow on a timescale that is astrophysically
``short'', for example, significantly below the age of the 
universe. Secondly, the instability must be able to
win the tug-of-war against all relevant dissipation mechanisms.

In this section we will discuss various estimates
that have been made, especially for damping mechanisms
that may counteract the growth of an unstable $r$-mode.
The estimates we present are made in the framework 
of Newtonian gravity. Since the $r$-mode instability is a
truly relativistic effect this obviously makes these results 
somewhat ad hoc. But we must remember that there are as yet no 
fully relativistic calculations of 
pulsation modes of rapidly rotating neutron stars. 
In absence of such results we   
expect Newtonian estimates to provide 
useful indications. One would not expect
more detailed models, eg. in full general relativity,
to completely change the Newtonian picture (we comment on some
recent results regarding relativistic $r$-modes in section~7.2).
After all, the relativistic corrections are likely 
to be of the order of (say) 20\% and there are far larger 
uncertainties associated with other aspects of this problem
(eg. the microphysics).

We will adopt the standard strategy for estimating instability
timescales\cite{il91}. First 
we assume that the true mode-solution is well represented
by the solution to the non-dissipative perturbation
equations (as described in Section~2.3). Then we use these solutions to 
evaluate the effect of the various dissipation 
mechanisms and add their respective contributions
to the rate of change of the mode energy $dE/dt$.
Finally, we can verify that the first assumption 
is justified by checking that the estimated
growth/damping time is considerably longer than the 
oscillation period of the mode.

We estimate the timescale $t_d$ 
of each dissipation mechanism by assuming that the eigenfunctions 
are proportional to $\exp(t/t_d)$. Then we recall that the mode-energy 
follows from the square of the perturbation and deduce that
\be
{dE \over dt} = -{2E \over t_d}
\ee
where
\be 
E \approx {1 \over 2} \int \rho |\delta\vec{u}|^2 dV \approx 
{l(l+1) \over 2} \omega_r^2 \int_0^R \rho a^4 |T_{ll}|^2 da 
\label{rotenerg}\ee
is the
energy of an $r$-mode measured in the rotating frame.
Using (\ref{normal}) we can  deduce that
(for an $n=1$ polytrope)
\be 
E \approx 10^{51} \alpha^2 M_{1.4} R_{10}^2 P_{-3}^{-2} \mbox{ erg} 
\ee

We focus our discussion 
on the $l=m$ modes because they have been shown 
to lead to the strongest instability\cite{aks99,lom98}.
Note also that we only keep the leading order terms
in (\ref{rotenerg}) and the dissipation integrals
below. For simplicity, we also use
 the Cowling approximation, an assumption that
only introduces an inaccuracy of a few percent\cite{pbr81,sa82},
cf. Table~{\ref{rfrequ}.
This is rather good precision in an astrophysical 
context, and  there are far larger uncertainties in our current 
understanding of the $r$-mode instability.

Ideally one would like to address issues regarding the 
relevance of the supranuclear equation of state for an active
mode-instability. However, to use a realistic equation of state in 
a Newtonian calculation is not particularly meaningful. 
The reason for this is that there is no one-to-one
correspondence between Newtonian and relativistic stellar
models in the sense that, for a given 
equation of state, the same central density 
leads to stars with comparable masses and radii in the two 
theories. Thus we believe that a calculation for realistic
equations of state in the Newtonian framework can be misleading.
Still, it is interesting to try to understand whether the 
overall properties of the equation of state, eg. the stiffness, 
affect the instability. This issue can be studied
by considering different polytropic models, and in the following
we consider results both for $n=1$ polytropes and 
constant density stars
(represented by the limit $n\to 0$). These 
results ought to be indicative for realistic
neutron stars given that the average adiabatic index of 
most realistic equations of state lies in the range $n=0-1$.

\subsection{Gravitational radiation reaction}

The first step in establishing the relevance of
the $r$-mode instability corresponds to showing that 
the modes grow on an astrophysically interesting timescale.
To estimate the gravitational-radiation
reaction we use the standard post-Newtonian 
multipole-formulas (see Thorne\cite{th80}). The gravitational-wave luminosity associated 
with a pulsation mode
(measured in the rotating frame) is then estimated using
\begin{equation}
\left. {dE\over dt}\right|_{\rm gw} = - \omega_r \sum_{l=2}^\infty N_l
\omega_i^{2l+1} \left( | \delta D_{lm} |^2 + | \delta J_{lm} |^2 \right) \ ,
\label{gwlum}\end{equation}
where 
\begin{equation}
N_l = {4\pi G \over c^{2l+1} } { (l+1)(l+2) \over l(l-1)[(2l+1)!!]^2 } \ .
\end{equation}
The first term in the bracket of (\ref{gwlum}) represents radiation due to
the mass multipoles. These are determined by
\begin{equation}
\delta D_{lm} = \int \delta \rho a^l Y_{lm}^* dV \ .
\label{mass}\end{equation}
The second term in the bracket of (\ref{gwlum})  corresponds to
the current multipoles, which follow from 
\begin{equation}
\delta J_{lm} = {2 \over c} \sqrt{ {l\over l+1}} 
\int a^l (\rho \delta \vec{u} +
\delta \rho \vec{\Omega}) \vec{Y}_{lm}^{B*} dV \ ,
\label{curr}\end{equation} 
where $\vec{Y}_{lm}^{B}$ are the magnetic multipoles\cite{th80}.

An order count based on the above formulas suggests
an interesting result: For $l=m$ $r$-modes, the dominant 
contribution to the gravitational radiation comes 
from the first term in (\ref{curr}). That this is the case 
can be seen as follows. Recall that the $l=m$ modes
have toroidal displacement to leading order, while the 
spheroidal perturbations enter at order $\Omega^2$.
Furthermore, we know that if the toroidal component
corresponds to the $l$th multipole, the spheroidal
components will correspond to $l+1$. 
This means that, as a gravitational-wave source 
the $r$-modes are quite unusual.  Since they are primarily 
perturbations of the velocity field in the star, with little disturbance 
in the star's density,  the gravitational 
radiation that they emit comes primarily from the time-dependent 
mass currents (we see that $\delta D_{lm}\sim\Omega^2$ while
$\delta J_{lm}\sim \Omega$).  This is the gravitational analogue of 
magnetic multipole radiation.  In fact, the $r$-mode instability 
is unique among expected astrophysical sources of gravitational radiation  
in radiating primarily by gravitomagnetic effects.

After inserting the leading order
$r$-mode eigenfunction in the relevant current 
multipole term,  we find 
\begin{equation}
\left. {dE\over dt}\right|_{\rm gw}^{\rm current} \approx - 4 l^2 N_l \omega_r^3 
\omega_i^{2l+1} \left| \int_0^R \rho a^{l+3} T_{ll} da \right|^2 
\label{egw}\end{equation}
for the rate of energy loss due to gravitational waves. This then leads to
an estimated growth timescale for the instability 
\be
t_{\rm gw}^{\rm current} \approx  - t_{cm} M_{1.4}^{-1} R_{10}^{-2l} P_{-3}^{2l+2} \ \mbox{ s}
\label{gwest}\ee
where the negative sign indicates that the mode is unstable.

\begin{table}
\tcaption{Estimated growth times due to current and mass-multipole
gravitational radiation. The results are for the single $l=m$
$r$-mode that exists for isentropic stars, and are given both for constant
density stars ($n=0$) and $n=1$ polytropes. We do not show mass multipole
results for the $\rho=$~constant case, since $\delta \rho =0$ in that case.}
\centerline{\footnotesize\smalllineskip
\begin{tabular}{|c|c|c|c|}
\hline
& \multicolumn{2}{c}{current multipole ($t_{cm}$ in s)} & 
{mass multipole ($t_{mm}$ in s)} \\
$l$ & $n=0$ & $n=1$ & $n=1$ \\
\hline
2 & 22  & 47  & $3.6\times10^5$ \\
3 & $4.7\times10^2$ & $1.3\times10^3$   & $4.3\times10^6$ \\
4 & $1.0\times10^4$ & $3.2\times10^4$   & $8.5\times10^7$ \\
\hline 
\end{tabular}} \label{gwtimes}
\end{table}

The estimated current multipole timescales are given (for
$l=m$ $r$-modes of an isentropic star) in Table~\ref{gwtimes}.
The listed results are consistent with 
original ones obtained by several authors\cite{aks99,ks99,lom98,lmo00}
From the tabulated results we can deduce two things. First of all we see that 
the timescale increases by roughly one order of magnitude
with each $l$. Thus, higher multipoles lead to 
significantly weaker instabilities and the $l=m=2$ $r$-mode
will be the most important. Secondly, we can see that the results for
constant density stars and $n=1$ polytropes differ by roughly a factor of two. 
This provides a useful illustration of the uncertainties associated with
 the supranuclear equation of state.

It is worth pointing out that the conclusion that the current multipoles
dominate the gravitational-wave emission is not necessarily  
true for all the  $r$-modes. 
It is certainly true for isentropic stars, since such  stars have  
no ``pure'' $r$-modes for $l \ne m$. But in the general non-isentropic case
we have to consider the infinite set of modes that exist for all $m\ne 0$.
An order count in (\ref{gwlum}) 
immediately shows that the situation is not so simple for
the $l \ne m$ modes. In general, the mode-solution for these modes
have contributions to the toroidal eigenfunction described by the 
$[l,m]$ multipole, but the spheroidal part and the density 
perturbation (proportional to $\Omega^2$) correspond to 
$[l\pm 1,m]$. Consequently, the current multipoles
and the mass multipoles contribute to (\ref{gwlum})
at the same order in $\Omega$ for many of the 
non-isentropic modes. Thus mass multipole radiation 
could play an important role.
It is also relevant to stress that the mass
multipoles are far from irrelevant for the isentropic 
$l=m$ modes. From (\ref{mass})
we find that the leading contribution to the mass multipole radiation 
follows from
\begin{equation}
\left. {dE\over dt}\right|_{\rm gw}^{\rm mass} \approx - 4 l^2 N_l \omega_r^3 
\omega_i^{2l+1} \left| \int_0^R {\rho^2 g a^{l+3}\over \Gamma p} \zeta_{l+1l} da \right|^2 \ , 
\end{equation}
where we have introduced
\be
\delta p_{lm} = \rho g a \zeta_{lm} \ . 
\ee
This leads to
\be
t_{\rm gw}^{\rm mass} \approx  - t_{mm} M_{1.4} R_{10}^{-2l-6} P_{-3}^{2l+6} \ \mbox{ s}
\ee
where the estimated values for $t_{mm}$ are listed in Table~\ref{gwtimes}.
Here it is interesting to note that the 
mass and current multipole results scale 
with the size of the star in rather different ways. 
This means that the relative importance of 
the two is significantly different 
for stars of different size.

\subsection{Viscous damping}

The estimates given in Table~\ref{gwtimes} 
show that the $r$-modes 
grow rapidly enough to be of potential significance.
Still, they must also overcome damping effects due to different
physical mechanisms, most of which are poorly understood and therefore
difficult to model in a realistic fashion.

As in the case of the unstable $f$-modes the main dissipation 
in a hot newly born neutron star is due to viscosity.
For the $r$-modes  to be relevant they must 
grow fast enough that they are not completely
damped out by viscosity. 
To assess the strength of the viscous damping of the 
$r$-modes we implement the
approximations that have been used to study
unstable $f$-modes in the Newtonian
context\cite{il91}. For the simplest 
neutron star models two kinds of viscosity, bulk and shear viscosity, are normally 
considered. These are due to rather 
different physical mechanisms and we discuss them separately below. 

\subsubsection{Shear viscosity}

At relatively low temperatures (below a few times $10^9$~K) 
the main viscous dissipation mechanism in a fluid star arises from
momentum transport due to particle scattering. In the standard
approach these scattering events are modelled in terms of a 
macrosopic shear viscosity. In a normal fluid star neutron-neutron scattering 
provides the most important contribution. In a superfluid the situation is 
a bit more subtle. Electron-electron scattering leads to the 
dominant contribution to the shear viscosity, 
but one must also account for exotic effects like the scattering
off of vortices in the superfluid, cf. Section~4.5.

The effect of shear viscosity on the $r$-modes can be estimated from\cite{lom98,aks99}
\be
\left. {dE\over dt}\right|_{\rm sv} = - 2\int \eta \delta\sigma^{ab} \delta 
\sigma^*_{ab} dV 
\label{shear}\ee
where the shear $\sigma_{ab}$ follows from
\be
\delta \sigma_{ab} = {{i\omega_r}\over 2} 
\left( \nabla_a \xi_b + \nabla_b\xi_a -2 g_{ab}\nabla_c\xi^c \right)
\ee
After working out the angular integrals we are left with
\begin{equation}
\left. {dE\over dt}\right|_{\rm sv}
= -\omega_r^2 l(l+1) \left\{  \int_0^R \eta 
 a^2 | a \partial_a T_{ll}|^2 da 
 + (l-1)(l+2) \int_0^R \eta a^2 |T_{ll}|^2 da \right\} \ .
\label{shear}\end{equation}

Above the transition temperature at which the neutron star becomes 
superfluid (several times $10^9$~K), the appropriate viscosity coefficient
(due to neutron-neutron scattering) is
\begin{equation}
\eta_{nn}  = 2\times 10^{18} \rho_{15}^{9/4} T_9^{-2} {\rm g/cms} \ .
\label{shearcoeff}\end{equation}
where 
\begin{eqnarray}
\rho_{15} &=& \rho / 10^{15} {\rm g/cm}^3 \\
T_9 &=& T /  10^9 {\rm K}
\end{eqnarray}
This leads to an estimated dissipation time-scale due to shear 
viscosity\cite{aks99,lom98,ks99} 
\be
t_{\rm sv} \approx M_{1.4}^{-5/4} R_{10}^{23/4} T_9^2
\times \left\{ \begin{array}{ll} 1.2 \times 10^8 \mbox{ s} \quad n=0 \\
 6.7\times10^7 \mbox{ s} \quad n=1 \end{array} \right.
\label{sv1}\ee
To obtain these results the star has been assumed to be isothermal.
This should be a reasonable approximation apart from  for the first 
few moments
of a neutron stars life.
These estimates of the shear viscosity are expected to
be relevant for the first months of the
life of a hot young neutron star.

Once the star has cooled below the superfluid transition 
temperature, the above viscosity coefficient must be 
replaced by 
\be 
\eta_{ee}  = 6\times 10^{18} \rho_{15}^2 T_9^{-2} 
{\rm g/cms} \ .
\ee
which follows from an analysis of electron-electron
scattering. This leads to\cite{aks99,lom98,ks99} 
\be
t_{\rm sv} \approx M_{1.4}^{-1} R_{10}^{5} T_9^2
\times \left\{ \begin{array}{ll} 3.6 \times 10^7 \mbox{ s} \quad n=0 \\
2.2\times 10^7 \mbox{ s} \quad n=1 \end{array} \right.
\label{sv2}\ee

The above shear viscosity estimates were based on 
standard Navier-Stokes theory as developed by, for
example, Landau and Lifshitz\cite{landau}.
However,  this 
theory is somewhat pathological in that it allows
dissipative signals to travel faster than light. A causal theory of 
dissipation has been developed following initial work by 
Israel and Stewart\cite{israel}. 
This alternative description differs from  
Navier-Stokes theory in some potentially important
ways. The main feature of the Israel-Stewart
description is that
there is a 
coupling between rotation and the gradients of 
both temperature and momentum. Intuitively one might
expect these additional effects to be small, but there are
two situations where they may be relevant. 
For high frequency oscillations the large time-derivatives may lead
to significant corrections to the Navier-Stokes
results. Similarly, rapid rotation can lead to a strong 
coupling. The latter effect could certainly be relevant 
for the $r$-modes (or indeed any other mode
in a rapidly spinning star). 
Inspired by this possibility,
Rezania and Maartens\cite{rm99} have 
estimated the relevance of the coupling between
vorticity and shear viscosity for the $r$-modes.
Their results show that the correction due to the vorticity 
coupling can be large,  especially for low temperatures. 

\subsubsection{Bulk viscosity}

At high temperature (above a few times $10^9$~K) 
bulk viscosity is the dominant dissipation
mechanism. Bulk viscosity arises because
the pressure and density
variations associated with the mode oscillation
drive the fluid away from beta equilibrium.
It corresponds to an estimate of the extent to which energy is
dissipated from the fluid motion as the weak interaction
tries to re-establish equilibrium.
The mode energy lost through bulk viscosity is carried away 
by neutrinos. 

In the previous section we found that we could calculate the shear 
viscosity timescale from the leading order contribution to the 
$r$-mode fluid motion, cf. (\ref{shear}). 
This is not the case for the bulk viscosity.
To assess its relevance we need the 
Lagrangian density/pressure perturbation.
But for $r$-modes these perturbations vanish to leading order in rotation.  
This means that the mode calculation
must be carried at least to order $\Omega^2$ if we want to
estimate the bulk viscosity  timescale. In particular, we need
\be
\left. {dE\over dt}\right|_{\rm bv} = - \int \zeta  | \delta \sigma |^2
\ee
where $\delta \sigma$ is the expansion associated with the mode, 
defined by
\be
\delta \sigma = -i \omega_r {\Delta \rho \over \rho} 
= -i \omega_r {\Delta p \over \Gamma p} 
\label{expand}
\ee
and
\be
\Delta p = \delta p + {dp \over da} \xi^a =  \rho g a\sum_{lm} 
\left[ \zeta_{lm} - S_{lm} \right] Y_{lm} \ ,
\ee
where we have used the fact that 
$dp/da = -\rho g$ for a Newtonian model.
The detailed calculation of the relevant eigenfunctions
has been described by, for example, Saio\cite{sa82}.
 
After doing the angular integrals we get (for $l=m$)
\be
\left. {dE\over dt}\right|_{\rm bv} = 
- \int_0^R \zeta \left( {\rho g a^2 \over \Gamma p} \right)^2
\left[ \zeta_{l+1l} - S_{l+1l} \right]^2 da \ .
\label{bulk}\ee

In the case when $\beta$-equilibrium is regulated by the 
so-called modified URCA reactions
the relevant viscosity coefficient is
\be
\zeta_{mU}  = 6\times 10^{25} \rho_{15}^2 T_9^6  
\left( {\omega_r \over1 {\rm Hz} } \right)^{-2} {\rm g/cms}\ .
\label{modurc}\ee
This leads to an estimated bulk viscosity timescale for $n=1$ 
polytropes\cite{aunp} 
\be
t_{\rm bv} \approx 2.7 \times10^{11} M_{1.4} R_{10}^{-1} P_{-3}^2 T_9^{-6} 
\mbox{ s}
\label{bulkest}\ee
This result, which was obtained within the Cowling approximation 
agrees well with a calculation including the perturbed gravitational 
potential\cite{lmo00}.

We note that it has been speculated that the proton fraction may be large enough to 
make direct URCA reactions possible in the cores of neutron stars.
If this is the case, the bulk viscosity coefficient becomes
significantly larger than (\ref{modurc}). However, direct URCA processes
will only be relevant if the proton fraction is 
above several percent\cite{durca,zdunik}. Since this is unlikely 
to be the case in the surface regions
where the $r$-modes are mainly located we will not consider the 
direct URCA bulk viscosity in detail here.

Let us also point out that, since very hot
stars are no longer transparent to neutrinos, the
above estimates for the bulk viscosity will not remain relevant 
at very high temperatures~\cite{lai}. This may be of some
relevance for the very early phase of $r$-mode
growth in a newly born neutron star.

Another point, that may turn out to be of crucial importance, was 
recently raised by Jones~\cite{pbj}. He pointed out that the bulk 
viscosity result would be significantly different if the 
presence of hyperons in the neutron star core was accounted for. 
Not only is the associated viscosity coefficient stronger
than assumed above, the temperature dependence is also different in the 
hyperon case (the coefficient scales as $T^{-2}$ rather than $T^6$). 
This makes the hyperon bulk viscosity relevant at low
temperatures. Jones argues that the $r$-mode instability
is almost completely suppressed in the case when hyperons are present
throughout the entire star. This is, however, not a particularly
realistic assumption. In order to assess the extent to which 
hyperons affect the estimated instability time-scales calculations 
that allow for the presence of exotic particles (as predicted
by modern equation of state results)
need to be performed. At the time of writing, no such results
are available.

Since there has been some debate in the literature 
regarding the estimated timescales for the bulk viscosity
it is relevant that we try to identify the reasons why
various given estimates differ. First of all, let us discuss the 
constant density estimate of Kokkotas and Stergioulas\cite{ks99}. 
As is easy to 
see from (\ref{expand}), one cannot rigorously calculate  
the expansion associated with a mode in a constant density star.
After all, the assumption $\rho={\rm constant}$ immediately leads to $\Delta \rho =0$.
Still, one can estimate of the relevant effect (following
Cutler and Lindblom\cite{cl87}) in the following way:
Assume that the Lagrangian variation in the pressure takes the following form
\be
{\Delta p \over p} \approx {\epsilon l \over \Gamma} \left( {a \over R}
\right)^l e^{i\omega_r t} Y_{lm} 
\label{best}\ee
where $\epsilon$ is a dimensionless normalisation constant. 
This formula 
then leads to the anticipated behaviour 
as $\Gamma \to \infty$, i.e. as 
$\rho \to \mbox{ constant}$, and the formula also agrees with the
result for $f$-mode oscillations in compressible stars (to within a
factor of two). Kokkotas and Stergioulas\cite{ks99} use this approximation also
for the $r$-modes, and arrive at an estimated 
bulk viscosity timescale (actually using $\Gamma = 5$ in (\ref{best}))
\be
t_{\rm bv} \approx 2.4 \times 10^{10} M_{1.4}^{-1} R_{10}^5 T_9^{-6} P_{-3}^2 \mbox{ s}
\ee
Compared to (\ref{bulkest}) this is an  
underestimate of the strength of bulk viscosity dissipation by roughly
one order of magnitude.

An estimate that was equally far away from the correct value was
given by Lindblom, Owen and Morsink\cite{lom98}. They tried to approximate the bulk 
viscosity using only first order (in $\Omega$) results. 
In particular, they assumed
that $\Delta \rho \approx \delta \rho$. This assumption is, however, 
not warranted for the $r$-modes. As we discussed 
in Section~2.3, 
one of the fundamental $r$-mode 
properties is that the fluid essentially moves on 
isobars in the perturbed
configuration. Hence, we will have $|\Delta p| << |\delta p|$, which results
in Lindblom, Owen and Morsink overestimating the strength of the bulk viscosity by roughly one order of magnitude. 

Finally,  the first published 
second order (in $\Omega$) results (due to Andersson, Kokkotas
and Schutz\cite{aks99}) also differ considerably from
(\ref{bulkest}), 
essentially in the scaling with 
the rotation rate. This was due to a typographical error in the 
numerical code used to 
calculate the mode eigenfunctions.  
Once this mistake is corrected, we arrive at (\ref{bulkest}).

\subsection{The $r$-mode instability window}

Armed with the estimates obtained in the previous two sections
we can address the main issue of interest: Will the 
$r$-mode instability be relevant for astrophysical
neutron stars?
First of all, it is easy to see that the $r$-modes will only
be unstable in a certain range of temperatures. To have an 
instability we need
$t_{\rm gw}$ to be  smaller in magnitude than both $t_{\rm sv}$ and 
$t_{\rm bv}$. From the various estimates we immediately
see that
the dissipation due to shear viscosity kills the mode
at low temperatures, while
the bulk viscosity dominates at high temperatures. In fact, 
by comparing (\ref{sv2}) 
to (\ref{gwest}) we deduce that shear viscosity will completely suppress the 
$r$-mode instability at core temperatures below $10^5$~K.
Similarly, bulk viscosity will prevent the mode from growing 
in a star that is hotter than a few times
$10^{10}$~K (but see the proviso
in section~4.2.2 regarding the relevance of the bulk viscosity 
at high temperatures).
In the intermediate
range there is a window of opportunity where the growth time due
to gravitational radiation is short enough to overcome the
viscous damping and drive the $r$-mode unstable. 

\begin{figure}[h]
\hbox to \hsize{\hfill \epsfysize=6cm
\epsffile{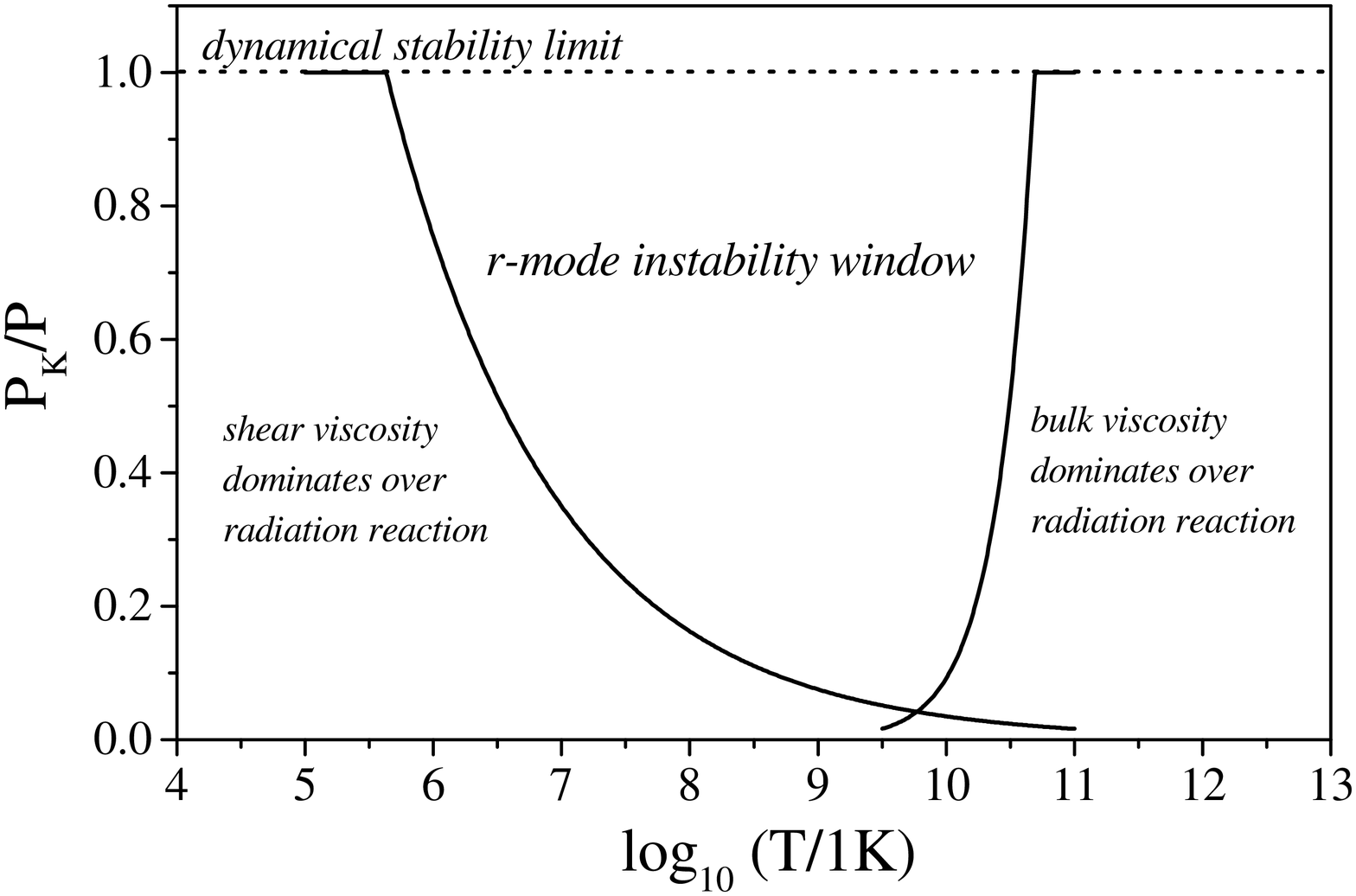} \hfill}
\fcaption{The critical rotation rates at which 
shear viscosity (at low temperatures) and bulk 
viscosity (at high temperatures) balance gravitational 
radiation reaction due to the $r$-mode 
current multipole. This leads to the notion of a 
``window'' in which the $r$-mode instability is active. 
The data in the figure is for the  $l=m=2$ $r$-mode of a canonical neutron star 
($R=10$  km and $M=1.4 M_\odot$ and 
Kepler period $P_K \approx 0.8$  ms).  } 
\label{win1}\end{figure}

This instability window
is usually illustrated by the critical rotation period, above which the
mode is unstable, as a function of temperature\cite{il91,lom98,aks99}. 
We find the relevant critical rotation rate by solving for the zeros
of
\be
{1\over 2E} {dE \over dt} = {1\over t_{\rm gw}} + 
\sum {1\over t_{\rm diss}} = 0
\ee
for a range of temperatures. A typical result (for canonical neutron star
parameters) is shown in Figure~\ref{win1}.
 For canonical values, the Kepler limit corresponds to $P_K=0.8$~ms.
In the figure we show the 
critical period curve for an  $n=1$ polytrope.
From this data we can 
deduce
that a rotating neutron star with a core temperature of $10^9$~K (i.e. a 
few months old) is unstable at rotation periods shorter than 25~ms.
There is, of course, a large uncertainty associated with this prediction.
If we, for example, assume that the instability sets in at roughly
the same $P_K/P$ for different equations of state (which is supported
by the $n=0$ and 1 results), then 
the uncertainty in the  Kepler period ($P_K\approx 
0.5-2$~ms) for realistic equations of state
suggests that the instability will be relevant at rotation 
rates faster than $P\approx 10-40$~ms at some 
core temperature.

Figure~\ref{win3} shows the different strengths of the normal fluid
shear viscosity (due to neutron-neutron scattering) and the superfluid
counterpart (due to electron-electron scattering). It is  
interesting to note that the superfluid is, rather 
counterintuitively,  more dissipative that the
normal fluid. We also indicate the strength of the $l=m=2$
$r$-mode mass multipole.  
As a comparison we should recall that the $m=4$ $f$-mode (believed to 
be the ``most unstable'' spheroidal mode)
becomes unstable
above $P_K/P \approx0.95$.  In other words, the $r$-mode instability 
provides a much stronger constraint on neutron star rotation than the 
instability in the $f$-mode does. In fact, even the comparatively weak 
radiation reaction associated with the $r$-mode mass 
multipoles
leads to an instability that completely dominates the $f$-mode one.

\begin{figure}[h]
\hbox to \hsize{\hfill \epsfysize=6cm
\epsffile{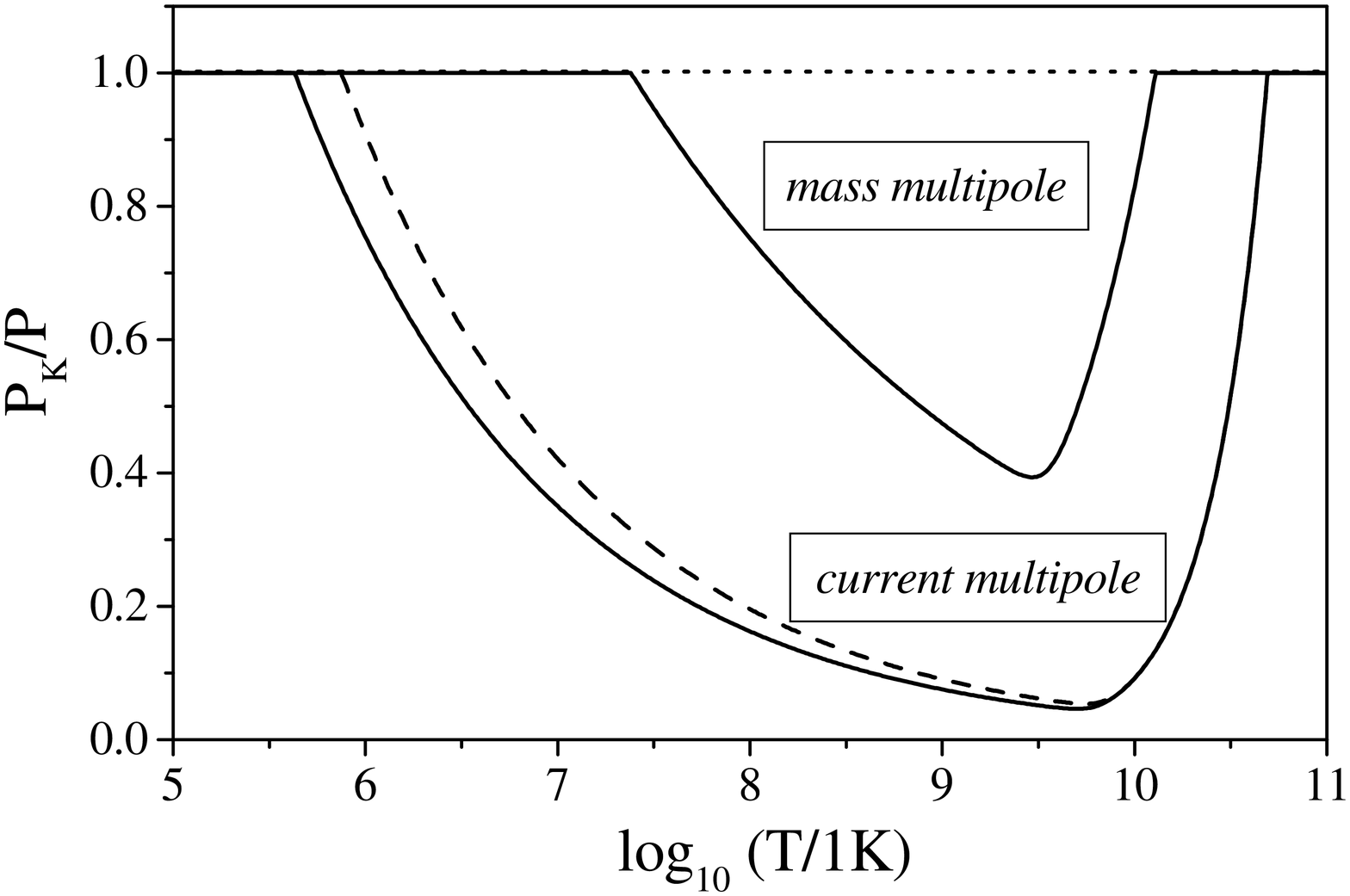} \hfill}
\fcaption{The $r$-mode instability window for a canonical neutron star 
(the data
is for an  $n=1$ polytrope). The lower two curves illustrate the difference 
between shear viscosity due to neutron-neutron scattering (solid line)
and electron-electron scattering (dashed curve). The latter should be 
relevant
for a superfluid star (i.e. below a few times $10^9$~K). It should be noted
that the superfluid shear viscosity is stronger than that of a 
normal fluid. We also show the 
instability window that arises if we only include the 
mass multipole radiation from the $l=m=2$ $r$-mode. It is useful to compare 
these results to those for the $m=4$ $f$-mode, that suggest that this mode is
unstable for $P_K/P>0.95$ or so.  } 
\label{win3}\end{figure}

Finally, 
 we illustrate the relative importance of the 
higher $l=m$ $r$-modes in Figure~\ref{win3}. 
It should be noticed that, even though the 
growth time increases with roughly one order of magnitude for each $l$
the higher multipole $r$-modes still provide a severe constraint
on neutron star rotation rates.

\begin{figure}[h]
\hbox to \hsize{\hfill \epsfysize=6cm
\epsffile{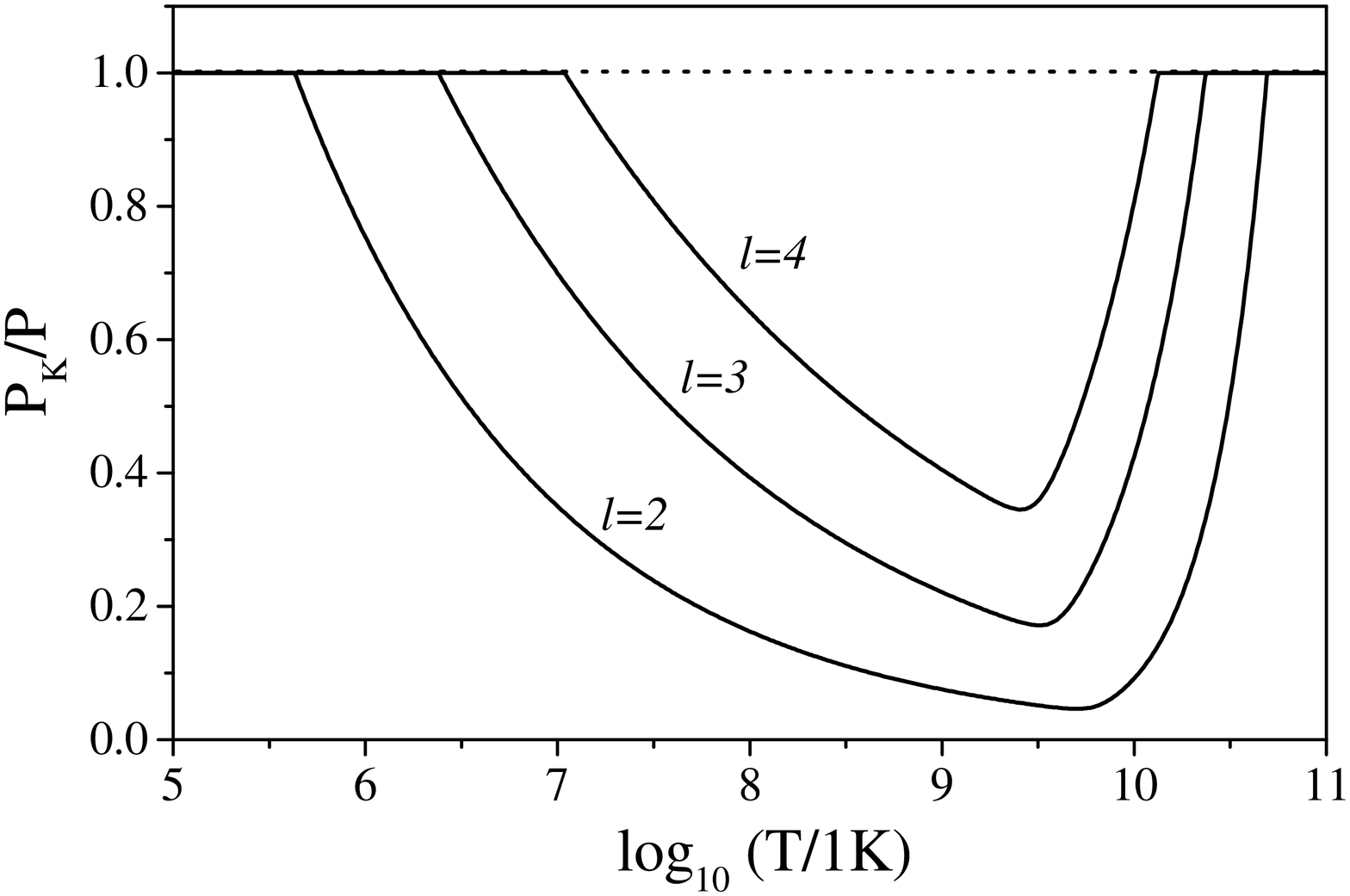} \hfill}
\fcaption{The $r$-mode instability window for a canonical neutron star (the data
is for an  $n=1$ polytrope) and different $l$-multipoles.  } 
\label{win3}\end{figure}

The results in Figures~\ref{win1}-\ref{win3} are encouraging and 
perhaps indicative 
of the astrophysical role of the $r$-mode instability\footnote{We note in 
passing that the $r$-modes may also be unstable in the fastest rotating white
dwarfs\cite{akst99,wdlee,wdhisc}; the so-called DQ Herculis stars. }. But more than anything
else they emphasize the 
urgent need for an improved understanding of the various elements in our model.
We will now discuss various attempts to improve 
on the detailed
physics included in the $r$-mode calculations. 

\subsection{Superfluid mutual friction}

As a neutron star cools below a few times $10^9$~K the extreme
density in the core is expected to lead to the formation of
a superfluid\cite{sflu1,sflu2}. This means that, while it may not be wholly
inappropriate to model a newly born neutron star as a
simple ``ball'' of a more or less perfect fluid, a model 
of an old neutron star must certainly be  more complicated.
In many ways it is useful to think of an 
old (which in this context means: older than a few hours/days)
neutron star as a ``layer cake''. A  simple, yet reasonable,  
picture of an old neutron star consists of i) a core in which 
superfluid neutrons (in a $^3 P_2$ condensate) coexist 
with superconducting protons (above $\rho \approx 1.5 \times 10^{14}$~g/cm$^3$), ii) a crust: a lattice
of nuclei permeated by  superfluid neutrons in a $^1 S_0$ state 
(the superfluid reaches out to the neutron drip density $\sim 10^{11}$~g/cm$^3$), and
iii) a fluid ocean.
A canonical neutron star would have a core
of 9~km, a crust of 1~km and a thin ocean 50-100 meters deep. 
The superfluid constituents play a crucial 
role in determining the dynamical properties of a rotating neutron
star. In particular, the interplay between the lattice nuclei  
and the superfluid  in the inner crust is a key agent in the
standard model for neutron star glitches. Intuitively,  the
presence of a superfluid should  have a considerable effect also on 
the various modes of pulsation. To some extent this is already
indicated by the results
in Figure~\ref{win3}, and when the presence of new dissipation mechanisms
is accounted for it turns out  that a superfluid 
may be more dissipative than a normal fluid! The most important
 new effect is the so-called mutual friction 
that arises from scattering
of electrons off of the neutron vortices (recall that a superfluid
mimics large scale rotation by forming a large number 
of vortices). This scattering 
is greatly  enhanced by the strong magnetic field (due to the 
entrained protons) within the 
vortices.  

The motion associated with a neutron star pulsation mode is
of large scale compared to both the neutron vortex and 
superconducting proton magnetic flux tube separations
($\sim 10^{-3}$~cm and $\sim 10^{-10}$~cm, respectively). This means
that it is sufficent to include the averaged dynamics of many vortices
and flux tubes in a model for superfluid stellar pulsation. 
This  leads to models consisting of two distinct, 
but dynamically coupled, 
fluids representing the superfluid neutrons and the ``protons'' (a fluid that 
contains all charged components in the star). The dynamics can then be 
described in terms of two velocity fields, described by two sets of equations
similar to (\ref{euler}) and (\ref{masscon}) with additional 
coupling terms (see
equations (1) to (4) in Lindblom and Mendell\cite{lm9x}).
It is interesting to note that this superfluid model leads to a 
new set of pulsation modes already in the nonrotating case\cite{epst,sflee}. 
In order to discuss the difference between
these modes and the familiar $f/p$-modes
it is useful to introduce a variable $\delta \beta$,  
which represents the ``departure
from $\beta$-equilibrium'' in the oscillating fluid.
Then the standard
fluid modes all have small $\delta \beta$. For these modes
the superfluid neutrons and the protons move more or less together. 
In contrast, the superfluid modes have $\delta \beta$ large, and the neutrons 
and protons are largely countermoving.
 
Superfluid mutual friction has been shown to completely suppress the 
instability associated with the $f$-mode in a rotating star\cite{lm95}. 
Plausibly, 
it is a dominant damping agent also on the unstable $r$-modes.
We can assess the relevance of mutual friction in a hand-waving way
following Mendell\cite{mendiss}: For a longitudinal wave travelling perpendicular
to the rotation axis the influence of mutual friction can be estimated as
\be
t_{\rm mf} \approx {\Gamma_{\rm mf} \over \Omega} {1 \over \lambda^2} t_{\rm sv}
\ee
Here $\lambda$ is the wavelength of the oscillation (which we will take
to be $\lambda \approx R \approx 10$~km), and 
\be
\Gamma_{\rm mf} < 10^{10} T_9^{-2} \mbox{ cm}^2/\mbox{s}
\ee
Comparing this to our estimated shear viscosity timescale we see that
\be
t_{\rm mf} > 200 \mbox{ s} 
\ee
and that this timescale is independent of the temperature. 
This  is  in clear contrast with the shear and bulk viscosities, both
of which are sensitive to changes in the core temperature. 
This result suggests that, even though  mutual friction may be important
it will not completely suppress the $r$-mode instability. In fact, we would 
expect a detailed calculation to lead to timescales considerably longer
than this upper limit. This is since mutual friction 
would have the largest effect on waves that travel perpendicular
to the rotation axis (as in our example) while the velocity field
of a pulsation mode has a considerable component parallell
to the axis. 

In the ``macroscopic'' picture, the  strength of mutual friction can be 
expressed in terms of  the ``entrainment parameter''
\be
\epsilon = {\rho_p \over \rho_n} \left( {m_p^* \over m_p} - 1 \right)
\ee
where  $\rho_n$ and $\rho_p$ are the neutron and proton  densities, 
respectively, 
and $m_p$ and $m_p^*$ are the 
bare and effective proton masses. Estimates of the ratio
of these two masses suggest that\cite{lm00}
\be
0.3 \le  {m_p^* \over m_p} \le 0.8
\ee
Steps towards more detailed studies of superfluid $r$-modes 
have been taken by Lindblom and Mendell\cite{lm00}. 
They have calculated the modes in a superfluid star and shown that 
they  are identical to those of a 
normal fluid star to lowest order in $\Omega$. Therefore the 
growth timescale due to gravitational radiation reaction remains 
essentially unchanged in a superfluid star, cf. (\ref{gwest}). 
For the  specific two-fluid stellar model considered by Lindblom and Mendell,
the shear viscosity timescale turns out to be roughly a factor of 3 shorter
than in the normal fluid case, cf. (\ref{sv1}). 
As far as the mutual friction is concerned, the calculations of
Lindblom and Mendell do not provide a complete answer to the 
question of whether this mechanism will suppress the 
$r$-mode instability or not. 

Lindblom and Mendell find that the typical result for $r$-mode
dissipation due to mutual friction is\cite{lm00}
\be
t_{\rm mf} \approx 2\times10^5 P_{-3}^5 \mbox{ s} \ .
\ee
This means that the $r$-mode instability window in (say) Figure~\ref{win1}
would be essentially unaffected by the inclusion of mutual friction.
However, as manifested in Figure~6 of Lindblom and Mendell, 
the result is sensitive to 
changes in the entrainment parameter. In particular, there are 
critical values of $\epsilon$ for which the mutual friction 
timescale becomes very short (and for which the unstable $r$-mode would 
be completely suppressed). It is interesting to note
that $\delta \beta$ is a key parameter 
in deciding the importance of mutual friction. 
The critical values of $\epsilon$ for which the mutual friction timescales
are short correspond to modes for which $\delta \beta$ is 
comparatively large. As already mentioned, the protons and neutrons are 
essentially countermoving for large $\delta \beta$. 
Given that mutual friction tends to 
damp out the relative motion between 
the neutrons and the protons,  it is natural that
modes with large
$\delta \beta$ should be strongly affected by it. 
The results of Lindblom and Mendell seem to suggest that there are
values of the entrainment parameter for which the $r$-modes are
in ``resonance'' with a superfluid mode (with large $\delta \beta$). 
This would explain why these parameter values lead to  
significantly increased dissipation.
Further studies of this problem are clearly needed 
if we are to understand the $r$-modes in a superfluid star
fully. Investigations of this problem should be strongly encouraged, 
both in Newtonian theory and general relativity\cite{comer99}.

\subsection{Ekman layer in neutron stars with a crust}

In addition to the core superfluid, we need to consider
effects due to the presence of a solid crust in an 
old neutron star. The melting temperature of the crust 
is usually estimated to be of the order of  $10^{10}$~K (for a
non-accreting neutron star), 
so the crust may form shortly after the neutron stars birth\cite{haensel}. 

That the presence of a solid crust will have a crucial effect 
on the $r$-mode motion can be understood as follows: Based on the 
perfect fluid mode-calculations we anticipate  the transverse motion
associated with the mode at the crust-core boundary
to be large. However,  if the crust is assumed to be rigid
the fluid motion must 
essentially fall off to zero at the base of the crust
in order to satisfy a no-slip
condition (in the rotating frame of reference). 
We can estimate the relevance of the crust using 
viscous boundary layer theory\cite{bu00}. The region 
immediately beneath the crust
then corresponds to a so-called Ekman layer.
The thickness of the boundary layer ($\delta$) can be deduced
by balancing the Coriolis force and shear viscosity\cite{g64}:
\be
\delta \sim \left( {\eta \over \rho \Omega} \right)^{1/2}
\ee 
where $\eta$ is the shear viscosity coefficient (\ref{shearcoeff}).
As is easy to show, this would correspond to a few centimetres for
typical parameters of a rapidly rotating neutron star. 
In other words, $\delta << R$ and the boundary layer approximation is 
warranted. 

The 
dissipation timescale due to the presence of the Ekman layer 
roughly follows from
\be
t_{\rm Ek} \approx {t_{\rm sv} \over \sqrt{Re}} 
\ee 
where $Re = \rho_b R_b^2 \Omega/\eta$ is the Reynolds 
number (the ratio
between the Coriolis force and viscosity), 
and $R_b$ and $\rho_b$ are the location of, and density in, the Ekman
layer, respectively.  
For typical neutron star parameters one would expect the 
base of the crust to lie at $\rho_b \approx 1.5\times10^{14} \mbox{ g/cm}^3$.
We would then have $Re\sim 10^{13}$, and 
 we see that the presence of a solid crust leads to a 
dissipation channel that is many orders of magnitude more
effective than 
the standard shear viscosity. 

A more detailed estimate of this effect
has been made by Bildsten and Ushomirsky\cite{bu00}.
The required dissipation rate follows from an
integral  over the surface area at the crust-core 
boundary
\be
\left. { dE \over dt}\right|_{\rm Ek} = 
- R_b^2 \int {|\delta \vec{u}|^2 \over 2}  \left( {\rho_b \omega_r
 \eta \over 2} \right)^{1/2} \sin \theta d\theta d\varphi 
\ee 
For canonical neutron star parameters and an $n=1$ polytrope we
can estimate that the base of the crust corresponds to $R_b\approx 9.4$~km.
When combined with the estimated mode-energy, this leads to 
an estimated damping timescale (note that this estimate in valid only 
for a $R=10$~km and $M=1.4M_\odot$ star)
\be
t_{\rm Ek} \approx 1.4\times 10^3 T_9 P_{-3}^{1/2} \mbox{ s}  
\label{ekman1}\ee

\begin{figure}[h]
\hbox to \hsize{\hfill \epsfysize=6cm
\epsffile{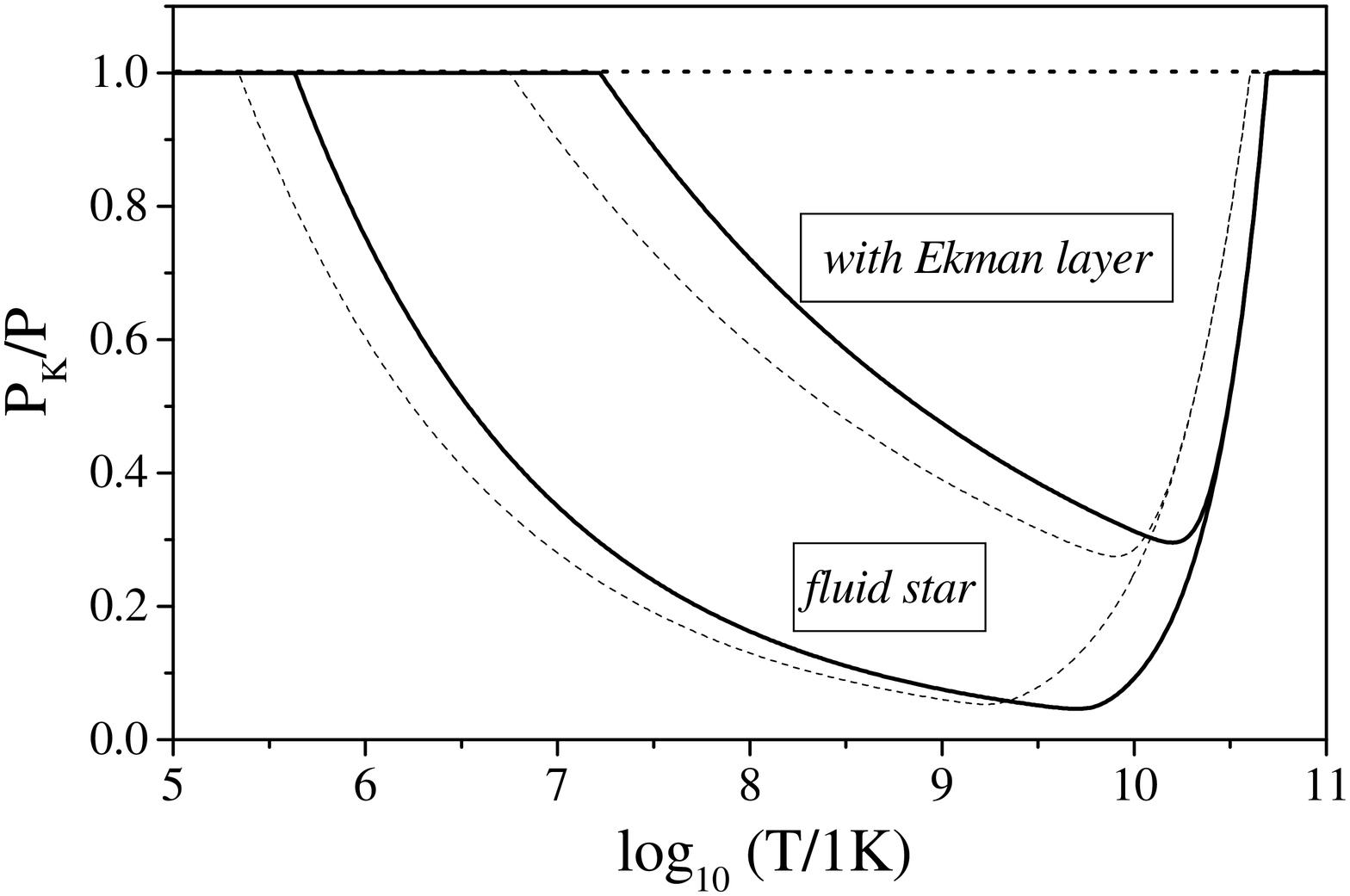} \hfill}
\fcaption{The $r$-mode instability window for a canonical fluid neutron star 
(lower solid curve) is compared to that for a star with an 
Ekman layer operating at
the base of the solid crust (upper solid curve). Also shown (as dashed curves)
are the corresponding instability windows for a constant density star.
It should be noted that the strength of the $r$-mode instability is 
only weakly dependent
on the stiffness of the equation of state.   } 
\label{win4}\end{figure}

This estimate was arrived at by appealing to an analogy between the
problem at hand and the standard one of an oscillating plate in a 
viscous fluid\cite{landau}.
One would not expect the change in geometry and inclusion
of the 
Coriolis force in the $r$-mode problem  
to change the order of magnitude of the effect.
 That this assumption is correct has  been shown by 
Rieutord\cite{rieu00}, who used an exact solution 
to the incompressible problem to show that when the 
angular dependence of the 
velocity in the Ekman layer is accounted for the dissipation timescale
for the $l=m=2$ $r$-mode is  a factor of 1.74 shorter than (\ref{ekman1}). 
The resultant estimate of the dissipation due to the presence
of the Ekman layer is
\be
t_{\rm Ek} \approx 830  T_9 P_{-3}^{1/2} \mbox{ s}  \ .
\label{tek2}\ee
We compare the $r$-mode instability window obtained from this estimate
to the normal fluid result in Figure~\ref{win4}. Clearly, a
solid crust has an influence that is far greater than any previously 
considered dissipation mechanism. For example, our estimate suggests 
that all neutron stars with a crust are stable at rotation periods
longer than roughly 5~ms.

The crust-core interface has been the focus of 
several recent studies. These studies add further dimensions
to the problem.  The interplay between the $r$-modes in the 
fluid core and modes in the solid crust is particularly 
interesting. It has long been known that the crust supports 
toroidal modes of oscillation (whose frequencies depend
directly on the crust shear modulus)\cite{mcdermott,strohmayer}.
Detailed calculations by Yoshida and Lee\cite{crust1} (see also
Levin and Ushomirsky\cite{crust2}) show that as the spin
of the star increases the unstable $r$-modes will undergo a series
of so-called avoided crossings with the crust modes. 
In a nonrotating star, the fundamental $l=2$ crust mode 
has a frequency
\be
\omega_c (\Omega=0) \approx 10^{-2} \sqrt{GM \over R} \ .
\ee
In the rotating case the mode-frequency changes in such a way that
\be
\omega_c(\Omega) \approx \omega_c (\Omega=0) + {m\Omega \over l(l+1)}
\ee
(in the rotating frame). Now it is easy to see that 
the toroidal mode becomes comparable
to the $r$-mode frequency at a rotation frequency\cite{strohmayer,ls96}
\be
\Omega_{\rm cross} \approx 5\times 10^{-2} \Omega_K
\ee 
Hence, avoided crossings between the modes occur at very 
low rates of rotation. This is conceptually interesting and it may have 
 repercussions on many of our estimates regarding the $r$-mode
instability. Most importantly, it casts  doubt on extrapolations of
the slow-rotation results into the regime of fast rotation. 

In addition to studying this new feature of the problem, Levin
and Ushomirsky\cite{crust2} show that the assumption of a 
rigid crust, which was made
in the above estimates of the Ekman layer dissipation rate,  
is likely not warranted. They show that the $r$-mode typically extends
into the crust. We can quantify the extent to which this affects the estimated
dissipation timescale (\ref{tek2}) in terms of the ratio $\Delta= |\Delta v |/|v|$, 
where $\Delta v$ is the difference between the core fluid velocity immediately
beneath the crust and the velocity induced in the crust and $v$ is the 
fluid velocity in absence of the crust. The rigid crust assumption
corresponds to $\Delta=1$. In contrast, the toy model calculations
of Levin and Ushomirsky\cite{crust2} suggest that a typical 
results should be $\Delta\approx 0.1$. This then affects the 
Ekman layer timescale by a factor of $1/\Delta^2$, so the true dissipation
may well be a factor of at least 100 weaker than (\ref{tek2}).  

Despite some recent advances in our understanding of the
effects that a solid crust may have on the $r$-modes it is 
clear that several crucial issues remain
to be investigated in detail.
For example,  the inner crust of a neutron star (out to the neutron
drip density) will likely be permeated by superfluid neutrons. 
It is not at all clear at the moment whether one should 
expect these neutrons to be strongly pinned to the 
crust nuclei or not. But if the superfluid is at all
free to move relative to the crust it will 
likely lead to a weaker Ekman layer
dissipation on the $r$-modes.

\subsection{Strange/quark stars}

Ever since Witten\cite{witten} suggested that strange matter (a conglomerate
of strange, up and down quarks) would be the most stable 
form of matter, it has been 
speculated that strange stars might exist in the universe\cite{alcock,hzs,colpi}.  It has been suggested that some (perhaps all) 
pulsars are actually strange stars. The observational evidence for this
is, however, rather tenuous. The main reason for this is 
that strange stars
are held together by both the strong interaction
and gravity, and at a mass
of one and a half solar masses, gravity dominates over the 
strong interaction. Thus it is not 
easy to distinguish a strange star from a neutron star,   
particularly not since the strange star is expected to 
be ``covered'' by a thin
nuclear crust\cite{alcock}. Still, the  
existence of strange stars
remains a conjecture which is consistent with, for example, 
millisecond pulsar observations.  

As was pointed out by Madsen\cite{jm99}, the $r$-mode 
instability may provide the means for distinguishing 
between strange stars and neutron stars. 
The main reason for this is that the viscosity coefficients
are rather different in the two cases. While the shear viscosity 
in a strange star would be comparable to that of a neutron star, 
the bulk viscosity would be many orders of magnitude stronger
than its neutron star counterpart. This has interesting
effects on the $r$-mode instability.

For the shear viscosity the relevant coefficient would be
\be 
\eta \approx 1.7\times10^{18} \left({ 0.1\over \alpha_s  } \right)^{5/3}
\rho_{15}^{14/9}T_9^{-5/3} \mbox{g/cms}
\ee
where $\alpha_s$ is the fine-structure constant for the strong interaction.
This leads to
\be
t_{\rm sv} \approx 7.4\times 10^7 \left( {\alpha_s \over 0.1} \right)^{5/3}
M_{1.4}^{-5/9} R_{10}^{11/3} T_9^{5/3} \mbox{ s}
\ee

Meanwhile, the situation is more complicated for the 
bulk viscosity (which now is a result of the change in 
concentration of down and strange quarks in 
response to the mode oscillation). The relevant 
viscosity coefficient takes the form\cite{jm99,jm00}
\be
\zeta = {\alpha T^2 \over \omega_r^2 + \beta T^4}
\ee
where the coefficients
$\alpha$ and $\beta$ are given by Madsen. From this 
we can immediately deduce that the bulk viscosity 
becomes less important  at both very low and very high temperatures. 
For low temperatures  we find that
\be
t_{\rm bv}^{\rm low} \approx 7.9 M_{1.4}^2 R_{10}^{-4}
P_{-3}^2 T_9^{-2} m_{100}^{-4} \mbox{ s} 
\ee
where $m_{100}$ represents the mass of the strange quark in units
of 100~MeV.
Meanwhile, at high temperatures we cannot readily
write down an expression with the appropriate
dependence on $M$, $R$, $P$ etcetera.
Instead, we have to perform numerical 
calculations for each stellar model. Typical 
results (for canonical parameters) are shown in 
Figure~\ref{win6}. From this figure we immediately
see that the $r$-mode instability would not - contrary
to the case for neutron stars - be active in 
strange stars with a core temperature of $10^9$~K.
In a strange star the $r$-modes are unstable at lower
temperatures (between $10^5-5\times10^8$~K) and
also at temperatures above a few times $10^9$~K. 
This is interesting since it means that the instability
would be active for a brief period 
after a strange star is born. Then the mode
would become stable until thousands 
of years later when the star has cooled sufficiently 
to enter the low-temperature instability window.

\begin{figure}[h]
\hbox to \hsize{\hfill \epsfysize=6cm
\epsffile{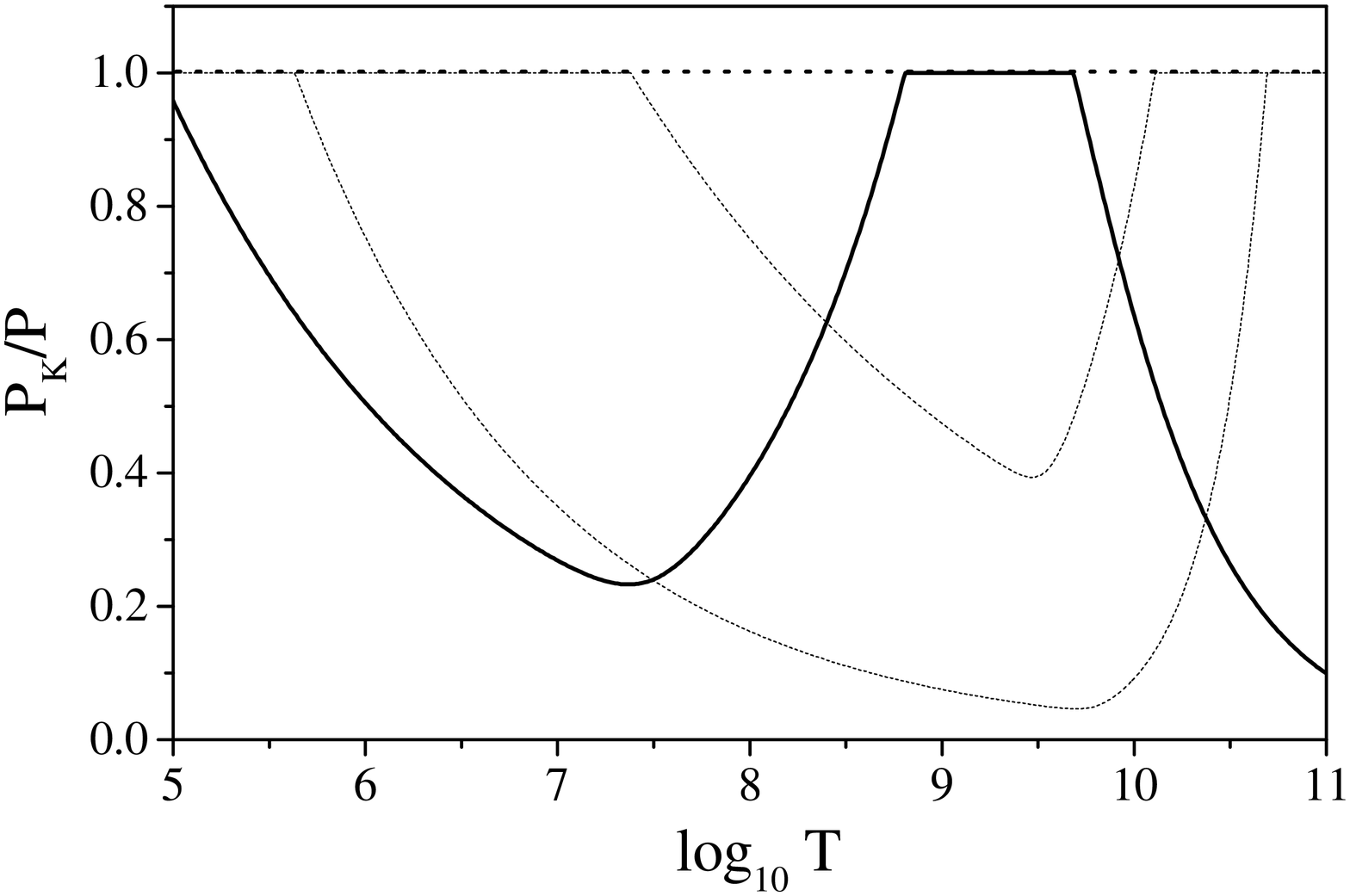} \hfill}
\fcaption{The $r$-mode instability window for a strange star
(solid line). As comparison we show the corresponding
instability results for normal fluid and crusted
neutron stars (dashed curves) from Figure~\ref{win4}. The decreased 
importance of bulk viscosity in hot strange stars is notable.} 
\label{win6}\end{figure}

Similar estimates would be valid for a neutron star with a quark core, 
a so-called hybrid star\cite{glend} and Madsen\cite{jm00} has commented
on the viscosity of some rather exotic
states of matter that may be relevant for hybrid star cores.

\section{Is the $r$-mode instability astrophysically relevant?}

In the previous sections we have described the nature
of the $r$-modes of a rotating neutron star. 
We have discussed the associated gravitational-wave
driven instability and shown that it may operate
in a neutron star with core
temperature in a window between roughly $10^5-10^{10}$~K.
This means that the  instability should be relevant for hot young 
neutron stars that are born spinning
at a rate above perhaps 5\% of the dynamical Kepler
limit.

We now want to discuss the possible astrophysical 
consequences of the $r$-mode instability. By necessity
this means that we will extrapolate the available 
results considerably. Astrophysical neutron stars are
much more complex than the simple models
for which the properties of the $r$-modes have so far
been studied. Nevertheless, the
speculations that we describe in this section are interesting and
potentially of great importance. After all, they
may eventually turn out to be largely correct!
We must, of course, keep in mind that our understanding
of much of the involved physics is (at best) 
rudimentary. 

\subsection{A phenomenological spin-evolution model}

If we want to make the discussion quantitative rather
than qualitative we must model the actual
evolution of an $r$-mode as it grows and spins down
the star. Given the current models, this is a 
challenging task and any model we devise is likely 
to be questionable.
Still, it is important that we try to gain insight 
into this issue. After all, once we have an initial 
prediction we can try to identify the crucial 
elements of the model and work to improve on them. 
It is also relevant 
to address many important astrophysical questions
at an ``order of magnitude'' level.

While the early growth phase 
of an unstable mode can be described by linear theory, 
an understanding of  many effects (such
as coupling between different modes) that will
become important and may eventually dominate the dynamics
require a nonlinear calculation. 
While such studies are outstanding we can only 
try to capture
the essential features of the behaviour. 
Intuitively one would expect a growth of an unstable mode to be halted
at some amplitude. As the mode saturates  it seems
plausible that the excess
angular momentum will be radiated away and the star
will spin down.
This general picture is supported by detailed
studies of instabilities in rotating 
ellipsoids\cite{press,miller,detweiler1,detweiler2}.

We model the evolution of a spinning star
governed by the $r$-mode instability using the 
phenomenological two-parameter model devised by Owen et al.\cite{o98}
That is, we consider
the spin rate $\Omega$ and the mode-amplitude $\alpha$
as our key quantities.
Recall that the latter is defined by expressing the 
 velocity
field (measured by an observer that is co-rotating with the star)
as (for $l=m$ modes)
\begin{equation}
\delta \vec{u} \approx \alpha \Omega R \left( { a \over R} \right)^l \vec{Y}_{ll}^B 
e^{i\omega_r t}
\label{normalise}\end{equation}

Assume that the total angular momentum of
the system can be decomposed as 
\begin{equation}
J = I\Omega + J_c
\label{angtot}\end{equation}
where the first term represents the bulk rotation
of the star and the second is the canonical
angular momentum of the $r$-mode. 
This representation makes sense since $J_c$ 
corresponds to the
second order change in angular momentum
\footnote{To be more precise: There is an ambiguity
	in deciding whether $J_c$ is the second-order change in angular
	momentum, because that second-order change also includes an
	arbitrary second order addition of an solution to the
	time-independent linearized equations that simply adds
	differential rotation to the equilibrium star.  This ambiguity
	can be  resolved by requiring that the
	perturbation be canonical and that the Lagrangian displacement 
	have no second-order part \cite{fs78a,fs78b}.  
       Under these conditions  $J_c$ is the second order change in the angular
momentum.   For a growing mode, requiring
	that the second-order part grows exponentially in the manner
	$e^(t/\tau)$, (with $\tau$ the growth time of the mode energy)
	eliminates the ambiguity in a physically more appropriate way,
	but the consistent solution requires one to include the
	also the second order radiation-reaction term.  } 
due to the presence of the mode\cite{fs78a,fs78b} (in absence
of viscosity or radiation).
In the  case of $r$-modes we have
\begin{equation}
J_c = - {3 \Omega \alpha^2 \tilde{J} MR^2 \over 2}
\end{equation}
where we have defined
\begin{equation}
\tilde{J} = {1\over MR^4} \int_0^R \rho a^6 da= 
\left\{ \begin{array}{ll} 3/28\pi \quad n=0 \\
1.635\times 10^{-2} \quad n=1 \end{array} \right.
\end{equation}

We now assume that angular momentum is only 
dissipated via gravitational waves.
The gravitational-wave luminosity 
follows from the $l=m=2$ current multipole
formula, and we get
\begin{equation}
\dot{J}_{\rm gw} = 3\Omega \alpha^2 {\tilde{J}
MR^2 \over t_{\rm gw}}
\end{equation}
Taking a time-derivative of (\ref{angtot})
and defining 
\be
\tilde{I}=I/MR^2 =\left\{ \begin{array}{ll} 2/5 \quad n=0 \\
0.261 \quad n=1 \end{array} \right.
\ee
we have
\begin{equation}
\left( \tilde{I}
- {3\over 2} \alpha^2 \tilde{J} \right)
{d\Omega \over dt} - 3\Omega \alpha \tilde{J}
{d\alpha \over dt} = {1\over MR^2}\dot{J}_{\rm gw} 
\label{evl1}\end{equation}
Here we have neglected the dependence of the 
moment of inertia on the spin rate (terms of order 
$\Omega^2$). It is easy to include such terms and
show that they have a minor effect.

A second equation follows from the expression
for the energy of the mode (as measured in the rotating system), cf.
(\ref{rotenerg}),
\begin{equation}
E = {1\over 2} \alpha^2 \Omega^2 MR^2 \tilde{J}
\end{equation}
and the fact that the mode grows according
to
\begin{equation}
{dE \over dt} = - 2E \left( {1\over t_{\rm gw} }
+{1\over t_{\rm diss} }
\right)
\end{equation}
where 
\begin{equation}
{1 \over t_{\rm diss}} = {1\over t_{\rm sv} }+{1\over t_{\rm bv}} + \mbox{other dissipation terms}
\end{equation}
These two equations combine to give
\begin{equation}
\Omega {d\alpha \over dt} + \alpha {d\Omega
\over dt} = -\alpha \Omega \left( {1\over t_{\rm gw} }
+{1\over t_{\rm diss} }
\right)
\label{evl2}\end{equation}

We can now combine (\ref{evl1}) 
and (\ref{evl2}) to get ``evolution equations''
for the mode amplitude and the bulk rotation rate:
\begin{equation}
{d\alpha \over dt} = -{\alpha \over t_{\rm gw}}
- \alpha {A_- \over A_+} 
{1\over t_{\rm diss} }
\end{equation}
and
\begin{equation}
{d\Omega \over dt} = -{ 1\over A_+} 3\alpha^2\Omega \tilde{J}
{1\over t_{\rm diss} }
\end{equation}
where
\begin{equation}
A_\pm = \tilde{I}  \pm {3\over 2} \alpha^2 \tilde{J}
\end{equation}
These equations govern the mode-evolution
in the phase when the amplitude 
of the $r$-mode grows, as well as in the 
late phase when the temperature has 
decreased sufficiently to again make the mode
stable. 

Since they should be adequately described by 
linear perturbation theory, 
the very early and late parts of the 
evolution of an unstable mode are  
comparatively well understood.
Unfortunately, it is during the nonlinear phase 
that the main spin-down is
anticipated to occur. Hence, we need to model 
the nonlinear regime in a  ``meaningful'' way. 
As our starting point we adopt the simple view that the
mode-amplitude will saturate at a critical value
$\alpha_s\le 1$. The assumed upper limit of the saturation 
amplitude, $\alpha_s=1$ corresponds to the $r$-mode
carrying a considerable part of the 
angular momentum of the system. We can readily
estimate that 
\be
\left| {J_c \over I\Omega } \right| \approx {3\alpha^2 \tilde{J}
\over 2 \tilde{I} } \approx 0.1 \alpha^2 
\ee
Hence, a value of $\alpha_s \approx 1$ should be considered
as very large (cf. also (\ref{height})).
We  assume that
$\alpha$ stays constant throughout the saturated
phase and
that $\Omega$ evolves according to (\ref{evl1}) with $\dot{\alpha}=0$.
This means that we have
\begin{equation}
{d\Omega \over dt} = 
{1\over A_-} {3\Omega \alpha^2 \tilde{J}
\over t_{\rm gw}}
\end{equation}

Since the viscosities are sensitive to 
temperature changes we need to model also 
the thermal evolution of the star 
in some suitably simple way. To do this we account for two
main processes: 
i) cooling due to (say) the modified URCA
process, and ii) reheating due to energy
deposited from the mode to the heatbath
via the shear viscosity. In stars with a solid crust
we also include energy dissipated into
heat in the Ekman layer.
The bulk viscosity
leads to the creation of neutrinos that can safely
be assumed to
escape from the star on a very short time-scale
at the temperatures we consider.

We will typically initiate the evolution of the star at
some temperature (recall that neutron stars are born above
$10^{11}$~K and cool to $10^9$~K in a few months). 
To relate the initial temperature to the thermal
energy available in the heat bath we can use\cite{stbook}
\begin{equation}
E_T = 3.5\times 10^{47} M_{1.4} \rho_{15}^{-2/3} T_9^2 
\mbox{ erg}
\label{therm1}\end{equation}
Then we model the subsequent  thermal evolution
of the star
by  assuming that  $E_T$ evolves according to
\begin{equation}
{dE_T \over dt} =
-\dot{E}_\nu + \dot{E}_{\rm visc}
\end{equation}
where
the luminosity due to the modified URCA 
process is\cite{stbook}
\begin{equation}
\dot{E}_\nu^{\rm URCA} = 1.1\times 10^{40} M_{1.4} \rho_{15}^{1/3}
T_9^8
\mbox{ erg/s}
\end{equation}
This mechanism is anticipated
to dominate the cooling in the relevant 
range of temperatures, but one should not 
forget the possibility of more rapid cooling
due to for example the direct URCA reactions\cite{durca}.
The possible outcomes should a faster cooling mechanism 
be operating have been discussed by Andersson, Kokkotas
and Schutz\cite{aks99} and Yoshida et al.\cite{yosh00}

The reheating due to shear viscosity is 
readily determined from
\begin{equation}
\dot{E}_{\rm sv} = {2\alpha \Omega^2 MR^2 \tilde{J} \over t_{\rm sv}}
\end{equation}
The heat generated 
due to the presence of
an Ekman layer in an older neutron star with a crust
follows from a similar expression.

\subsection{The spin-evolution of young neutron stars}

Given the general features of the $r$-mode instability
we expect that it could be relevant 
in newly born neutron stars. Hence it is interesting to 
consider the observational data regarding the 
initial spin rate of young pulsars. 
The best studied young pulsar is the Crab (PSR0531+21), 
whose initial period is estimated (assuming the standard
magnetic braking model) to have been about 19~ms\cite{glend}.  
Even the recently discovered young 16 ms X-ray pulsar in the supernova 
remnant N157B probably had an initial period no 
shorter than a few ms\cite{marshall} (assuming a braking index typical of young
pulsars). These estimates are in clear contrast with  
the shortest known period 
of a recycled pulsar of 1.56~ms, and with the 
theoretical lower limit on the period of about 0.5 to 2 ms\cite{nslr}, 
depending on the equation of state. 
The available
data essentially suggests that neutron stars are formed 
spinning slowly, at perhaps
$\Omega/\Omega_K < 0.1$. From a theoretical
point of view, this is rather surprising
since, assuming that angular momentum is conserved in 
the collapse event that forms the neutron star, 
young pulsars ought to be spinning close to the 
Kepler limit. 

One possible explanation for the slow 
initial spin rate of newly born neutron stars
was proposed by Spruit and Phinney\cite{sp98}, who argue
that 
magnetic locking between the core and the envelope of
the progenitor star may prevent the 
collapsing core from spinning rapidly.
Their estimates suggest that the core would actually
rotate far too slowly to lead to (by conservation of angular momentum)
young neutron stars spinning as fast as they do. 
To explain the observed rotation rates, Spruit
and Phinney propose that the neutron star spin is
due to the birth kicks that also (or alternatively) 
produce the large 
linear velocities observed for pulsars. 

Even if correct, Spruit and Phinney's model still
suggests that some pulsars are born spinning rapidly. 
Theoretically, it is easy to estimate that birth kicks that lead to 
velocities larger than 500 km/s can also produce rotation at
(or above) the Kepler limit.
Furthermore, the 16~ms pulsar in N157B is clear
evidence that some pulsars form
with periods shorter than say 10~ms. 
For such young neutron stars, the secular $r$-mode
instability may 
play a role in determining the rotation rate.
It is this possibility that we
focus on here.

As soon as we bring the spin-evolution model of the
previous section to bear on the problem we see that the
key parameter is the saturation amplitude $\alpha_s$.
Provided that $\alpha_s$ is sufficiently large, the 
$r$-modes will spin a young neutron star down 
appreciably. If we assume that the neutron star is
born with core temperature well above $10^{10}$~K 
and that it initially spins at the Kepler limit, 
the $r$-mode instability comes into play within a 
few seconds as the star cools and enters the 
instability window. The mode then grows from some
small initial amplitude to the saturation level in a few
minutes. Once the mode has saturated, the star spins down. 
At some point the 
star has cooled (or spun down) sufficiently that the
$r$-mode is again stable. Then the mode amplitude decays
and the star presumably enters a phase where magnetic
 braking takes over and
dominates the spin-evolution.
Examples of this scenario are shown in
Figure~\ref{evol1}.

The final spin period depends on several factors.
The most important of these are the saturation
amplitude, the cooling rate and whether a crust 
forms during the evolution. If we consider a simple
perfect fluid model the $r$-mode spin-down leads to 
$P \approx 12-22$~ms for $\alpha_s$ in the range $0.01-1$
(and canonical neutron star parameters).
This result is clearly consistent
with observations for many young pulsars (in particular
the Crab).
 
\begin{figure}[h]
\hbox to \hsize{\hfill \epsfysize=6cm
\epsffile{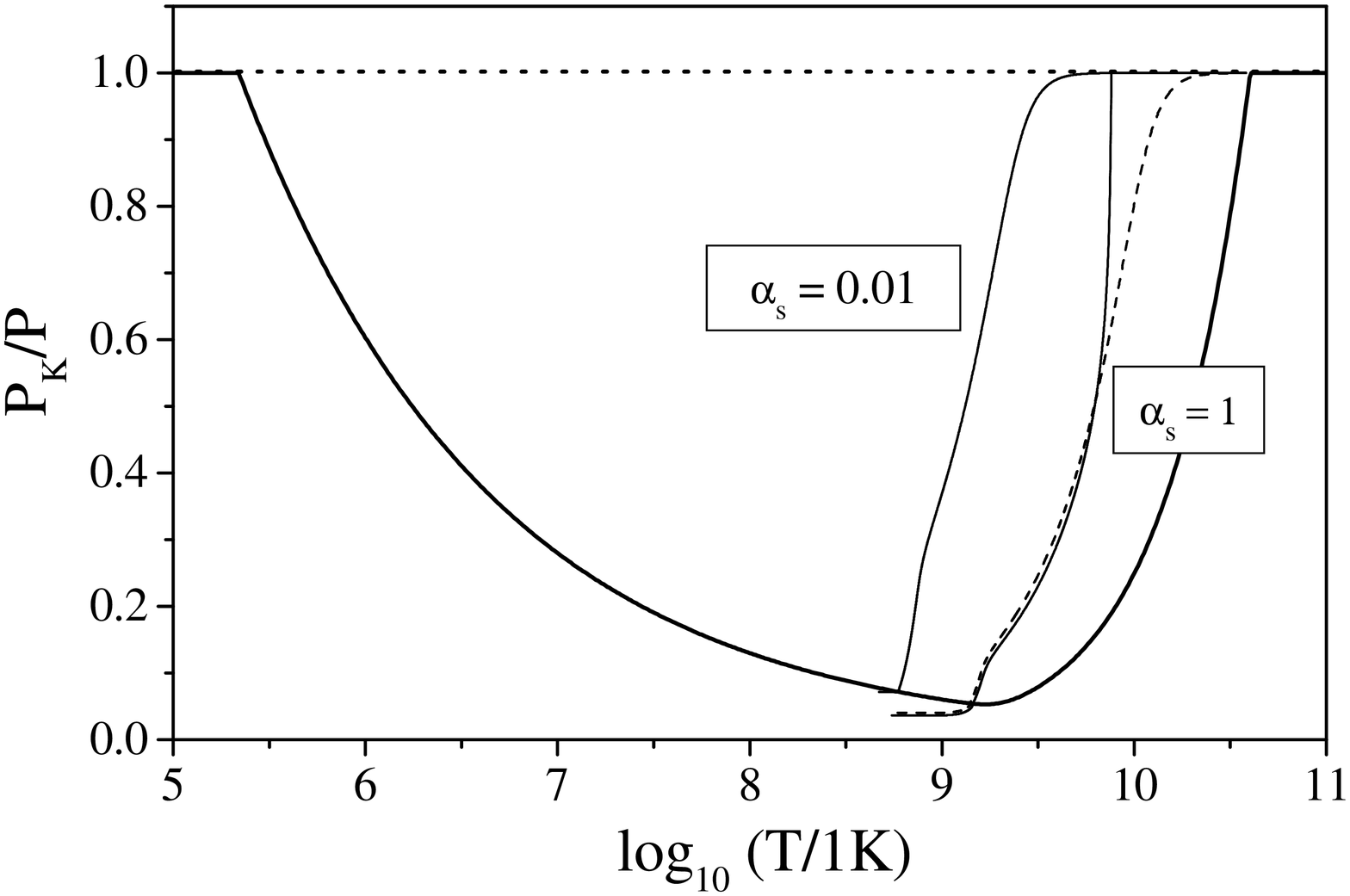} \hfill}
\fcaption{Period evolutions resulting from the 
$r$-mode spin-down scenario. We assume that the star is isothermal
and that it cools entirely due to the modified URCA process. 
Two different evolutions corresponding to an $r$-mode saturated at 
$\alpha_s=0.01$ and 1, respectively, are shown as 
solid lines.
The dashed curve indicates the evolution that would 
follow if the $r$-mode was initially excited to the saturation level
rather than given a very small initial amplitude
($\alpha = 10^{-6}$) at the onset of the
instability. It is notable the initial amplitude 
has little effect on the 
period at the end of the spin-down phase.
}\label{evol1}
\end{figure}

As we discussed in Section~4.5, the $r$-mode instability may be strongly
suppressed in neutron stars with a solid crust. 
Given that the  melting temperature of the crust 
could be as high as  $10^{10}$~K,
the crust may form shortly after the neutron star is born
and we need to discuss the effect that this may have on the
$r$-mode scenario. The interplay between a growing mode and the 
formation of the crust leads to complicated
questions that require much further study. For example, 
it is not at all clear to what extent the melting temperature
of the crust is affected by large scale surface waves in the star. 
Here we will not address such questions in detail (but see comments in Section~6). Instead we 
consider what may be a ``worst case scenario'' as far 
as the instability is concerned.
We assume that the $r$-mode will not be able to prevent
the crust from forming (and that the melting temperature is 
at the high end of the anticipated range). The main dissipation 
mechanism is then due to the presence of the Ekman layer
(at least above the superfluid transition temperature).
Resulting spin evolutions are shown in Figure~\ref{evol2}. 
In this scenario,  
the period reached after the spin-down phase lies in the range
$P\approx 2.5-4.5$~ms. Just as in the perfect fluid case, this is 
an interesting prediction since  it agrees quite well with the data
for the recently discovered 16~ms pulsar PRS~J0537-6910 in N157B.

\begin{figure}[h]
\hbox to \hsize{\hfill \epsfysize=6cm
\epsffile{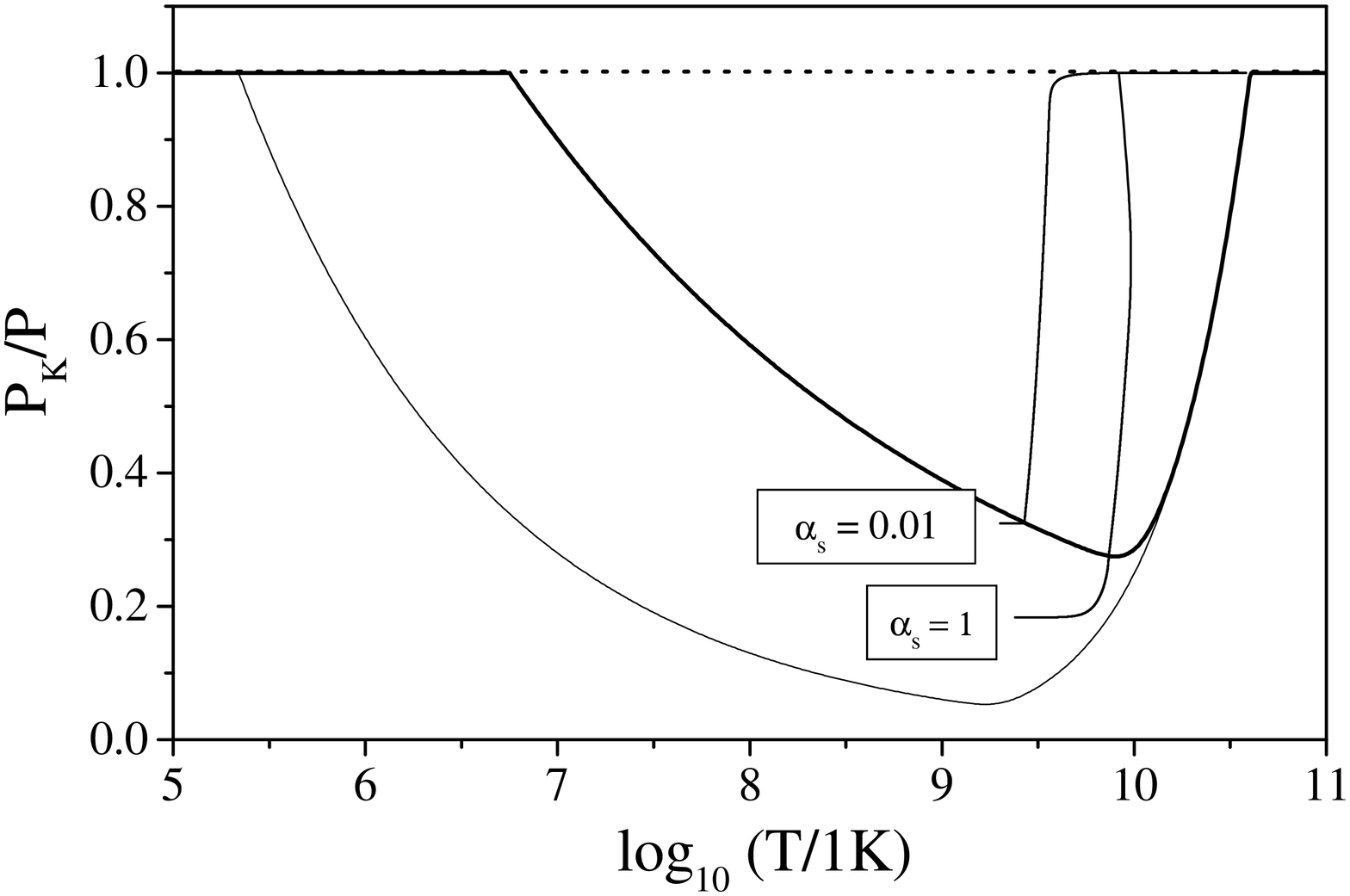} \hfill}
\fcaption{Spin-down evolutions for a neutron star in which a
solid crust forms before the $r$-mode has grown appreciably. The 
presence of an Ekman layer at the base of the crust leads to a 
strong dissipation of the mode energy and the final spin-period
after the instability phase is much shorter than in the 
crustless case.}
\label{evol2}
\end{figure}

We have thus seen that the $r$-mode instability can lead to 
young neutron stars being spun down to rotation rates that 
would agree quite well with extrapolations from 
current observations. Furthermore, we have predicted
that hot young neutron stars
 may follow two different evolution routes\cite{ajks00}.  
Which scenario applies depends sensitively 
on the early cooling of the star and the crustal formation 
temperature. To illustrate this, we consider the evolution of a neutron star 
just after its birth in a supernova explosion. 
 We might expect to  model its 
$r$-mode evolution using 
the normal (crust-free) fluid viscous damping 
times for stellar temperatures above the melting temperature
of the crust ($T_m$), and the 
viscous boundary layer damping time for temperatures below $T_m$.  
However, the situation is probably a little more complicated than this. 
Recall that the latent heat (i.e. the Coulomb binding energy) of a typical 
crust is 
\be
E_{\rm lat} \approx 10^{48} \ \mbox{ erg}
\ee
As is clear from (\ref{rotenerg}) the energy in the $r$-mode (which 
grows exponentially on a timescale $t_{\rm gw} \sim 20-50$~s) will
easily exceed $E_{\rm lat}$, provided that   
the time taken for the star 
to cool to $T_m$ is sufficently long. This would then 
prevent the formation of the crust, even when $T < T_m$.  
Then the star will spin down in the manner shown in Figure~\ref{evol1}.
This phase will end either because the star 
leaves the instability window 
or because the mode energy has fallen below the crustal binding 
energy. Supposing that the $r$-mode energy is distributed in such a way that
(say) 10\% is located in the region where the crust will form 
(the outer km or so of the star), we reach balance, i.e. 
$E_{\rm lat} \approx 0.1 E_{\rm mode}$, 
when 
\be 
P_{-3} \approx 10 \alpha_s
\ee 
In other words, for a saturation amplitude of order unity
the crust would be unable to form until the that has spun down 
to a rotation rate of roughly 100~Hz. 
Again, this would be  consistent with the extrapolated 
initial spin rates of many young pulsars. On the 
other hand, if the mode is not given time to grow very large 
it will not prevent crust formation at $T_m$. 
The the $r$-mode instability 
would  spin the star down as in Figure~\ref{evol2}.
A likely key parameter (in addition to the cooling rate, the melting 
temperature and the extent to which a large amplitude 
mode can melt the crust, see section~6) is 
whether the supernova
collapse leads to a large initial  $r$-mode amplitude or not.

Besides explaining the slow rotation of young pulsars, the 
$r$-mode instability has the consequence that  young neutron stars
can only reach  
rotation periods shorter than (say) 3-5~ms if they are
recycled by accretion in 
a binary system.  The 
alternative model, that these stars are formed by accretion-induced 
collapse of a white dwarf, is inconsistent with the $r$-mode
scenario\cite{aks99}. The collapse would form 
a star hot enough to spin down because of the instability. 
However, the situation would be quite different in 
the case of strange stars\cite{jm99}. As shown in Figure~\ref{win6}, 
the instability window for 
 strange matter differs significantly from the
neutron star one. This means that the
$r$-mode instability would be less effective (apart from at very high 
temperatures) for
strange stars. Furthermore, a strange star
cools much slower than a neutron star.
Assuming that the strange star cools on a timescale $t\approx
10^{-4}T^{-4}_9$ yrs, we find that the strange star reaches $P\approx 3$~ms
after $4\times10^5$ yrs. This means that 
an observation of a young millisecond pulsar spinning faster than (say) 3~ms 
may
indicate the presence of a strange star\cite{jm99}.

\subsection{Are the gravitational waves observable?}

Having suggested that the $r$-mode instability may 
spin a newly born neutron star down to a fraction 
of its initial spin rate in a few months we want to
know whether the
gravitational waves that carry away most of the stars initial
angular momentum are detectable. This is a particularly 
relevant question given the generation of large-scale
interferometers that is about to come into operation\cite{gwreview}.

We assess the detectability of the 
emerging gravitational waves in the standard way.
First of all, we note that  
the frequency of the emerging gravitational waves is
(for the main $l=m=2$ $r$-mode)
\begin{equation}
f_{\rm gw} = {2\Omega \over 3\pi} \ .
\end{equation}
From the gravitational-wave luminosity we readily deduce that
the dimensionless strain amplitude follows from \cite{o98}
\begin{equation}
h(t) = 7.54\times10^{-23} \alpha \tilde{J} M_{1.4} R_{10}^3
\left( {15 \mbox{ Mpc} \over D} \right)
\end{equation}
where $D$ is the distance to the source, here assumed to be in the
Virgo cluster. At this distance one would expect to see several 
neutron stars being born per year. The typical $r$-mode scenarios
shown in Figures~\ref{evol1} and \ref{evol2} 
then lead to the gravitational-wave strains 
illustrated in Figure~\ref{gw1}. Clearly, this amplitude is 
not sufficiently strong that it can be observed without 
a detailed data analysis strategy. To assess the possible improvement
that would follow if such a strategy could be developed for the $r$-mode
signal we will use the standard matched-filtering approach. 
This gives us an idea of the improvement that a specially tailored 
data analysis approach may bring to the $r$-mode detection 
problem, even though it must be recognized that 
matched filtering is unlikely to 
be possible for this kind of signals\cite{brady,owen}.

\begin{figure}[h]
\hbox to \hsize{\hfill \epsfysize=6cm
\epsffile{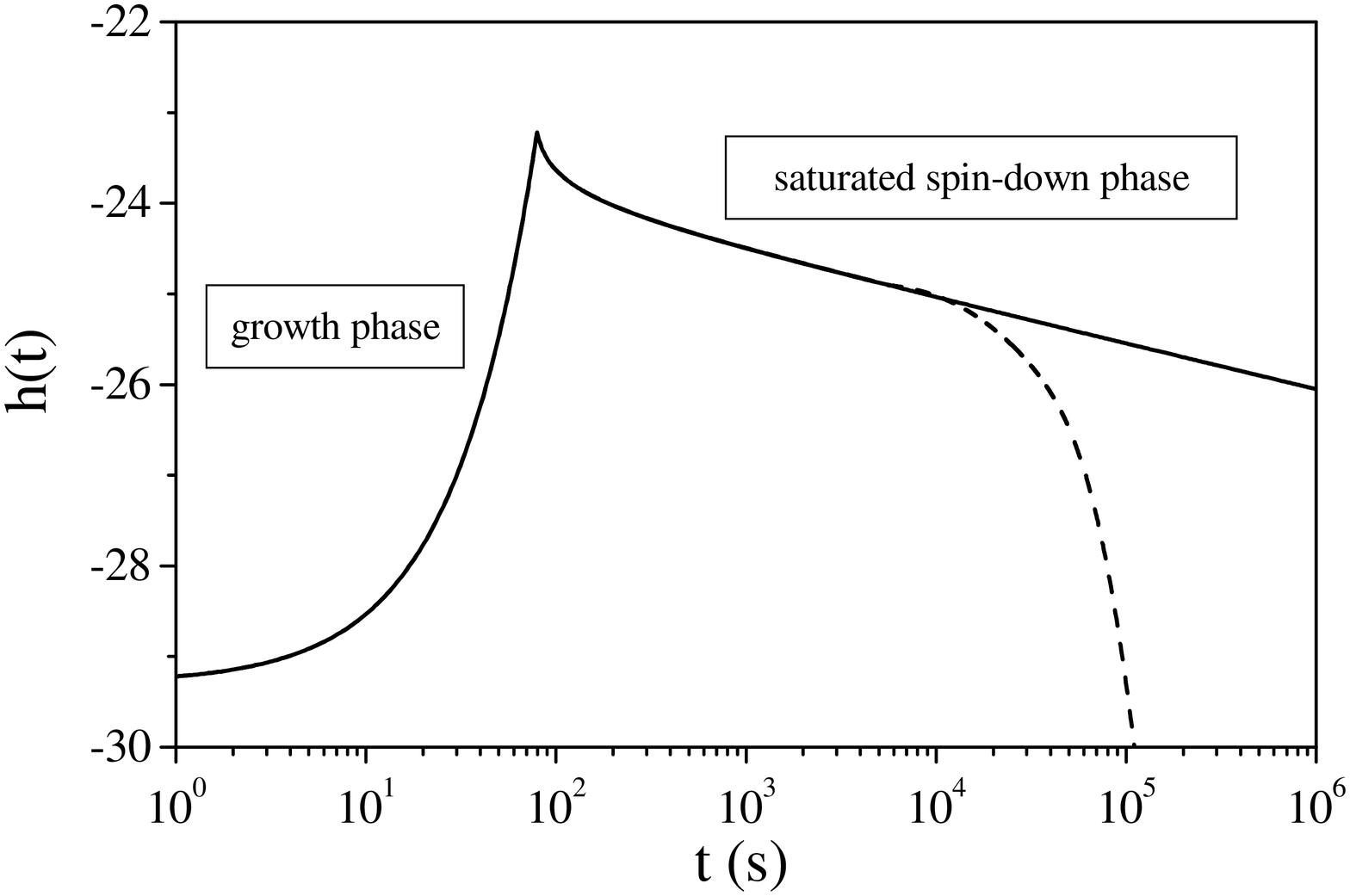} \hfill}
\fcaption{ The dimensionless gravitational-wave amplitude following from a typical 
$r$-mode scenario. The two phases of evolution (the early growth phase of the mode and 
the spin-down after saturation) are easily distinguished. The second 
phase  is the most likely to be observed with future detectors.
We show signals corresponding to both fluid neutron stars (thick solid line) 
and ones with a crust (dashed). In both cases we assume that the saturation amplitude
is $\alpha_s =1$. It should be noted that the signal lasts $10^7$~s in the fluid case but only
$10^5$~s for the crusted star. 
}\label{gw1}
\end{figure}

With the Fourier transform of the gravitational-wave signal defined by
\begin{equation}
\tilde{h}(f) = \int_{-\infty}^\infty h(t) e^{2\pi ft} dt
\end{equation}
the signal-to-noise ratio attainable by optimal filtering
 can be determined from
\begin{equation}
\left( {S \over N}\right)^2 = 2 \int_0^\infty f { |\tilde{h}|^2  \over
h_{rms}^2 } df  
\end{equation}
where $h_{rms}=\sqrt{fS_h (f)}$, $S_h$ being the spectral density of the 
strain noise in the detector.  Now introducing the characteristic amplitude
$h_c$ we can write
\begin{equation}
\left( {S \over N}\right)^2 = 2 \int_0^\infty {df \over f}
\left( { h_c  \over
h_{rms} } \right)^2  
\end{equation}
where 
\begin{equation}
h_c = f |\tilde{h}| \approx  h \sqrt{ f^2 \left| {dt \over df}\right| }
\end{equation}
The last relation follows via the stationary phase approximation.
It can be interpreted as meaning that the detectability of an 
almost periodic signal improves as the square root of the number of
cycles radiated in the time it takes the frequency to change by $f$.
In Figure~\ref{gw2} we compare this characteristic amplitude  
to estimated sensitivity curves for the new generation of interferometric
detectors. We immediately see that the $r$-mode instability would be 
a promising source for an advanced generation of detectors such as LIGO~II.
One can estimate that the gravitational waves from a hot young neutron star
will be detectable with a good signal-to-noise ratio\cite{o98,brady,owen}. This makes the $r$-mode
instability one of the most promising sources of detectable gravitational waves. 
It is unlikely that the $r$-modes will be observed by the first generation
detectors, however. For this to happen requires a unique event in 
our Galaxy or the local group. One would typically expect a supernova
in the galaxy every 30 years or so, which means that we would be extremely
lucky to see such an event. 
Anyway,  the chances of future  detection 
are promising and they
may  improve as
our understanding of the instability becomes more detailed and
better theoretical templates for the gravitational-wave signal can be
constructed.

\begin{figure}[h]
\hbox to \hsize{\hfill \epsfysize=6cm
\epsffile{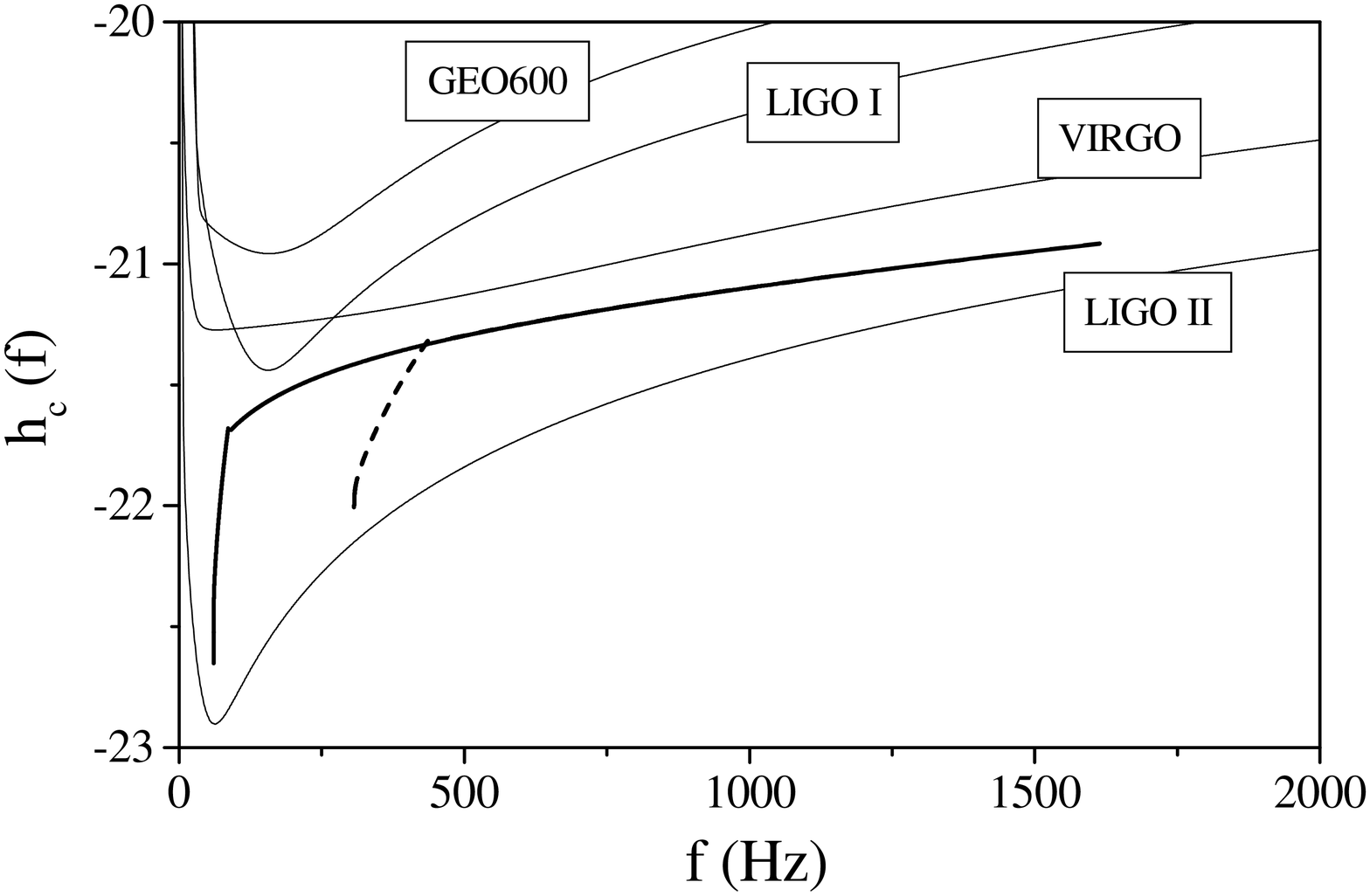} \hfill}
\fcaption{ The effective amplitude achievable after matched filtering (which
is somewhat unrealistic for the $r$-mode instability)  compared to the expected
sensitivity of the new generation of interferometric
detectors. The two gravitational-wave signals are the same as in Figure~\ref{gw1}.
}\label{gw2}
\end{figure}

In addition to signals from individual neutron stars, 
there will  be a stochastic background of gravitational waves
from cosmological neutron stars spinning down 
due to the instability. Estimates
show that this background will be difficult to detect, but  it could 
perhaps be relevant for an advanced generation of 
detectors\cite{o98,ferrari98}.

\section{Accreting neutron stars}

In the last few years the Rossi X-ray Timing Explorer
has provided a  wealth of observational data 
regarding  accreting neutron stars.
These observations present a  
challenge for theorists in that  neutron stars in 
low-mass X-ray binaries (LMXB) seem to be confined to a rather narrow 
range of spin frequencies, perhaps 260-590~Hz\cite{vanderklis00}.
Different models have been proposed to 
explain this surprising result.
The first model (due to White and Zhang\cite{white97}) is based 
on the standard magnetosphere model for accretion 
induced spin-up, while two other models are
inspired by the idea
that gravitational radiation balances the accretion 
torque. In the first such model for the LMXB 
(proposed by Bildsten\cite{bildsten98} and recently refined by  
Ushomirsky, Cutler and Bildsten\cite{ushomirsky00}), the 
gravitational waves are due to an accretion induced
quadrupole deformation in the deep neutron star crust. The second 
gravitational-wave model relies on the  
 $r$-mode instability
to dissipate the accreted angular momentum from the neutron
star\cite{akst99}.

The idea
that gravitational waves from 
unstable $r$-modes provide the agent that
balances the accretion torque
was first analysed in detail by Andersson, Kokkotas and 
Stergioulas\cite{akst99} 
(but see also Bildsten\cite{bildsten98}). Originally, 
it was thought that an accreting star in which 
the $r$-modes were excited to a significant level
would reach a spin-equilibrium, very much 
in the vein of early suggestions by
Papaloizou and Pringle\cite{papaloizou} and  
Wagoner\cite{wagoner}. 
Should this happen, the neutron stars in LMXB 
would be prime
sources for detectable gravitational waves. 
That the associated (essentially periodic) 
gravitational waves would be detectable 
can be seen in the following way: 
Assume that the
$r$-mode instability is active and
provides a limit on the spin of Sco~X1, which is the  strongest 
X-ray source in the sky and therefore a prime candidate
for this kind of speculation. If we assume that the 
average accretion rate onto the neutron star is $\dot{M}\approx 3\times
10^{-9}M_\odot/\mbox{yr}$ and that the accretion torque is balanced by 
gravitational radiation we deduce that
$h\approx 3.5\times10^{-26} \rightarrow h_c \sim 10^{-21}$
after two weeks worth of signal has been accumulated\cite{akst99}.

However,  this idea is probably not viable\footnote{Although it 
should be noted that the Wagoner scenario might be relevant 
if accretion at extreme rates is possible. The role of the unstable $r$-modes
during such hypercritical accretion has been discussed by 
Yoshida and Eriguchi\cite{hyper}. }. 
In addition to generating gravitational waves that dissipate 
angular momentum from the system, the $r$-modes
will heat the star up (via the shear viscosity and the Ekman layer).
Since the viscosity gets weaker as the 
temperature increases, the mode-heating triggers a
thermal runaway\cite{spruit,levin} and in a few months the $r$-mode would 
spin an accreting neutron 
star down to a 
rather low rotation rate. , This essentially 
 rules out the $r$-modes in galactic LMXB as a source
of detectable gravitational waves, since they will 
only radiate for a tiny fraction of the systems
lifetime. 

This mechanism could still be of great astrophysical 
significance\cite{ajks00}. Hence, we want to model how the 
potential presence of an unstable $r$-mode affects the spin-evolution
of rapidly spinning, accreting neutron stars. To do this we
use the phenomenological two-parameter model described in section~5.1.
Then the following picture emerges: 
After accreting and spinning up for something like $10^7$ years, 
the star reaches the period at which the $r$-mode 
instability sets in. For a canonical neutron star 
this corresponds to a period
of 1.5~ms (at a core temperature of $10^8$~K). It is notable
(albeit likely coincidental) that this value is close to the 1.56~ms 
period of the fastest known pulsar PSR1937+21. 
Once the $r$-mode becomes unstable (point A in 
Figure~\ref{evol3}), viscous 
heating (mainly due to  the energy released in the Ekman layer) 
rapidly heats the star up to a few times $10^9$~K. The $r$-mode 
amplitude increases
until it reaches the saturation level 
(amplitude $\alpha_s$), at
which unspecified nonlinear effects halt further growth (point B in 
Figure~\ref{evol3}). 
Once the mode has saturated, the neutron star rapidly spins down.
When the star has spun down to the point where the
mode again becomes stable (point C in 
Figure~\ref{evol3}), the  amplitude starts to decay and the 
mode eventually plays no further role in the spin evolution of 
the star (point D in Figure~\ref{evol3}).
Two examples
of such $r$-mode cycles (corresponding to $\alpha_s=0.1$ and 1, respectively) 
are shown in Figure~\ref{evol3}.

\begin{figure}[t]
\hbox to \hsize{\hfill \epsfysize=6cm
\epsffile{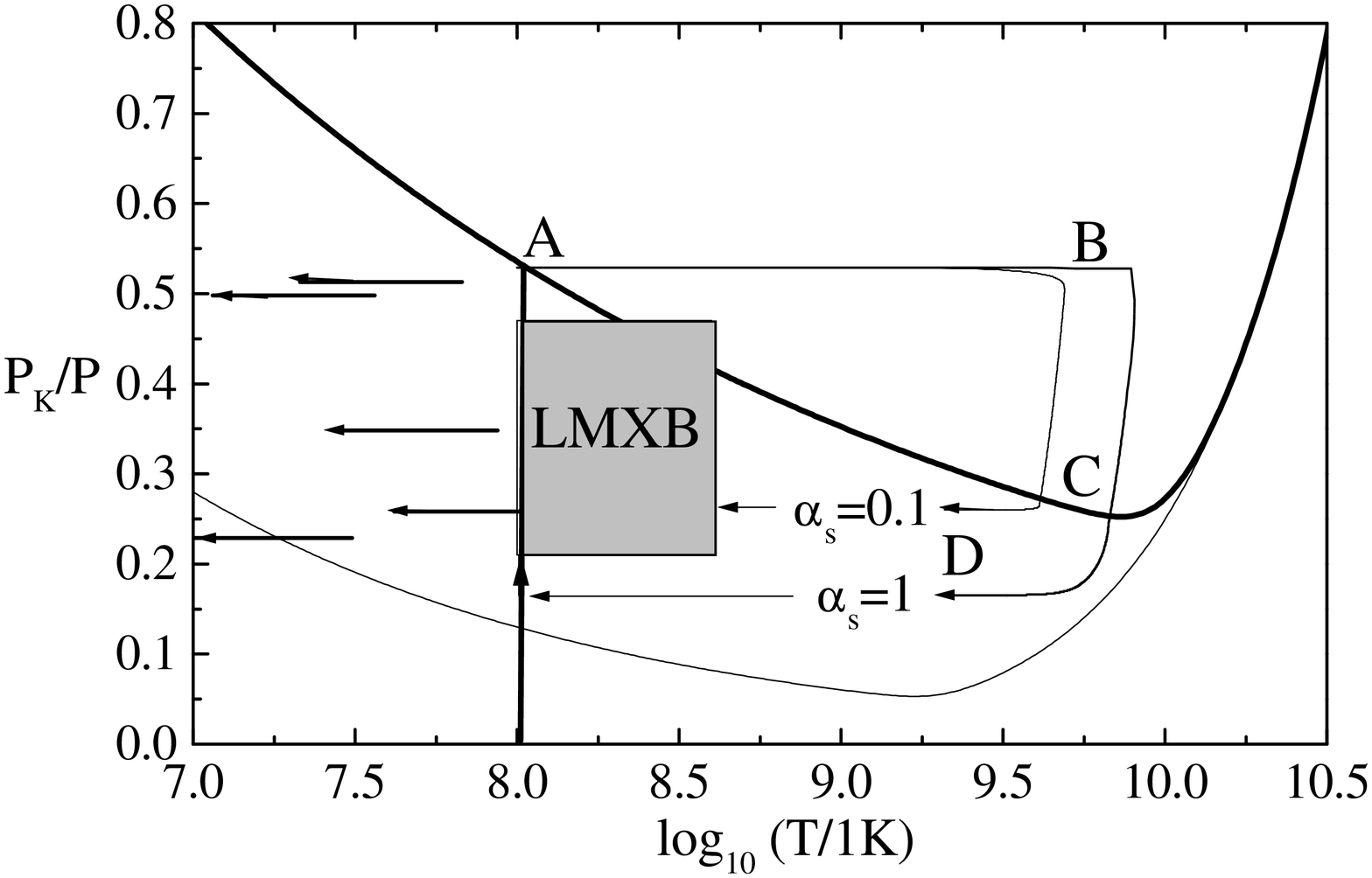} \hfill}
\fcaption{
The $r$-mode instability window relevant for 
old neutron stars. We show results for the simplest (crust-free)
model (thin solid  line), as well as for a
star with a crust (thick solid line).
 We illustrate two typical $r$-mode cycles (for mode saturation amplitudes
$\alpha_s=0.1$ and 1), resulting
from thermo-gravitational runaway after the onset of instability.
Once accretion has spun the star up to the critical period
(along the indicated spin-up line (thick vertical line)) the 
$r$-mode becomes unstable and the star evolves along the path
A-B-C-D. After a month or so, the mode is stable and the star 
will cool down until it again reaches the spin-up line.   
For comparison with observational data, 
we indicate the possible range of spin-periods inferred from current
LMXB data (shaded box) 
as well as
the observed periods and estimated upper limits
of the temperature 
of some of the most rapidly spinning millisecond pulsars
(short arrows). }
\label{evol3}
\end{figure}

This simple model leads to some interesting quantitative
predictions. First of all, the model suggests that an accreting star 
will not spin up beyond  1.5~ms. This value obviously depends on 
the chosen stellar model, but it is
independent of the 
$r$-mode saturation amplitude and only weakly dependent on the accretion
rate (through a slight change in core temperature). 
In fact, the accretion rate only affects the time it takes
the star to complete one full $r$-mode cycle. As soon as the 
mode becomes unstable the
spin-evolution is dominated by gravitational radiation and viscous 
heating. Once the star has 
gone through the brief phase when the $r$-mode is active it has spun 
down to a period in the range 3-5~ms (corresponding to 
$0.01\leq \alpha_s \leq 1$). 
 At this point the 
mode is again stable and continued accretion may resume to spin the star up.
Since the star must accrete roughly 
$0.1M_\odot$ to reach the instability point, and 
the  LMXB companions typically have masses in the 
range $0.1-0.4M_\odot$, it can pass through several 
``$r$-mode cycles'' during its lifetime.  

We can readily confront these results with current observations\footnote{Here it is worth mentioning that Brown and Ushomirsky\cite{hotobs} have discussed 
how  observations of X-ray transients may provide a constraint
on the $r$-mode amplitude in such systems.}  . 
We note that the model predicts that an accreting neutron star
 must, once it has
been spun up beyond (say) 5~ms, remain in the
rather narrow range of periods $1.5-5$~ms until it has stopped
accreting  and magnetic braking  slows it down.
Since a given star can go through several
$r$-mode cycles before accretion is halted one would expect
most neutron stars in LMXB and the
millisecond pulsars to be found within this range.
 As is clear from Figure~\ref{evol3}, this 
prediction agrees  well with the range of rotation 
periods inferred from observed kHz quasiperiodic oscillations
in LMXB.
The observed range shown in Figure~\ref{evol3} corresponds to
rotation frequencies in the range
260-590~Hz\cite{vanderklis00}. The model also
agrees with the observed data for 
millisecond pulsars\cite{ajks00}, which are mainly found in the
range $1.56-6$~ms.
In other words,  the $r$-mode runaway model
is in agreement with current observed data for rapidly
rotating neutron stars. 

However, many crucial pieces of physics have not yet been included
in this model. For example, we based the discussion on the 
simplest estimate (\ref{tek2}) of the Ekman layer dissipation, rather
than more refined ones attempting to include the crust-core
interaction in a detailed way\cite{crust1,crust2}. Furthermore, 
we assumed that the heat released in the Ekman layer will not be able to 
melt the crust. This latter issue has recently been studied in some detail
by Lindblom, Owen and Ushomirsky\cite{crust4} (see also
Wu, Matzner and Arras\cite{crust3}), who solve the
appropriate heat equation incorporating thermal conductivity
for the core fluid and the crust as well as neutrino cooling 
due to the modified URCA process. They find that, for a star spinning 
at the Kepler limit, an $r$-mode with an amplitude exceeding
$\alpha \approx 5\times 10^{-3}$ will be able to melt the crust. 
This means that if the $r$-mode instability is active in a young neutron
star, the formation of the crust may be significantly delayed 
(as we argued in section~5.2). 
But it is not yet clear what will happen if the $r$-modes do indeed manage
to melt an existing crust. As soon as the crust melts the Ekman layer (that led
to the excessive heating) disappears and the material will rapidly cool
down to a level where the crust would begin to form again. As argued
by Lindblom et al\cite{crust4}, the likely outcome is a solid-fluid
mixed state which will be very difficult to model in detail. 

We conclude this section by discussing briefly the detectability
of the gravitational waves that are radiated during the
relatively short time when the $r$-mode is saturated and the star
spins down. Since the $r$-mode
is active only for a small fraction of the lifetime of the system
(something like 1 month out of the $10^7$~years it takes to complete
one full cycle) the event rate is far too
low to make galactic sources relevant.  However, it interesting to note that
the spin-evolution is rather similar to that of a
hot young neutron star 
once the $r$-mode has reached its saturation amplitude.
This means that we can analyse the detectability of the 
emerging gravitational waves using the framework of 
section~5.3.
We then find that these events can be observed from rather
distant galaxies. For a source in the Virgo cluster 
 these gravitational
waves could be detected with a signal to noise ratio of a few 
using LIGO~II.  However, even at the distance of the
Virgo cluster these events would be quite rare. 
By combining a birth rate for LMXB of $7\times 10^{-6}$ per
year per galaxy with the fact that the 
the volume of space out to the Virgo cluster contains $\sim 10^3$ galaxies,
and the possibility that each LMXB passes through (say) four $r$-mode cycles
during its lifetime we deduce that one can only hope to see
a few events per century in Virgo. In order to see several events
per year the detector  must be sensitive enough to detect these
gravitational waves from (say) 150~Mpc. This would require
a more advanced detector configuration.

\section{Steps towards more detailed modelling}

From the discussion in the previous sections it should be clear that
the present results indicate that the unstable $r$-modes may have
great astrophysical significance, and that the associated gravitational-wave
signals may well be detected within a few years. These
ideas are  obviously very exciting, but it is important
to realize that our modelling must be improved in many directions
before any reliable conclusions can be drawn. In this section 
we discuss some recent work that provide extensions
(or alternatives) to the ideas that we have already 
described. 

\subsection{Inertial modes and rapid rotation}

As was mentioned in the introduction, the $r$-modes belong to a
larger class of pulsation modes which in fluid mechanics
are commonly described as ``inertial 
modes''\cite{rieu00}. These modes are all dominated by the 
Coriolis force and for slowly rotating stars their
frequencies scale linearly with $\Omega$.
In non-isentropic stars the rotation 
dominated modes are all toroidal in nature (the $r$-modes).
In contrast, the space of zero frequency modes of
isentropic spherical stars includes the spheroidal $g$-modes as well as 
the $r$-modes.  This large degenerate subspace of zero-frequency 
modes is then split by rotation,
and the inertial modes of isentropic stars generally
have a ``hybrid'' nature\cite{lf99} (their spherical limit is a mixture of 
toroidal and spheroidal perturbations).  

Detailed calculations of 
such hybrid modes in Newtonian stars have been made by 
Lockitch and Friedman\cite{lf99} as well as Yoshida and Lee\cite{yl99,yl00}
(see also Lee et al\cite{lsvh92} and the studies of Maclaurin
spheroids discussed below). These studies show that the 
low frequency mode spectrum 
of a rotating star is tremendously rich. Furthermore, many 
of the hybrid modes are affected by the CFS instability mechanism. 
Estimated growth times and viscous damping rates for these unstable modes
suggest that they are likely to lead to weaker instabilities than 
that of the fundamental $r$-mode
\footnote{In this context, we should also mention that 
instabilities associated with
the $g$-modes in non-isentropic stars have  been discussed
by Lai\cite{lai99}.}.  
This can be understood from the fact that 
a mode with a significant spheroidal part will be affected
by (for example) bulk viscosity to a larger extent. 

A particularly interesting question, that provided the original 
motivation for the work of Lockitch and Friedman\cite{lf99},
concerns what happens to all the $r$-modes of a non-isentropic star
as ${\cal A}_s\to 0$, i.e. in the isentropic limit.
Recall that a non-isentropic star has an infinite set of $r$-modes 
for each $l$ and $m\neq0$, while an isentropic star retains only a single
such mode for $l=m$.
 The general consensus
at the present time is that most of the non-isentropic $r$-modes will change
their nature by acquiring a spheroidal component as ${\cal A}_s\to 0$, i.e.
they become hybrid modes.
Some light on the detailed behaviour has recently been shed by Yoshida
and Lee\cite{yl00}. 
They suggest that the $r$-modes of a non-isentropic star
fall into one of three categories. The first class are
the fundamental $l=m$ modes (that have eigenfunctions $T_{ll}\sim a^{l-1}$
and lead to the strongest instability). These modes are largely unaffected
by changes in ${\cal A}_s$. Similar results were reported by Andersson, Kokkotas and Schutz\cite{aks99}. 
The second class of modes are the various $l=m$ overtones (whose eigenfunctions
have nodes inside the star). According to Yoshida and Lee\cite{yl00}, these modes
approach hybrid modes as the spin-rate of the star increases to a 
level where the slow-rotation approximation is no longer
appropriate. Finally, the $l\neq m$ $r$-modes of a non-isentropic star
have frequencies that vanish (in the rotating frame) as the spin-rate increases. 
Consequently, these modes 
may not be relevant in a rapidly rotating star. 

It turns out that 
the general behaviour of the modes in the isentropic limit
is rather similar to  what happens at fast rotation rates. 
Until recently, low-frequency modes had not been calculated for rapidly 
rotating stars. The situation has changed with calculations for the 
Maclaurin spheroids\cite{li98} (essentially reproducing and elucidating very early
work by Bryan\cite{bryan}) as well as for rapidly rotating polytropes\cite{yosh00}.  
These calculations are complicated by the fact that the mode structure
changes as the spin rate increases, and for rapidly rotating stars one must
include a large number of multipoles in a decription of the 
mode eigenfunctions. Interestingly, the polytrope
results of Yoshida et al.\cite{yosh00} indicate
that the slow rotation expansion (taken to order $\Omega^2$) provides
a good representation of the $r$-modes also in rapidly spinning stars.
This means that the current models for the spin evolution of young neutron 
stars etcetera may, in fact, be better than expected.

\subsection{Slowly rotating stars in general relativity}

One important step towards improving our understanding
of the $r$-mode instability corresponds to describing the modes
in general relativity.
The Newtonian picture of stellar pulsation, that we introduced in sections~2.1-2.3,
is readily generalised to the relativistic case.
The main difference is that the gravitational waves 
generated by the modes now appear as perturbations of the spacetime
metric. Consequently, the various modes 
are no longer ``normal modes'', but must satisfy
outgoing-wave boundary conditions at spatial infinity.
This provides an additional complication in general mode-calculations
(where the eigenfrequencies are now complex), but it is not relevant
for a slow-rotation study of low-frequency oscillations.
As can be deduced from (\ref{gwest}) the gravitational-wave dissipation 
will not 
enter until at very high orders in the slow-rotation expansion. Hence, 
the $r$-mode frequencies remain real to leading order, and we are  
(in principle) dealing with 
a ``standard'' normal-mode problem.
  
Considerable progress towards an understanding of the relativistic
$r$-mode problem has been 
made recently\cite{a97,k98,bk99,kh99,kh00}. Particularly relevant is
work by 
Lockitch, Andersson and Friedman\cite{laf00}
that puts the various facets of the problem in the appropriate context
and provides the first actual calculation of the relativistic $r$-modes.
The picture that emerges with this work is somewhat more complex
than the Newtonian one, but it can be understood if we recall
the difference between isentropic and non-isentropic 
Newtonian perturbations. First of all, one can  prove that
(apart from a set of stationary dipole modes\cite{lo99}) rotating 
relativistic isentropic stars have no pure 
$r$-modes (modes whose limit for a spherical star is purely toroidal).
This is in contrast with the 
isentropic Newtonian stars which retain a vestigial set of purely 
toroidal modes (the fundamental $l=m$ modes that lead to the strongest
instability). 
Instead, the relativistic corrections to the Newtonian $r$-modes with $l=m\geq 2$ 
are both toroidal and spheroidal. Thus these modes become
discrete hybrid modes of the corresponding relativistic models.
So far these modes have been computed  for slowly rotating isentropic stars to first 
post-Newtonian order ($\sim M/R$).

As in the Newtonian problem,  
non-isentropic stars are somewhat different.
In the slow-motion approximation in which they have so far been
studied,  non-isentropic stars have, remarkably, a
continuous spectrum. Kojima \cite{k98} has shown 
that purely axial modes would be described
by a single, second-order differential equation. The continuous 
spectrum is implied by the
fact that the corresponding eigenvalue problem is singular 
(the  coefficient of the highest derivative term of the
equation vanishes at some value of the radial coordinate).
The mathematical arguments for the continuous spectrum have been 
elucidated by  Beyer and
Kokkotas\cite{bk99}. However,   
as the latter authors point out, it is not yet clear that the continuous 
spectrum is physically relevant.
In addition to the possible continous spectrum one can show that discrete
$r$-mode solutions also exist. Such discrete mode-solutions  to Kojima's equation
were recently calculated\cite{laf00}.
These modes are the relativistic analogue to the Newtonian
$r$-modes in non-isentropic stars.

Even though much work remains before the $r$-modes
and their instability are understood in 
general relativity, the recent results provide much useful
information. In particular, even though the  nature
of the equations for isentropic
relativistic stars is rather different from the 
corresponding non-isentropic ones,
the particular modes that would be the analogues of
the vestigial $l=m$ $r$-modes of isentropic
Newtonian stars
are not too dissimilar\cite{laf00}. In fact, the  calculated
mode frequencies differ only at the level of a few percent. 
The toroidal components of the velocity field are also
virtually identical (although one must remember that the
relativistic
isentropic mode also has a significant spheroidal component).

\subsection{The role of the magnetic field}

Another important facet of neutron star physics that we have not yet
discussed is the magnetic field. Intuitively one would expect
the interplay between a large amplitude pulsation mode and  
the magnetic field to be interesting. The simple fact that this may
lead to observable effects is strong motivation 
for trying to incorporate the magnetic field in our models. 
To do this is, however, far from trivial and at the present time 
all results in this direction must be considered as rather uncertain.
On the other hand, the same is true for the $r$-mode spin-down scenario
(section~5.1)
so why should we shy away from speculating on the role of the 
magnetic field?

There have so far been three main discussions of magnetic fields
in the context of the $r$-mode instability. In the first of these, 
Spruit\cite{spruit}
outlined a possible scenario where an accreting neutron star with a
weak 
magnetic field is spun up to the point where the $r$-mode becomes
unstable
(see Figure~\ref{evol3}). As we discussed earlier, it is likely 
that the $r$-mode instability then leads to a thermo-gravitational
runaway that heats the star up and makes the mode-amplitude grow.
Spruit suggests that gravitational radiation reaction then induces
differential rotation in the star. If so, the interior 
magnetic field could wind
up until it reaches a critical point and  becomes unstable due to
buoyancy forces. When this happens, the star sheds a considerable
amount of electromagnetic energy, perhaps as a gamma-ray burst, 
which slows it down significantly. Left over after the turmoil
is a slowly rotating neutron star with an extreme magnetic field (of
the order of 
$10^{14}-10^{15}$~G): a
magnetar\cite{duncan,magnetar}. This idea is undoubtedly speculative 
but it raises some interesting questions, and may also
capture some features of the $r$-mode
runaway. Qualitatively, the most important issue regards 
whether the instability leads to differential rotation or not. 

All the models that we have discussed in the previous sections 
assume that the star
remains 
spinning uniformly. That this will actually be the case is far from
clear. An important counterexample is the evolution of an unstable
bar-mode in a rotating ellipsoid. As was first shown by Miller\cite{miller}, 
an unstable mode drives a Maclaurin spheroid towards a Dedekind
ellipsoid (more or less a rotating american football). En route 
the evolution is thought to proceed via a sequence of differentially rotating 
Riemann ellipsoids. By analogy one might therefore expect that 
the $r$-mode instability will generate differential rotation. 
To conclusively prove this will, however, not be easy
since it involves modelling how radiation reaction affects various
parts of the star. 

The question whether an unstable $r$-mode can be prevented from growing 
by the magnetic field was recently discussed by Rezzolla, Lamb and 
Shapiro\cite{rez1}. 
In an interesting approach to the problem, they considered the 
leading order result for the $r$-mode velocity field (basically the
time-derivative of our equations (\ref{thetaeq}) and (\ref{phieq})) 
as evolution equations
for the fluid elements. By evolving this system in time one finds
a differential drift, the magnitude of which depends on the 
latitude of the fluid element and the mode amplitude. 
Rezzolla et al study how this deduced 
differential drift affects the magnetic field of the star (assuming
that the field is frozen into the fluid elements). This leads to 
several suggestions. First of all, it is estimated that the 
$r$-mode oscillations cannot prevail if the star has a magnetic field
$\sim 10^{16}$~G or higher. 
In a star with a weaker magnetic field, $B\ge10^{10}$~G, the rotation
rate must be above $0.35 \Omega_K$ in order for the 
mode to survive. 
This means that the instability could
operate both in  young neutron stars ($B\sim 10^{12}$~G) and recycled ones
($B\sim10^8$~G) provided that they spin fast enough.
 Secondly, the differential drift due to the $r$-mode
twists the magnetic field and, just like in Spruit's model, this
affects the strength and nature of the field. Rezzolla et al
predict that, given an initial configuration with a strong poloidal
field and a much weaker azimuthal one, the magnetic field evolves
in such a way that after the $r$-mode spin-down the azimuthal
field has been strengthened to the point where it is 
4-8 orders of magnitude stronger than the (virtually unchanged) 
poloidal field. This would suggest that the $r$-mode instability 
generates strong azimuthal magnetic fields in young pulsars. 
This is an intriguing possibility, and it may be worthwhile
investigating whether it could be tested observationally.

These are interesting suggestions, but they come with a 
disclaimer. The derived differential drift is not determined consistently
 within perturbation theory. Rezzolla et al. use first order 
(in both $\alpha$ and
$\Omega$) results to deduce a nonlinear effect (of order $\alpha^2$). 
In absence of true nonlinear calculations it is not clear to what
extent the predicted drift will be present. 

Several other issues regarding the possible role of the magnetic field
are discussed in a recent study by Ho and Lai\cite{hl99}. 
First of all, they
discuss the effect of the standard magnetic  braking (that
dominates the spin-down in absence of the unstable $r$-modes). 
From the standard results one finds that magnetic braking proceeds
on a timescale
\be
t_{\rm B} \approx {6c^3 I \over B^2R^6\Omega^2} = 3\times 10^8 M_{1.4}
R_{10}^{-4}B_{12}^{-2}P_{-3}^2 \mbox{ s}  \ ,
\ee 
for a polytropic stellar model. Here $B_{12}=B/10^{12} \mbox{ G}$. By comparing this to the 
timescale of $r$-mode growth (\ref{gwest}) 
we see that magnetic braking will only 
be competitive for neutron stars with $B\sim 10^{15}$~G and higher.
In this context, Ho and Lai make an interesting observation. 
They note that, in absence of radiation, the canonical angular
momentum is conserved. This means that if we imagine a situation where
the star is spun down (or up) by some other agent, then the 
amplitude of the mode must change in order to keep $J_c$ 
constant. Hence, one would expect to find
$\alpha \sim 1/\sqrt{\Omega}$. This would not be important
under normal circumstances, but it could be highly relevant
if other spin-down torques dominate, eg.
for a young magnetar.

As we have already discussed in Section~3.1, a mode may be driven 
unstable by any radiative mechanism. Hence, it is interesting to 
try to quantify the extent to which electromagnetic radiation 
from the $r$-modes may affect the growth rate of the 
instability. Ho and Lai estimate this effect, and based on their
results we find 
\be
t_{\rm em} \approx - 4.8\times10^{10} M_{1.4} B_{12}^{-2} R_{10}^{-6}
P_{-3}^4 \mbox{ s} \ .
\label{emdrive}\ee 
Clearly, this effect would be relevant only for very strong
magnetic fields. Basically, the electromagnetic driving of
the mode would become competitive with  gravitational radiation for
$B\approx 10^{15}$~G, so again we deduce that the effect 
could be of importance for  magnetars. 
Finally, Ho and Lai also point out 
that the shaking of the magnetic field lines due to the
mode-oscillations
will generate Alfv\'en waves in the magnetosphere. These can also
drive the mode, and one can estimate that
\be
t_{\rm A} \approx - 2 \times 10^8 M_{1.4}R_{10}^{-2}B_{12}^{-2} \mbox{ s} \ .
\ee
Interestingly, this leads to a more significant driving 
of an unstable mode than direct electromagnetic radiation
(\ref{emdrive}).  

The above results suggest that the magnetic field may have great effect
on the $r$-modes in  magnetars, i.e. for stars with magnetic fields
above (say) $10^{14}$~G. For stars with weaker magnetic fields, 
such as recycled neutron stars in LMXBs the picture is not so clear. 
The outcome depends crucially on whether the $r$-mode leads to 
differential rotation/drift in the fluid or not. 
There is also the obvious caveat regarding neutron stars
with a solid crust. It could be that the magnetic field is
essentially frozen into the crust and that waves
in the fluid core do not significantly penetrate the crust. If this is
the case, the magnetic field may not have a large effect
on the $r$-modes at all. On the other hand, it is expected that the core
fluid will be electromagnetically locked into corotation with the crust
on a very short time-scale. This may then prevent oscillations in the 
core entirely.
Much further
work into the interplay between neutron star pulsation and the
magnetic field is needed before we can claim to have even a 
rudimentary understanding of this problem.

\subsection{Nonlinear effects}

It is clear that most of the results we have discussed in this article
may be strongly affected by nonlinear effects. Nonlinear
stellar pulsation theory is a relatively unexplored research area
that provides many conceptual and technical challenges.
Still, if we want to understand the actual evolution of an
unstable pulsation mode and possible effects it may have
on the bulk rotation of a star
we must explore the non-linear regime.

In the case of the $r$-mode scenarios discussed in Section~5, one main 
unknown parameter
 is the saturation amplitude $\alpha_s$. The values we
used in our discussion ($\alpha_s=0.01-1$) were chosen ad hoc, by assuming
that the $r$-mode would be able to grow to a ``large'' amplitude. But it
is not at all clear that this will be possible. Several nonlinear 
effects may set in at considerable smaller values of $\alpha$ and 
prevent further mode-growth. If this is the case, the various astrophysical
scenarios must be altered accordingly. Some candidates for 
nonlinear mode saturation are: i) coupling to other pulsation modes, that could lead to a cascade of energy from the unstable mode into other modes, ii)
turbulence, which is likely to play a role since the Reynolds number is very high in the $r$-mode problem, and iii) nonlinear effects that 
change the nature of the mode
in such a way that, for example, shocks develop.
At the present time, the only available results regarding the nonlinear 
mode saturation follow from an estimate of turbulence that arises in the 
Ekman layer at the base of the crust\cite{crust3}. 
These estimates suggest that the shear motion in the Ekman layer
becomes turbulent when the mode reaches an amplitude\footnote{It should be 
noted that the turbulence estimates are based on experimental data for
the flow of water above rugged surfaces. It is not clear to what extent 
such results
will be relevant for neutron star oscillations. One should also note 
that it is assumed that one can meaningfully use the standard $r$-mode
gravitational wave estimates even in the turbulent case. This corresponds 
to assuming that the turbulence is confined to the Ekman layer, which may 
be dubious since turbulence is by its very nature convective. }
\be
\alpha_c \approx 1.6\times10^{-3} \Delta^{-1} T_8^{-1}P_{-3}^{1/2}
\ee 
where $\Delta= |\Delta v |/|v|$ as in Section~4.5. We recall that a
typical value might be\cite{crust2} $\Delta \approx 0.1$. Hence 
turbulence would play a role provided that the mode can grow to 
an amplitude $\alpha \approx 0.1$. Since the turbulent dissipation 
rate increases as the cube of the amplitude while the gravitational-radiation
growth scales as $\alpha^2$ one can estimate the the mode will saturate
when\cite{crust3} 
\be
\alpha_s \approx 3.5 \times10^{-3} \Delta^{-3} P_{-3}^{-5} \left(
{5 \times10^{-3} \over C_D} \right)
\ee
where $C_D$ is the relevant drag coefficient. From this result we
can deduce that turbulence may saturate the $r$-mode growth at 
amplitudes
significantly below unity provided that $\Delta$ is relatively large. 
But if $\Delta \approx 0.1$ we get $\alpha_s \approx 3.5$ and
it seems likely that other nonlinear
mechanisms will provide the mode-saturation.  

A second issue of utmost importance regards the spin-evolution of the star
during  the instability phase, i.e. the back-reaction of the mode
onto the bulk rotation of the fluid. In the phenomenological 
spin-evolution model discussed in Section~5.1 it was assumed that, once
the $r$-mode had reached the saturation amplitude, all the excess
angular momentum was drained from the bulk rotation. That this will actually
be the case is far from clear, and would in fact seem rather unlikely. 
Nonlinear effects may well cause differential rotation in the star
(or a secular drift as in the study of Rezzolla et al\cite{rez1}), and if 
this is the case our current modelling will be inadequate in many ways.
A first attempt to gain insight into the non-linear back-reaction 
has been made by Levin and Ushomirsky\cite{lu99}. 
The considered the toy-problem 
of an $r$-mode in a shell, and find that the $r$-modes do indeed generate
differential rotation in the star.
 
It seems to us that we need numerical studies using
fully nonlinear hydrodynamics to investigate these
difficult issues. This is obviously a very challenging
task. Ideally, one would want to study the onset of instability 
and follow the mode through to saturation and then see what effect
the instability has on the spin of the star. Computationally, 
this means that one would need a numerical evolution that resolves
the mode-oscillation (on a millisecond timescale)
and tracks the star through hundreds of seconds. 
This is certainly not possible given current technology. As we
understand it, the best one can hope for at present is 
to study the star over (say) a few tens of rotation periods.
This will not allow a complete study of the problem, but it may
still provide answers to many crucial questions. 
 Efforts in this direction are underway, based on either
direct numerical evolution of the general-relativistic
hydrodynamical equations on a fixed background spacetime\cite{sfk,sf2000} 
or as an integration of the Newtonian hydrodynamical
equations and the use of an "accelerated" gravitational radiation
reaction force\cite{rez2}. As we were finishing this review 
the first results of these efforts became available. 
Simulations due to Stergioulas and Font indicate that the $r$-modes
may be able to grow to a surprisingly large amplitude.
In fact, there are no significant signs of mode-saturation until at 
unrealistically large amplitudes. These results are very promising, 
and it seems likely that continued efforts in this direction
will provide important insights into the nonlinear physics
of the $r$-mode instability. 

\section{Concluding remarks} 

In this review we have discussed the $r$-modes in rotating 
neutron stars and the associated gravitational-wave driven 
instability. This is a research area that has  attracted
considerable interest following the prediction that hot young
neutron stars may be spun down to a level comparable to 
that extrapolated from observations and the suggestion that 
the associated gravitational waves may be detectable 
with the new generation of interferometers.

Our aim in writing this review was to summarize most of the
ideas and suggestions that have been made regarding 
the $r$-mode instability and put them in
the appropriate context.
In doing this we obviously had to make sacrifices, but hopefully
we have managed to provide a useful introduction to the 
many relevant issues. What should be clear
is that our present understanding of  
instabilities in rotating neutron stars and their potential 
astrophysical relevance  is unsatisfactory in many respects.
All current suggestions are based on simplified, often phenomenological, models. 
The hope must be that 
future investment in this field will lead to a truly quantitative
 description of
``realistic'' neutron stars, and that 
the various proposed astrophysical scenarios will be described
in detail rather than  
``order of magnitude''. We have far to go before we reach
this goal, and it may well be that actual observations will 
beat us theorists to the answers of the many challenging questions
that we are just beginning to formulate. 

\nonumsection{Acknowledgements}
\noindent
We would like to thank many of our colleagues for stimulating 
discussions of the issues described in this review.
In particular,  we a grateful to John Friedman,
Ian Jones, Luciano Rezzolla, Bernard Schutz
and Nick Stergioulas for insightful comments.
This work was partly supported by PPARC grant PPA/G/1998/00606 to NA.

\nonumsection{References}
\noindent

\def\prl#1#2#3{{ Phys. Rev. Lett.\ }, {\bf #1}, #2 (#3)}
\def\prd#1#2#3{{ Phys. Rev. D}, {\bf #1}, #2 (#3)}
\def\plb#1#2#3{{ Phys. Lett. B}, {\bf #1}, #2 (#3)}
\def\prep#1#2#3{{ Phys. Reports}, {\bf #1}, #2 (#3)}
\def\phys#1#2#3{{ Physica}, {\bf #1}, #2 (#3)}
\def\jcp#1#2#3{{ J. Comput. Phys.}, {\bf #1}, #2 (#3)}
\def\jmp#1#2#3{{ J. Math. Phys.}, {\bf #1}, #2 (#3)}
\def\cpr#1#2#3{{ Computer Phys. Rept.}, {\bf #1}, #2 (#3)}
\def\cqg#1#2#3{{ Class. Quantum Grav.}, {\bf #1}, #2 (#3)}
\def\cma#1#2#3{{ Computers Math. Applic.}, {\bf #1}, #2 (#3)}
\def\mc#1#2#3{{ Math. Compt.}, {\bf #1}, #2 (#3)}
\def\apj#1#2#3{{ Astrophys. J.}, {\bf #1}, #2 (#3)}
\def\apjl#1#2#3{{ Astrophys. J. Lett.}, {\bf #1}, #2 (#3)}

\def\apjs#1#2#3{{ Astrophys. J. Suppl.}, {\bf #1}, #2 (#3)}
\def\acta#1#2#3{{ Acta Astronomica}, {\bf #1}, #2 (#3)}
\def\sa#1#2#3{{ Sov. Astro.}, {\bf #1}, #2 (#3)}
\def\sia#1#2#3{{ SIAM J. Sci. Statist. Comput.}, {\bf #1}, #2 (#3)}
\def\aa#1#2#3{{ Astron. Astrophys.}, {\bf #1}, #2 (#3)}
\def\apss#1#2#3{{Astrop. Sp. Sci.}, {\bf #1}, #2 (#3)}
\def\mnras#1#2#3{{ Mon. Not. R. Astr. Soc.}, {\bf #1}, #2 (#3)}
\def\prsla#1#2#3{{ Proc. R. Soc. London, Ser. A}, {\bf #1}, #2 (#3)}
\def\ijmpc#1#2#3{{ I.J.M.P.} C {\bf #1}, #2 (#3)}

\end{document}